\documentclass[preprint,12pt]{elsarticle}

\usepackage{graphicx}
\usepackage{amssymb}

\usepackage{aas_macros}
\usepackage{amsmath}

\usepackage{lineno}

\usepackage{url}

\journal{Advances in Space Research}

\begin{document}

\begin{frontmatter}


\title{Neutrinos and Cosmic Rays Observed by IceCube}



\author[Adelaide]{M.~G.~Aartsen} 
\author[Zeuthen]{M.~Ackermann\corref{CorrAuthor}} 
\ead{markus.ackermann@desy.de}
\cortext[CorrAuthor]{Corresponding author:}
\author[Christchurch]{J.~Adams} 
\author[BrusselsLibre]{J.~A.~Aguilar} 
\author[MadisonPAC]{M.~Ahlers} 
\author[StockholmOKC]{M.~Ahrens} 
\author[Geneva]{I.~Al~Samarai} 
\author[Erlangen]{D.~Altmann} 
\author[Marquette]{K.~Andeen} 
\author[PennPhys]{T.~Anderson} 
\author[BrusselsLibre]{I.~Ansseau} 
\author[Erlangen]{G.~Anton} 
% ...
\address[Adelaide]{Department of Physics, University of Adelaide, Adelaide, 5005, Australia}
% ...
\address[BrusselsLibre]{Universit\'e Libre de Bruxelles, Science Faculty CP230, B-1050 Brussels, Belgium}
% ...
\address[Christchurch]{Dept.~of Physics and Astronomy, University of Canterbury, Private Bag 4800, Christchurch, New Zealand}
% ...
\address[Erlangen]{Erlangen Centre for Astroparticle Physics, Friedrich-Alexander-Universit\"at Erlangen-N\"urnberg, D-91058 Erlangen, Germany}
\address[Geneva]{D\'epartement de physique nucl\'eaire et corpusculaire, Universit\'e de Gen\`eve, CH-1211 Gen\`eve, Switzerland}
% ...
\address[MadisonPAC]{Dept.~of Physics and Wisconsin IceCube Particle Astrophysics Center, University of Wisconsin, Madison, WI 53706, USA}
% ...
\address[Marquette]{Department of Physics, Marquette University, Milwaukee, WI, 53201, USA}
% ...
\address[StockholmOKC]{Oskar Klein Centre and Dept.~of Physics, Stockholm University, SE-10691 Stockholm, Sweden}
% ...
\address[PennPhys]{Dept.~of Physics, Pennsylvania State University, University Park, PA 16802, USA}
% ...
\address[Zeuthen]{DESY, D-15735 Zeuthen, Germany}

\begin{abstract}
The core mission of the IceCube Neutrino observatory is to study the origin and propagation of cosmic rays. IceCube, with its surface component IceTop, observes multiple signatures to accomplish this mission. Most important are the astrophysical neutrinos that are produced in interactions of cosmic rays, close to their sources and in interstellar space. IceCube is the first instrument that measures the properties of this astrophysical neutrino flux, and constrains its origin. In addition, the spectrum, composition and anisotropy of the local cosmic-ray flux are obtained from measurements of atmospheric muons and showers. Here we provide an overview of recent findings from the analysis of IceCube data, and their implications on our understanding of cosmic rays.


\end{abstract}


\end{frontmatter}

\vspace{1cm}


\section{Introduction}
\label{S:Intro}

\label{S:1}

The first detection of high-energy neutrinos of cosmic origin in 2013 by the IceCube Neutrino Observatory \cite{Aartsen:2013jdh} opened a new window to the non-thermal processes in our universe. Neutrinos interact only weakly with matter, and can escape energetic and dense astrophysical environments that are opaque to electromagnetic radiation. Moreover, at PeV energies, most of the universe is impenetrable to electromagnetic radiation, due to the scattering of high-energy photons ($\gamma$~rays) on the cosmic microwave background and other radiation fields. Neutrinos therefore promise to provide unique insights into a large number of extreme astrophysical phenomena, ranging from stellar explosions to the accretion onto massive black holes. They are key messengers in the search for the origin of the highest energy cosmic rays (CRs). High-energy neutrinos may be produced through the interaction of CRs with ambient matter or radiation fields.  Unlike the charged CRs they are neither deflected by magnetic fields, nor affected by matter or radiation fields on the way from the source to the Earth. They propagate undisturbed over cosmic distances, allowing us to observe an otherwise opaque high-energy universe and identify the sources in it.

\begin{figure}[h]
\centering\includegraphics[width=0.6\textwidth]{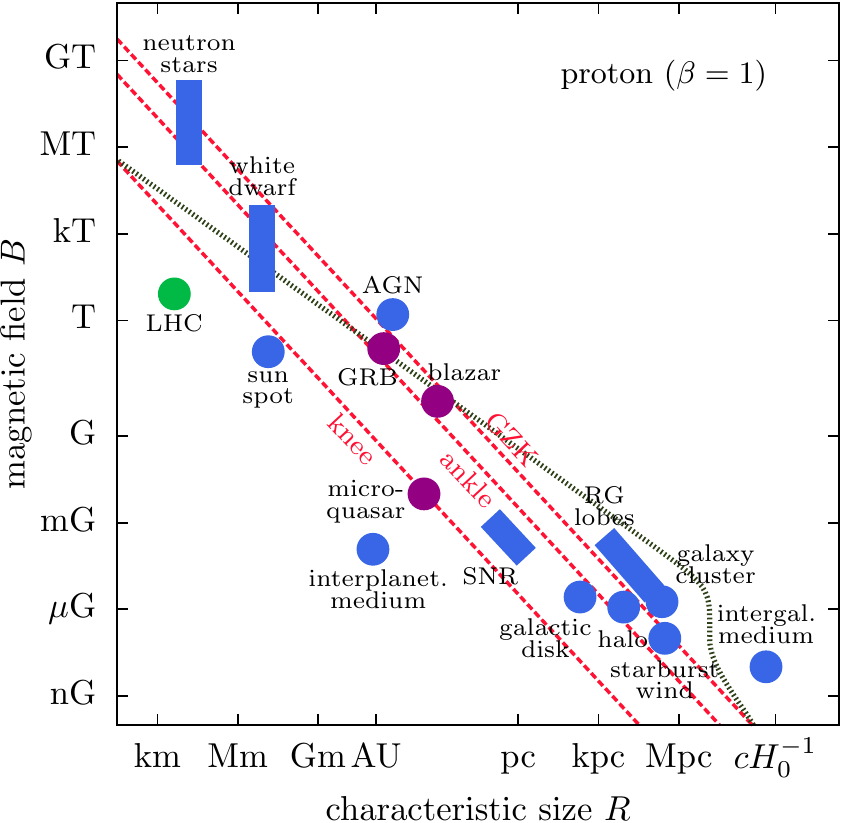}
\caption{A modern adaption of the so called ,,Hillas plot'' from \cite{Ahlers:Fermilab}. It displays upper limits on the reachable CR energy dependent on the size of the acceleration region and magnetic field strength. The red lines indicate the upper limits due to the loss of confinement in the acceleration region for CRs at the {\it knee}, {\it ankle}, and the {\it Greisen-Zatsepin-Kuzmin (GZK)} cutoff \cite{Greisen:1966jv,Zatsepin:1966jv}. The dotted gray line corresponds to a second upper limit that arises from synchrotron losses in the sources and interactions in the cosmic photon background.}
\label{fig:hillas}
\end{figure}

Many candidate source classes exist that fulfill the basic requirements of accelerating CRs to the highest observed energies of about 10$^{20}$~eV. An upper limit on the reachable cosmic-ray (CR) energy in gradual acceleration processes, like e.g. Fermi acceleration, was noted by Hillas in \cite{Hillas:1984}. Here, the size of the acceleration region has to be larger than the Larmor radius of the produced CRs, otherwise the particles are not confined for further acceleration. This notion led to the plot shown, in a modern adaptation, in Figure \ref{fig:hillas}. The potential sources of ultra-high-energy CRs are many, including gamma-ray bursts (GRBs, e.g. \cite{Waxman:1997ti}), young neutron stars and pulsars (e.g. \cite{Fang:2012rx}), the jets (e.g. \cite{Mannheim:1993}) and cores of active Galaxies (e.g. \cite{Stecker:1991}), and galaxy merger shocks in clusters (e.g. \cite{Kashiyama:2014}).

Cosmic neutrinos detected by IceCube can be used to probe the particle acceleration processes in these candidate source classes. Information can be deduced from the observed spectrum, the flavor composition and possible correlations of neutrino observations with known transients or sources. However, IceCube is more than ``just'' a cosmic neutrino detector. Using the surface array IceTop, and the thousands of muons from CR showers in the atmosphere that are registered every second in the in-ice array, IceCube can be used to study the spectrum, the composition and the isotropy of CRs that arrive at Earth at TeV and PeV energies. 
At PeV energies a transition in the CR spectrum and composition has been observed by many instruments (see review in \cite{Olive:2016xmw}). This so called  ``knee'', is commonly attributed to Galactic sources being unable to accelerate CRs to energies above a few PeV per nucleon. Consequently, the composition of CRs changes at PeV energies, being dominated by increasingly heavier nuclei as the energy increases.
The high statistics available in IceCube and the unique combination of a measurement of the electromagnetic and high-energy muon component of a CR air shower enable a precise measurement of both, spectral features and composition changes, in this energy range. 

Even though CRs at TeV to PeV energies are efficiently deflected in the Galactic magnetic fields, the observation of small anisotropies in their arrival directions can give important clues to the existence and location of CR sources in our Galactic neighborhood. Such anisotropies have been observed by several instruments on the northern hemisphere~\cite{Tibet:2005jun, Tibet:2006oct, 
SuperK:2007mar, Milagro:2009jun, MINOS:2011icrc, Bartoli:2013}.  IceCube data now provide the most accurate measurement of this anisotropy in the Southern hemisphere at TeV and PeV energies, completing our picture of the arrival patterns of CRs on the sky.  

After a short introduction to the IceCube neutrino telescope in section \ref{S:IceCube}, we summarize the findings and insights that have been obtained in the first five years of IceCube operation on the properties (section \ref{S:Spectrum}) and the origin of the cosmic neutrino flux (sections \ref{S:Sources} -- \ref{S:GalacticPlane}). In each section we will also discuss the implications of these findings for the sources of high-energy CRs. We then describe the CR spectrum and composition measurements in section \ref{S:IceTop}, and show and discuss recent results of the IceCube anisotropy measurement in section \ref{S:Anisotropy}, before concluding this review in section \ref{S:Conclusions}.

\section{The IceCube neutrino observatory}
\label{S:IceCube}

The IceCube Neutrino Observatory at the South Pole instruments approximately one cubic kilometer of the Antarctic ice sheet. It has been taking data in full configuration since spring 2011 with a duty cycle of more than 99\%. IceCube is more than an order of magnitude larger than any experiments operating in the North (Baikal Deep Under-water Neutrino Telescope \cite{Belolaptikov:1997ry}, Antares \cite{Ageron:2011nsa}). The planned KM3NeT and GVD detectors to be constructed in the Mediterranean sea and in the Lake Baikal in Siberia respectively, target a similar size as that of IceCube \cite{Adrian-Martinez:2016fdl, Avrorin:2014vca}. 

\begin{figure}[h]
\centering\includegraphics[width=\textwidth]{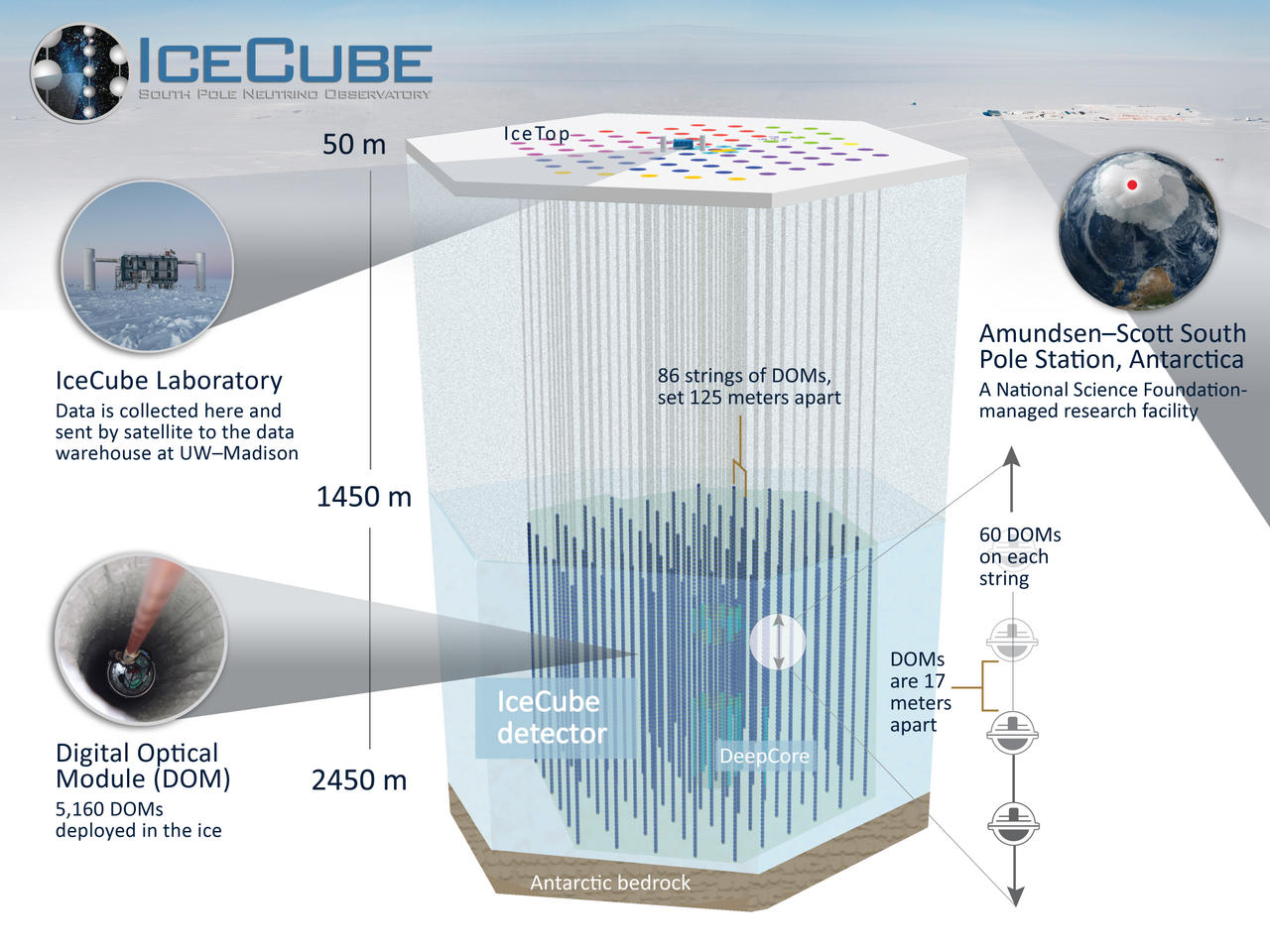}
\caption{The IceCube Neutrino Observatory is composed of the IceCube array, the surface array IceTop, and the low-energy sub-array called DeepCore. }\label{fig:icecube}
\end{figure}

IceCube consists of three components: the main IceCube array, the surface array IceTop, and a densely instrumented sub-array called DeepCore optimized for neutrinos with energies of a few tens of GeV. 
Optical sensors have been deployed at depths between 1450~m to 2450~m below the surface (see Figure~\ref{fig:icecube}). In total, 5160 digital optical modules (DOMs) are attached to 86 cables (“strings”) in a 3D hexagonal array
optimally arranged to detect the Cherenkov photons emitted by charged particles traversing the ice. 

All three components use the same instrumentation, design of DOMs and associated electronic readout \cite{Achterberg:2006md,Abbasi:2008aa,Aartsen:2016nxy}. 
The primary detector array is composed of 78 strings with a vertical separation of the DOMs of 17 m and an inter-string distance of about 125~m.
With this geometry, IceCube detects neutrinos from the entire sky with energies above 100~GeV.
Primary CRs interacting above the IceCube array, are detected with the CR air shower array IceTop that is operated in coincidence with the IceCube array \cite{IceCube:2012nn}. IceTop is composed of 162 water tanks filled with clear ice and arranged in pairs at stations on the surface.  Each station is $25$~m from the top of an IceCube string.
Finally, DOMs have also been deployed in the central and deeper part of the IceCube array forming DeepCore, a more densely instrumented volume that extends IceCube operation to the lower energy regime of 10~GeV \cite{Abbasi:2011ym}. Here, the vertical DOM-to-DOM spacing is 7 m and the inter-string spacing is between 72~m and 42~m.

IceCube records events at a rate ranging between 2.5~kHz and 2.9~kHz~\cite{Aartsen:2016nxy}. The overwhelming majority of these events are muons from CR air showers that penetrate the ice and reach the depth of IceCube. Only about one in a million recorded events is from a neutrino interaction. Yet, this rate is sufficient for the collection of an unprecedentedly large sample of high-energy neutrinos ($\sim10^5$ yr$^{-1}$, predominantly of atmospheric origin) that offer a unique testbed for extreme astro- and particle physics. 

Three main signatures can be distinguished for neutrino events in IceCube. ``Track-like'' events arise from muons produced in charged-current (CC) interactions of $\nu_{\mu}$. ``Shower-like'' events are generated in neutral-current (NC) interactions of all neutrino flavors, as well as in CC interactions of $\nu_{e}$ (all energies) and $\nu_{\tau}$ (E~$\leqslant$~100~TeV). High-energy $\nu_{\tau}$ can produce a specific identifying signature, the so called ``double-bang'' events~\cite{Learned:1994wg}. The hadronic shower at the $\tau$ generation vertex and the shower produced at the $\tau$ decay vertex can be separately identified when the tau track is longer than a few tens of meters ($\gamma c \tau_\tau\approx 50$~m at 1 PeV).

\begin{figure}[h]
\centering\includegraphics[width=0.385\textwidth]{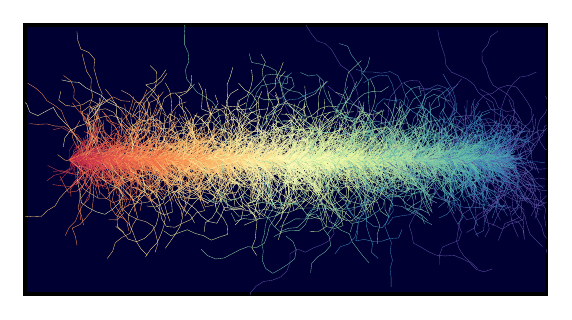}
\includegraphics[width=0.20\textwidth]{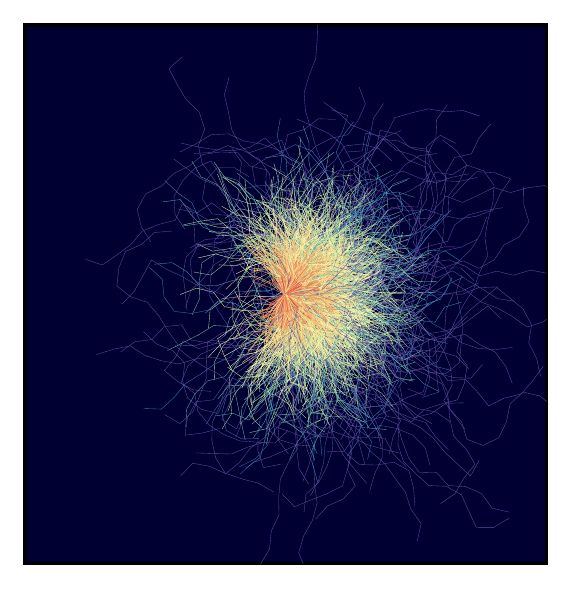}
\includegraphics[width=0.385\textwidth]{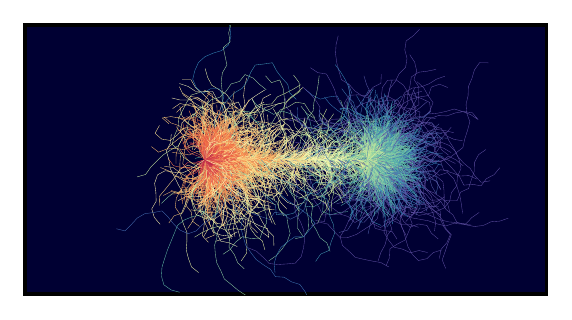}
\caption{Simulation of Cherenkov light propagation in the ice for the three event signatures observable by IceCube: a track-like event (left), a shower-like event (middle) and a double-bang event (right). Each track marks the path of a photon. The colors indicate the relative time of the photons with respect to each other. Early photons are red, late photons are blue.}\label{fig:eventsig}
\end{figure}

Figure \ref{fig:eventsig} shows the propagation of Cherenkov photons in a simulation of the Antarctic ice for each described signature.
Reconstruction of the physical properties of the neutrino that generated the  event --- direction, energy and flavor --- is challenging due to the complex optical properties of the natural medium \cite{SpicePaper}. Scattering and absorption of photons in the ice mainly arises from deposits of minerals, soot and ash over more than a hundred thousand years.
Therefore both, scattering and absorption lengths vary strongly with depth. Additionally, the flow of the antarctic ice shield introduces an anisotropy to the scattering. Melting and refreezing of the ice during DOM deployment locally changes the optical properties. In particular, for high energies above few tens of TeV, the reconstruction of event properties in IceCube is systematically limited due to these effects. Shower-like events can be reconstructed with an energy resolution of $\sim$~15\% \cite{Aartsen:ereco}, but the resolution of their arrival direction is rather poor at about 15$^{\circ}$.  On the other hand, the arrival direction of track-like events can be reconstructed with an accuracy better than 1$^{\circ}$, but the energy of the neutrino can only indirectly be inferred from the energy deposited in the instrumented volume. 

IceTop is located at an altitude of 2835 m above sea level, corresponding to an atmospheric overburden of 690~g~cm$^{-2}$. Each tank is instrumented with two DOMs operating
at different gains to provide a dynamic range from about 1/6 VEM (vertical equivalent muon\footnote{The energy deposited by a minimally ionizing muon vertically traversing the tank.}) to 1140 VEM. The IceTop surface array is triggered when six tanks in three stations register a signal in coincidence. The signal in the triggering tanks is typically dominated by the electromagnetic component of the air showers. For each trigger, both the surface detector and the in-ice signal are read out. IceTop has a small, central in-fill array with a threshold of about 100 TeV primary CR energy, while the regular-spaced array has a threshold of
300 TeV, recording air showers from primary CRs of energies up to about 2 EeV. Above this energy the rate becomes too low for analysis. The direction of events is reconstructed from the shower front arrival time, and has a resolution of $\sim$~0.3$^{\circ}$ at 10 PeV. The energy is determined by fitting the lateral shower profile and using the signal size at a perpendicular distance from the shower core of $125$~m. The resolution for protons at 30~PeV energy is 0.05 in log$_{10}$(E/GeV) \cite{IceCube:2012nn}.

\section{Spectrum and flavor composition of astrophysical neutrinos}
\label{S:Spectrum} 

\begin{figure}[th]
\centering\includegraphics[width=0.55\linewidth]{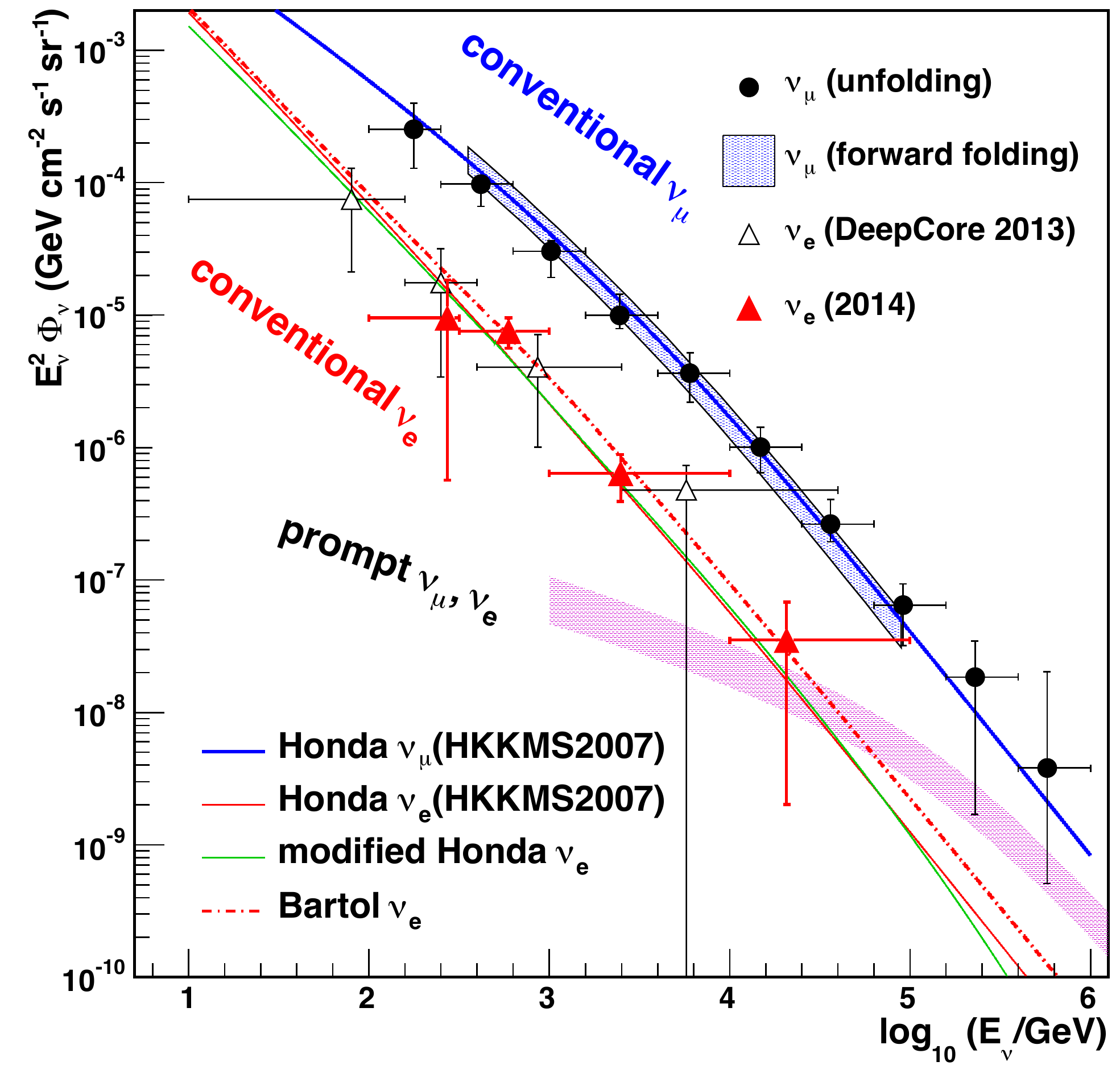}

\caption{Spectrum of atmospheric neutrinos measured by IceCube:   $\nu_\mu$ forward folding from~\cite{Abbasi:2011jx}; $\nu_\mu$ unfolding from~\cite{Aartsen:2014qna}; $\nu_e$ from cascades in DeepCore~\cite{Aartsen:2012uu}; $\nu_e$ with full IceCube~\cite{Aartsen:2015xup}.  The purple band
shows the prompt flux expected from Ref.~\cite{Enberg:2008jm} after modification to account for the knee of the primary spectrum.
}
\label{fig:AtmosNue}
\end{figure}

The majority of the neutrino events detected by IceCube are produced by CR interactions in the atmosphere.  Figure~\ref{fig:AtmosNue} from Ref.~\cite{Aartsen:2015xup} shows measurements of the flux of $\nu_e$ as well as the more numerous $\nu_\mu$.  The prompt component from decay of charmed mesons has not yet been detected. A hard astrophysical spectrum is expected to show up at high energy above the steeply falling atmospheric spectrum.

The first strong evidence for a cosmic neutrino component came from a search using data from May 2010 to April 2012~\cite{Aartsen:2013bka}, where two shower-like events from interactions within the detector with energies above $1$~PeV were discovered.  A follow-up search for events starting in the detector with more than $\simeq 30$~TeV deposited energy that utilized the same dataset identified 25 additional high-energy events~\cite{Aartsen:2013jdh}. The spectrum and zenith angle distribution of the events was incompatible with the hypothesis of an atmospheric origin at $>4\sigma$. 
IceCube has since collected independent evidence for an astrophysical neutrino signal by analyzing different event signatures as described below, including shower-like and starting events at lower energies as well as track-like events that interact outside the detector (called through-going events).


\subsection{Starting Events}

Neutrino interactions are identified in IceCube by searching for an interaction vertex within the instrumented volume. This search is sensitive to both shower-like and track-like events. Since the main background for this search is comprised of muons from CR air showers, the rejection strategy is to identify Cherenkov photons from a track entering the detector. For that, the outer parts of the instrumented volume are assigned to a ``veto'' region. An event is rejected if a certain number of Cherekov photons are found in this veto region at earlier times than the photons produced at the interaction vertex. For a more detailed description see \cite{Aartsen:2014gkd}.  Data recorded between May 2010 and Apr 2014 have been analyzed to obtain a starting event sample with an energy threshold of E$_{\nu}\sim$~30~TeV \cite{Aartsen:2015zva} including three shower-type events with energies in excess of 1 PeV. The first three years of this sample have also been used for an initial determination of the flavor ratio of astrophysical neutrinos \cite{Aartsen:2015ivb}.
If the size of the veto region is chosen to increase as energy decreases, neutrino-induced shower-like and track-like events above a few TeV can be isolated from the background effectively.  Using this approach, the starting event sample for two years (May 2010 to April 2012) has been extended to include lower-energy events down to E$_{\nu}\sim$~3~TeV \cite{Aartsen:2014muf}.

\subsection{Shower-type events}

An alternative strategy to distinguish neutrino-induced shower-type events from atmospheric backgrounds is to search for a spherical light pattern that fits the  characteristics of Cherenkov light emission from a short\footnote{For O(100~TeV) hadronic and electromagnetic showers there is only a few meters distance between interaction vertex and shower maximum in ice, which is small compared to the typical distance between strings of 125~m.} and well localized particle shower in or around the instrumented area. This allows identification of showers from neutrino interactions also in regions of the instrumented volume that serve vetoing purposes in the starting event searches and even to find showers nearby the instrumented volume. Data from May 2010 to Apr 2012 has been analyzed using this technique \cite{Aartsen:2015zvb} selecting 172 shower-type events above E$_{\nu}\sim$~10~TeV. Most of these events are not included in the previously described starting event sample of the same time period.

\subsection{Through-going muons}

Muons produced in CC neutrino interactions far outside the detector can still reach the instrumented volume to produce track-like events. Even at 1 TeV a muon can penetrate several kilometers of ice before it stops and decays. This allows observation of high-energy neutrino interactions from a much larger volume than the instrumented one, thereby substantially increasing the effective area of the detector.
However, these so called ``through-going'' muons from neutrino interactions are indistinguishable from single high-energy muons produced in atmospheric showers. For this reason the Earth must be used as a filter to separate neutrino-induced from CR-induced muons. Muons that arrive from zenith angles above $\sim 85^{\circ}$ must be produced in neutrino interactions, as muons produced in CR air showers could not penetrate far enough through the Earth and ice to reach the detector. Analyzing two years of IceCube data we found that the spectrum of neutrino-induced, upward muons shows a hardening above the steep atmospheric background consistent with an astrophysical flux~\cite{Aartsen:2015rwa}. The search for such muons has recently been extended to 6 years of IceCube data recorded between May 2009 and April 2015. The highest energy track found in this sample deposited 2.6~PeV of energy inside the volume of IceCube \cite{Aartsen:2016xlq}. The search for through-going muons is sensitive to cosmic neutrinos above an energy of about E$_{\nu}\sim$~200~TeV. At lower energies muons from the interactions of atmospheric $\nu_{\mu}$ dominate over the cosmic component.


\subsection{Showers from $\nu_{\tau}$ interactions}

A study to identify $\nu_{\tau}$ interactions was performed on the IceCube data recorded between May 2010 and Apr 2013, searching for a double pulse signature within single optical modules that would be characteristic of a double shower from the interaction of the $\nu_{\tau}$ and the decay of the $\tau$~\cite{Aartsen:2015dlt}. No such signature was found in 3 years of IceCube data, which is compatible with 0.54 expected events from simulations if cosmic neutrinos arrive at Earth with a flavor ratio of $\nu_{e}:\nu_{\mu}:\nu_{\tau} = 1:1:1$. While no $\nu_{\tau}$ signature was detected the analysis helped to constrain the measurement of the flavor ratio in combination with the observation channels introduced above.

\subsection{Combined results}

\begin{figure}[h]
\centering\includegraphics[width=0.55\linewidth]{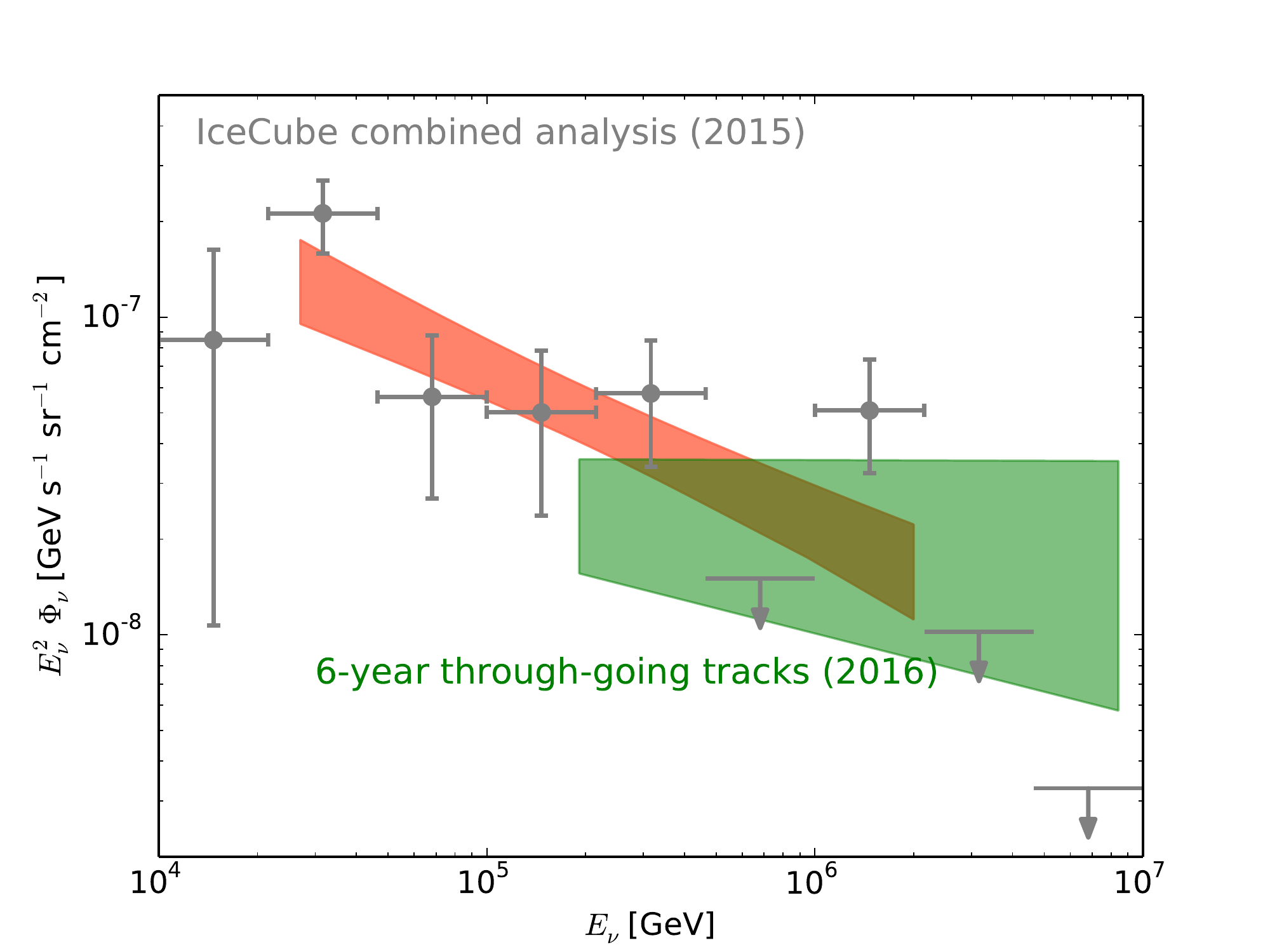}\includegraphics[width=0.45\linewidth]{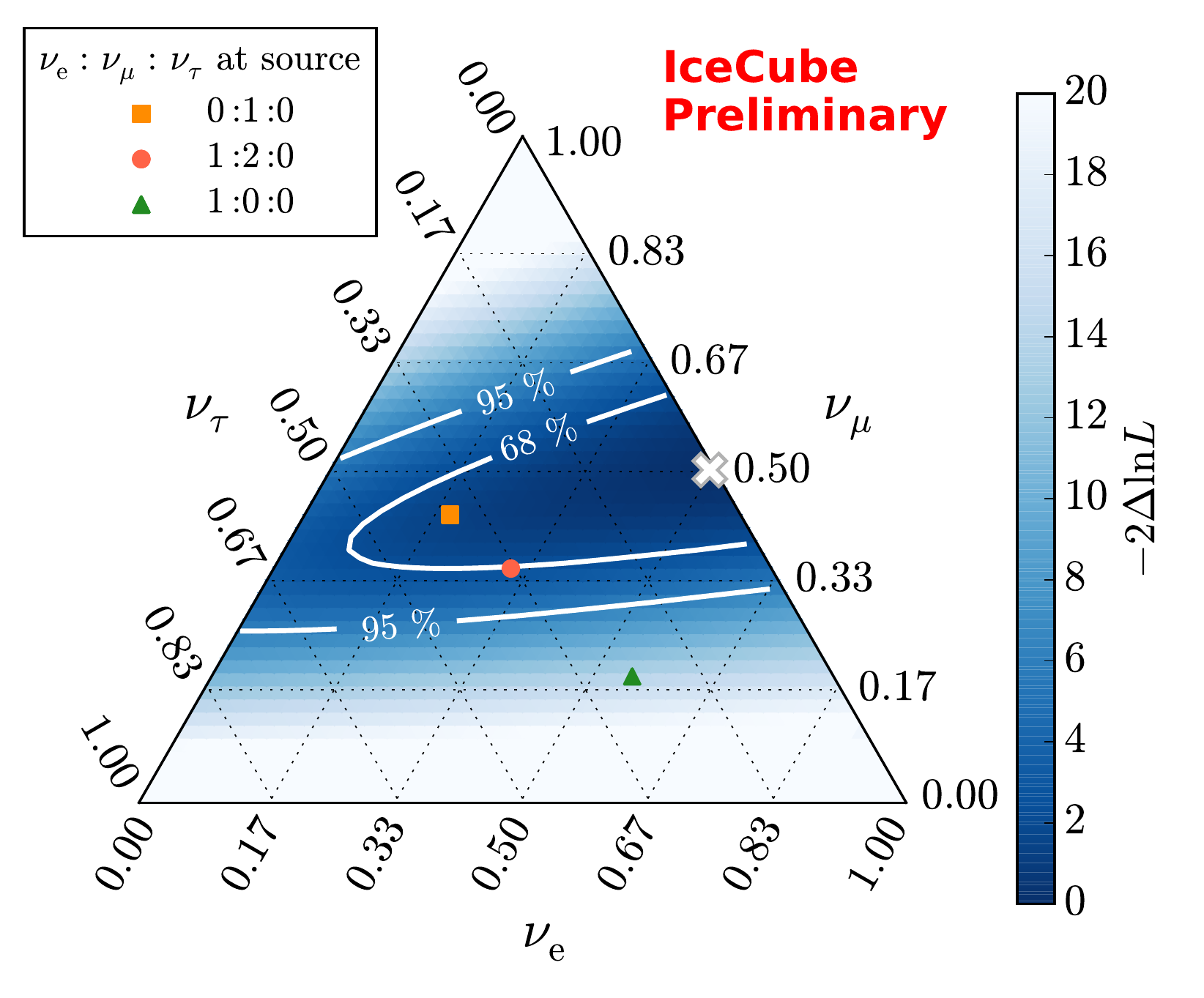}

\caption{(Left) Spectrum of cosmic neutrinos measured in a combined analysis of all detection channels. The red bar indicates the best fit with a power-law spectral hypothesis. The gray points display the result for a fit of the neutrino flux in individual energy bands. A new measurement based on 6 years of through-going muons (green bar) that is sensitive at higher energies indicates a harder spectrum above few hundred TeV. (Right) Flavor constraints on the cosmic neutrino flux from the combined analysis~\cite{Aartsen:2015zva3} in comparison to different scenarios expected for neutrino production in astrophysical sources.}
\label{fig:icecube-spectrum}
\end{figure}

The combined analysis of IceCube data of all the detection channels described above\footnote{Through-going muons are only included from the data taking periods between May 2009 and Apr 2012. An additional 3 years of through-going muons were analyzed only after the publication of the combined analysis. See also Figure \ref{fig:icecube-spectrum}.} results in a spectrum between 27~TeV and 2~PeV consistent with an unbroken power law with a best-fit spectral index of $-2.49 \pm 0.08$ \cite{Aartsen:2015knd,Aartsen:2015zva3}. A slightly improved likelihood is obtained if the data is fit with a harder spectrum with a spectral index of $-2.31$ and an exponential cutoff at 2.7~PeV. However, the improvement is not significant enough ($\sim 1.2 \sigma$) to claim the existence of such a cutoff, and both spectral models can describe the data reasonably well. The most recent analysis of high-energy muon tracks above 200~TeV shows a preferred spectral index of $-2.13 \pm 0.13$ \cite{Aartsen:2016xlq}. This result may be indicative of a spectral hardening (see Figure \ref{fig:icecube-spectrum} left) at high energies.

The energy flux of cosmic neutrinos above 10 TeV is $6.8 \times 10^{-10}$ ergs cm$^{-2}$ s$^{-1}$~sr$^{-1}$, based on the best fit power-law spectrum with exponential cutoff from the combined analysis. 
The spatial distribution of events on the sky is compatible with an isotropic distribution of sources, suggesting an extragalactic origin of a substantial fraction of the observed cosmic neutrinos. Using the combined analysis, their flavor ratio can also be constrained. Figure \ref{fig:icecube-spectrum} (right) shows the constraints on the relative contributions of the individual neutrino flavors to the cosmic neutrino flux. Typical astrophysical scenarios predict a flavor ratio at the production site of 
$\nu_{e}:\nu_{\mu}:\nu_{\tau} = 1:2:0$ in case the neutrinos are produced by the decay of pions. Standard neutrino oscillations change this into an expected ratio of  $\nu_{e}:\nu_{\mu}:\nu_{\tau} \approx 1:1:1$ on arrival at Earth. If the secondary muons lose most of their energy before they can decay, e.g. due to strong magnetic fields in the sources, the production flavor ratio would be $\nu_{e}:\nu_{\mu}:\nu_{\tau} = 0:1:0$ (,,muon-damped'' scenario). The opposite scenario is also possible, i.e. the muons are accelerated substantially before they decay, shifting the flavor ratio towards $\nu_{e}:\nu_{\mu}:\nu_{\tau} = 1:1:0$~\cite{Klein:2012ug}. In case the neutrinos are produced in the decay of neutrons a $\nu_{e}:\nu_{\mu}:\nu_{\tau} = 1:0:0$ flavor ratio would be expected. 
The neutron decay origin is excluded at more than $3\sigma$, while the other production scenarios mentioned above are compatible with current observations. 

%

\section{Neutrino sources}
\label{S:Sources}

\subsection{Search for individual neutrino sources}

In the case where the cosmic neutrino flux is dominated by bright indivdual sources, they should be detectable as a local excess of events on the sky with respect to the atmospheric neutrino and diffuse cosmic neutrino background. The sensitivity of a search for such features depends crucially on the precision by which the direction of the neutrinos can be reconstructed from the data, i.e. on the detector angular resolution.

Therefore, the best event signatures for this search are the through-going muons and track-like starting events with a median angular resolution of $\leq$~1$^{\circ}$.
The starting events are particularly important for the analysis of the Southern hemisphere sources where the strong background of muon bundles from CR air showers requires a very high energy threshold for the acceptance of through-going tracks.   

The most recent analysis~\cite{Aartsen:2016oji} combines seven years of IceCube data recorded between May 2008 to Apr 2015, corresponding to a livetime of 2431 days of through-going muons, and 1715 days of track-like starting events\footnote{Starting events were not available  before May 2010 while IceCube was under construction. The through-going muon sample contains data from the partially completed IceCube detector in its 40 and 59 string configurations.}. In total 422791 through-going muons from the Northern hemisphere, 289078 through-going muons from the Southern hemisphere and  961 starting tracks have been identified. The overwhelming majority of the muons from the North originate from atmospheric neutrinos, while most of the muons from the South arise from muons and muon bundles created in CR air showers.
The acceptance of background events vs. signal neutrinos has been optimized to achieve the optimal sensitivity for a detection for a range of potential point source spectra. The datasets are analyzed using a maximum likelihood technique to find one or more localized excesses over the diffuse backgrounds that correspond to the neutrino sources.

\begin{figure}[ht]
\centering\includegraphics[width=0.7\textwidth]{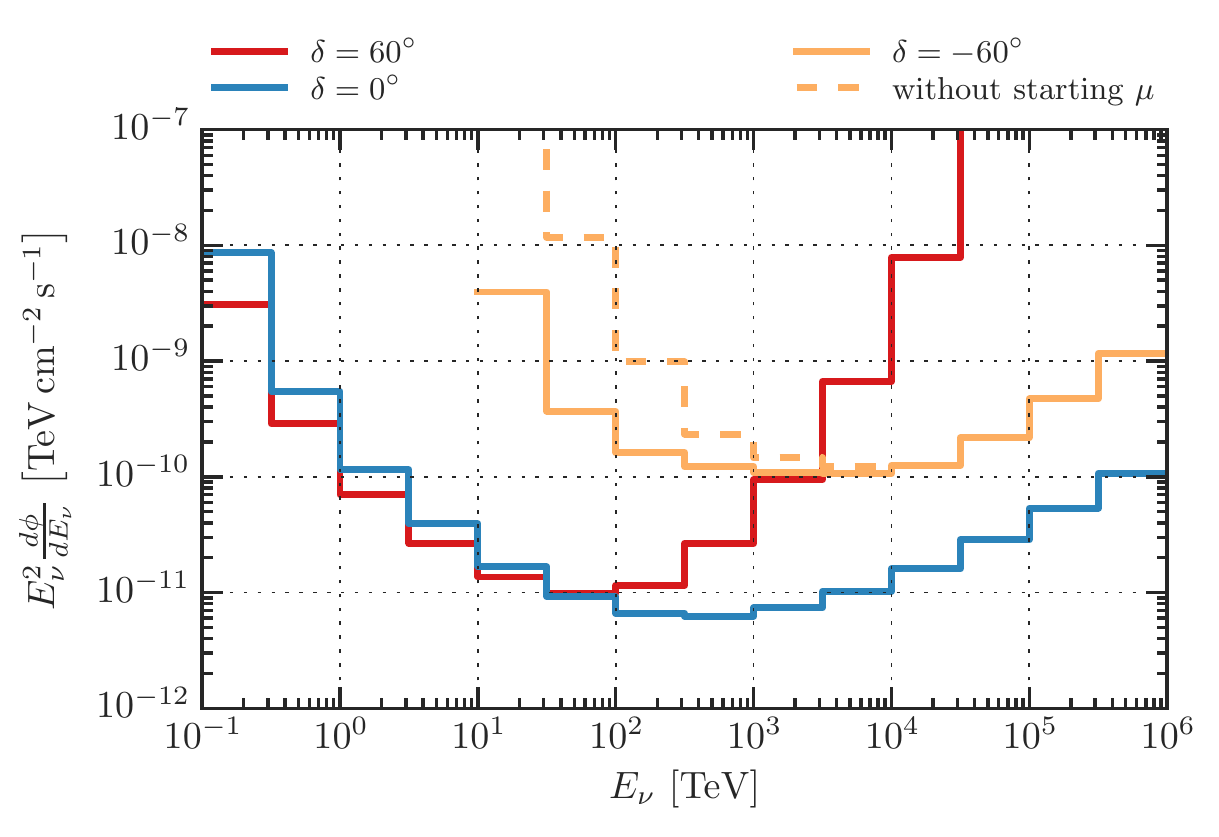}
\caption{Differential discovery potential of the point source search~\cite{Aartsen:2016oji} for various zenith angles. Shown is the neutrino flux from a point source over half a decade in energy that would lead to a $5\sigma$ discovery in the current search for 50\% of statistical realizations. The dashed line indicates the sensitivity if starting events would be ignored. A power-law spectrum with an index of 2 is assumed for the neutrino flux within a single energy bin.} 
\label{fig:PS_DiscPotential}
\end{figure}

Figure \ref{fig:PS_DiscPotential} presents the discovery potential for point sources at various declinations ($\delta$) achieved in this analysis. 
The most sensitive energy range changes with declination and is $> 1$~PeV for $\delta=-60^{\circ}$, between 100~TeV and 1~PeV at the horizon, and below 100~TeV at $\delta=60^{\circ}$. The generally lower discovery potential for sources at $\delta=-60^{\circ}$ is due to the limited overburden of ice above the detector which limits the amount of target material, and the high-energy threshold for accepting muons thus reducing the strong background from CR air showers. For sources at a declination of  $\delta=60^{\circ}$, neutrinos with energies $>$100~TeV are increasingly absorbed in the Earth, reducing the discovery potential at high energies. 

\subsection{Flux upper limits derived from IceCube data}
\label{sec:fluxlimits}
\begin{figure}[ht]
\centering\includegraphics[width=0.8\textwidth]{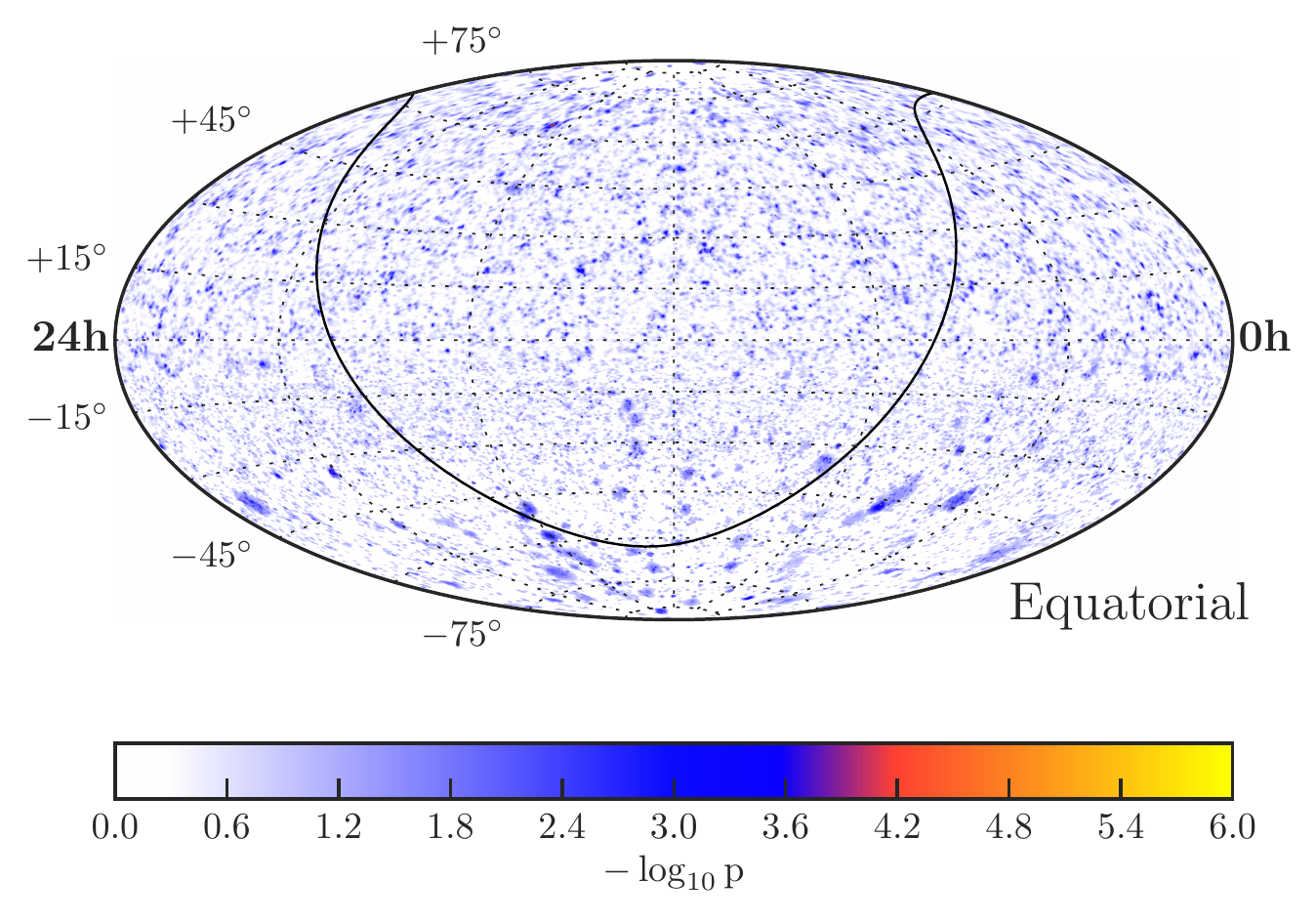}
\caption{Map of p-values representing the local probability that an excess of events at a given position in the sky is due to a fluctuation of the expected background \cite{Aartsen:2016oji}.}
\label{fig:PS_Result}
\end{figure}

No indication for a neutrino point source has been found in the IceCube data so far and Figure \ref{fig:PS_Result} summarizes the results of the search described above. The map shows the p-values for each point in the sky giving the local probability that an excess is a fluctuation of the background. To estimate the significance of the lowest observed p-values on each hemisphere, event samples have been generated with the right ascension coordinates randomized. These samples have been analyzed in the same way as the original dataset. The distribution of the minimum p-values in the samples can then be compared to that observed in the data. The fraction of randomized samples with a lower p-value than the lowest observed p-value in the data is 29\% for the Northern sky, and 17\% for the Southern sky. That is, the observations are compatible with fluctuations of the diffuse background.

\begin{figure}[ht]
\centering\includegraphics[width=0.7\textwidth]{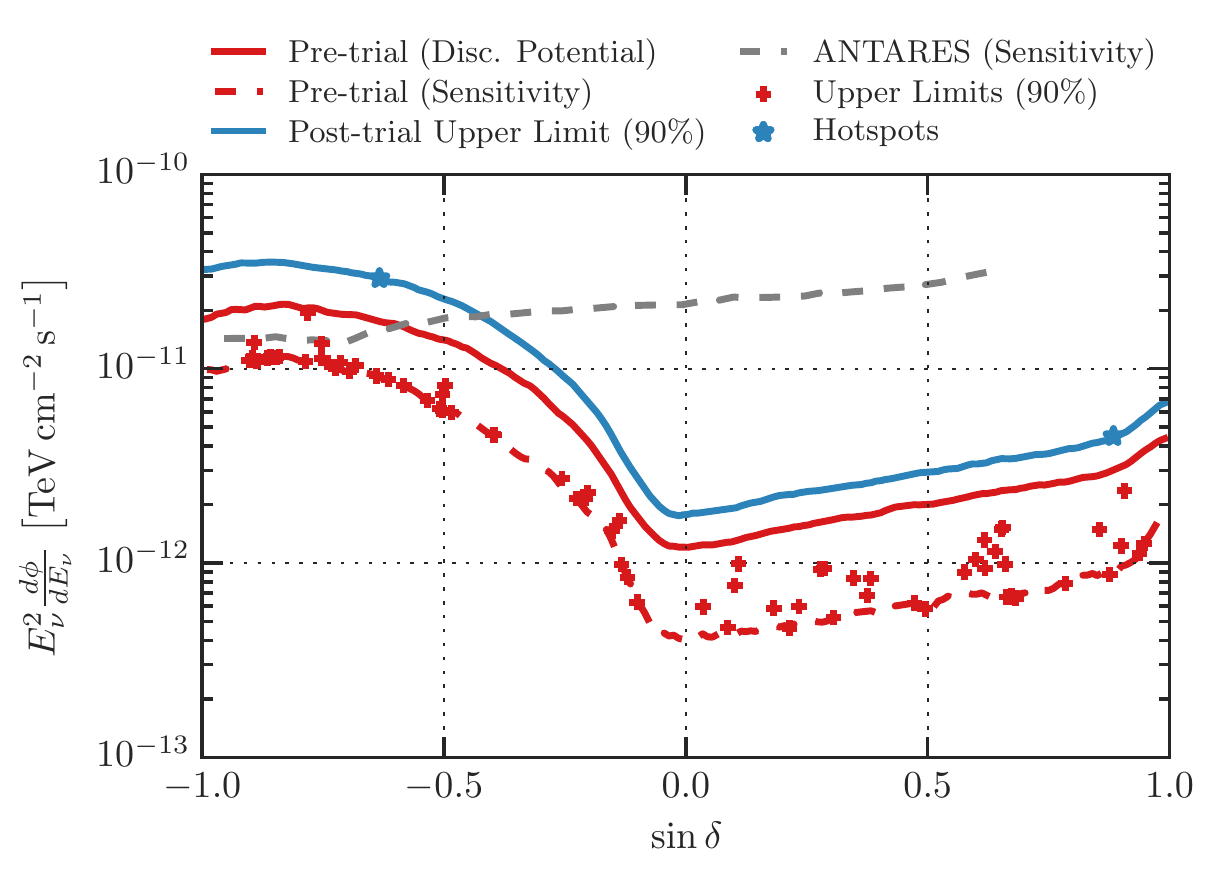}
\caption{Neutrino flux upper limits for various source candidates, sensitivities and discovery potential as a function of the source declination~\cite{Aartsen:2016oji}. The red dots indicate the 90\%~CL flux upper limits for individual candidate sources. The dashed red line represents the corresponding sensitivity at the respective declination.  The gray dashed line indicates the sensitivity of the Antares neutrino telescope \cite{Adrian-Martinez:2014wzf}. The blue line shows the flux upper limit that corresponds to the lowest observed p-value in each half of the sky as a function of declination (the actual declination of the observed spots is indicated by a star). A power-law spectrum with an index of 2 is assumed when generating the limits.}
\label{fig:PS_Limit}
\end{figure}

Additionally, the known locations of promising individual neutrino source candidates have been tested. These candidates have been selected based on model calculations and/or the observation of non-thermal emission features in the electromagnetic spectrum. None of the tested candidates shows a significant excess of neutrino events around its position.
In Figure \ref{fig:PS_Limit} the neutrino flux upper limits summarize the result from this non-observation. Also shown is the discovery potential, i.e. the flux that would lead to a $5\sigma$ discovery of a source in 50\% of the statistical representations (without any corrections for multiple trials). 


\begin{figure}[ht]
\centering\includegraphics[width=0.6\textwidth]{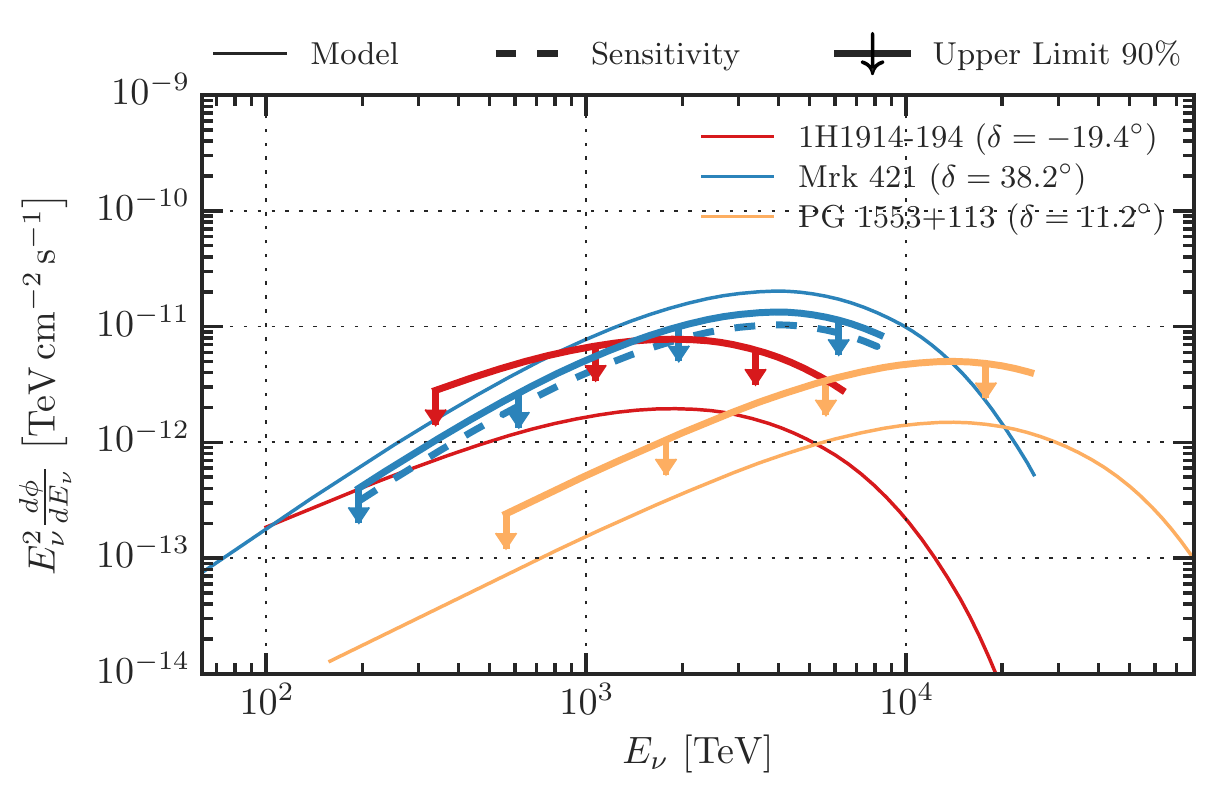}
\includegraphics[width=0.6\textwidth]{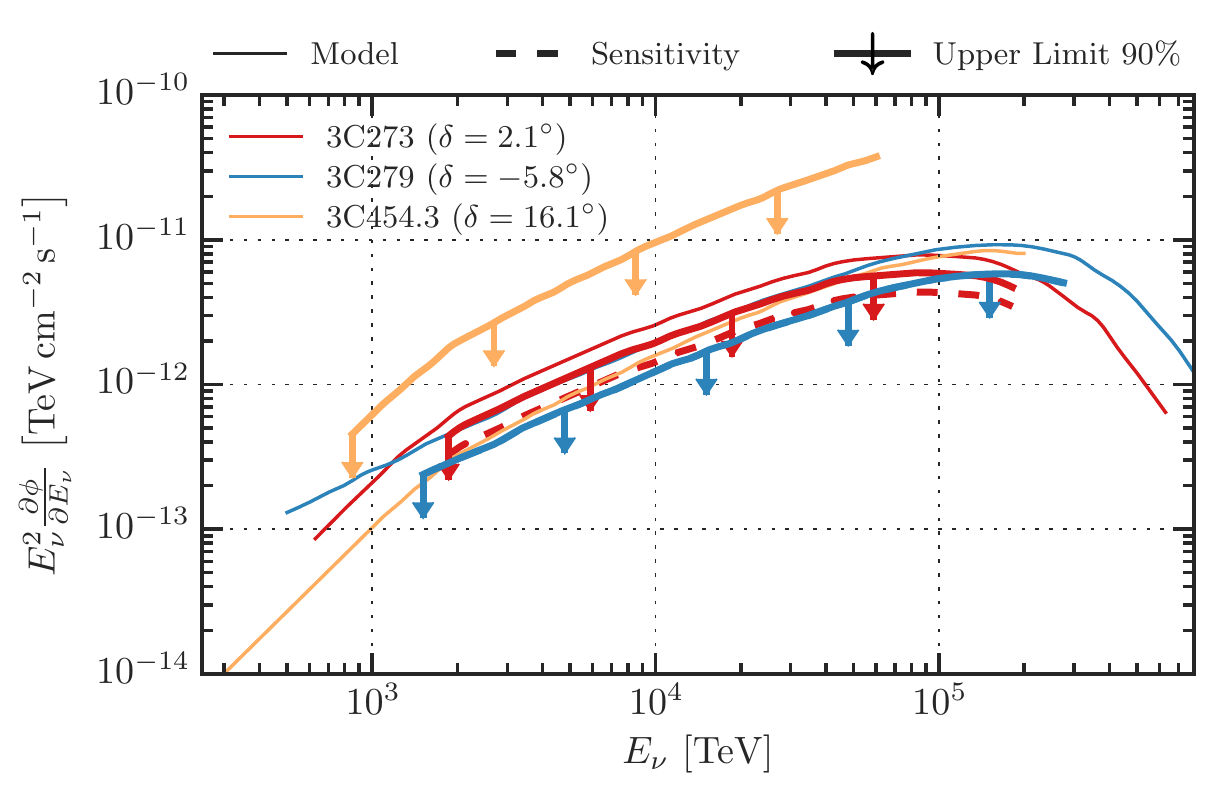}
\caption{Experimental upper limits on the neutrino flux~\cite{Aartsen:2016oji} in comparison to predictions for blazars in~\cite{Reimer:2015abc,Petropoulou:2015}.}\label{fig:PS_Interpretation}
\end{figure}

A comparison of the flux upper limits to a selection of individual source emission models is shown in Figure \ref{fig:PS_Interpretation}. The flux limits have to be calculated specifically for the predicted neutrino spectra based on the declination and energy dependent instrument response. The two panels show  examples of recent models of the neutrino emission from blazars \cite{Reimer:2015abc,Petropoulou:2015}. The predicted spectra are compared to the flux upper limits derived from IceCube data. For some of the sources the limits are on the level of the calculated flux and start to constrain the parameter space of such models. More details about the search presented above can be found in \cite{Aartsen:2016oji}. In addition, dedicated tests were performed to find transient sources \cite{Aartsen:2015wto} and sources that are spatially extended \cite{Aartsen:2014cva}\footnote{These searches so far did not use the full May 2008 -- Apr 2015 dataset described above.}, both yielding null results. 

\subsection{Constraints on astrophysical source populations}

The observation of an isotropic flux of astrophysical neutrinos seems to be at odds with the non-observation of individual neutrino point sources in the same data~
\cite{Aartsen:2016oji,Adrian-Martinez:2014wzf,Aartsen:2015wto,Aartsen:2013uuv,IceCube:2011ai,AdrianMartinez:2012rp,Abbasi:2012zw,Adrian-Martinez:2013dsk}.
However, the two results are consistent if the diffuse flux is dominated by many weak sources that are individually below the point source sensitivity~\cite{Lipari:2006uw,Becker:2006gi,Silvestri:2009xb}. This argument can be turned into a lower limit on the abundance of extragalactic neutrino sources, that we outline in the following.

The (quasi-)diffuse flux of neutrinos ($\phi$ in units of GeV$^{-1}$~s$^{-1}$ ~cm$^{-2}$~sr$^{-1}$) originating in multiple cosmic sources is simply given by the redshift integral~\cite{Ahlers:2014ioa}
\begin{equation}\label{eq:phinu}
\phi_\nu(E_\nu) = \frac{c}{4\pi}\int_0^\infty\frac{{\rm d}z}{H(z)}\mathcal{Q}_\nu(z,(1+z)E_\nu)\,.
\end{equation}
Here, $H(z)$ is the redshift dependent Hubble expansion rate and $\mathcal{Q}_\nu$ is the spectral emission rate density of neutrinos. To a first approximation, we decompose the emission rate density into $\mathcal{Q}(z,E) = \rho(z)Q_\nu(E)$ where $\rho$ is the source density and $Q_\nu$ is the emission rate per source. Note, this approximation assumes neutrino sources are standard candles and does not allow for luminosity distributions. However, these aspects can be included in a more detailed treatment.  

The Hubble expansion in the redshift integral of Eq.~(\ref{eq:phinu}) limits the contribution of sources beyond the Hubble horizon $c/H_0$. The redshift dependence of the source distribution can then be parametrized by the energy dependent quantity
\begin{equation}\label{eq:xiz}
\xi_z(E) = \int_0^\infty{\rm d}z\frac{H_0}{H(z)}\frac{\mathcal{Q}_\nu(z,(1+z)E)}{\mathcal{Q}_\nu(0,E)}\,,
\end{equation}
which is typically of $\mathcal{O}(1)$. For the special case of power-law spectra $\mathcal{Q}_\nu(E)\propto E^{-\gamma}$, this quantity becomes energy independent and, for simplicity, we will assume the case $\gamma=2$ in the following. For instance, $\xi_z\simeq2.4$ if we assume that the source evolution follows the star-formation rate (SFR)~\cite{Hopkins:2006bw,Yuksel:2008cu}
or $\xi_z\simeq0.5$ for a source distribution with no evolution in the local ($z<2$) Universe.

Based on the observed per-flavor diffuse flux at the level of $E^2\phi_\nu\simeq 10^{-8}$~GeV$^{-1}$~s$^{-1}$ ~cm$^{-2}$~sr$^{-1}$ we can then estimate the average neutrino point source luminosity via Eq.~(\ref{eq:phinu}) as $E^2Q_\nu(0,E) \simeq (4\pi H_0/c{\xi_z\rho_0})E^2\phi_\nu$.
On the other hand, for a homogeneous source distribution $\rho_0$ in the local Universe we expect that the brightest source contributes with a flux $E^2\phi^{\rm PS}_\nu \simeq 0.55(f_{\rm sky}\rho_0)^{2/3} E^2Q_\nu$, where $f_{\rm sky}\leq1$ is the effective sky coverage of the observatory (see Ref.~\cite{Ahlers:2014ioa} for details). This translates into a point source flux given by:

\begin{equation}\label{eq:brightest}
E^2\phi^{\rm PS}_\nu \simeq 1.9\times10^{-12} \left(\frac{\xi_z}{2.4}\right)^{-1}\left(\frac{f_{\rm sky}}{0.5}\right)^{\frac{2}{3}}\left(\frac{\rho_0}{10^{-8} {\rm Mpc}^{-3}}\right)^{-\frac{1}{3}}\frac{\rm TeV}{{\rm cm}^{2}\,{\rm s}}\,.
\end{equation}

\begin{figure}[h]
\centering\includegraphics[width=0.7\textwidth]{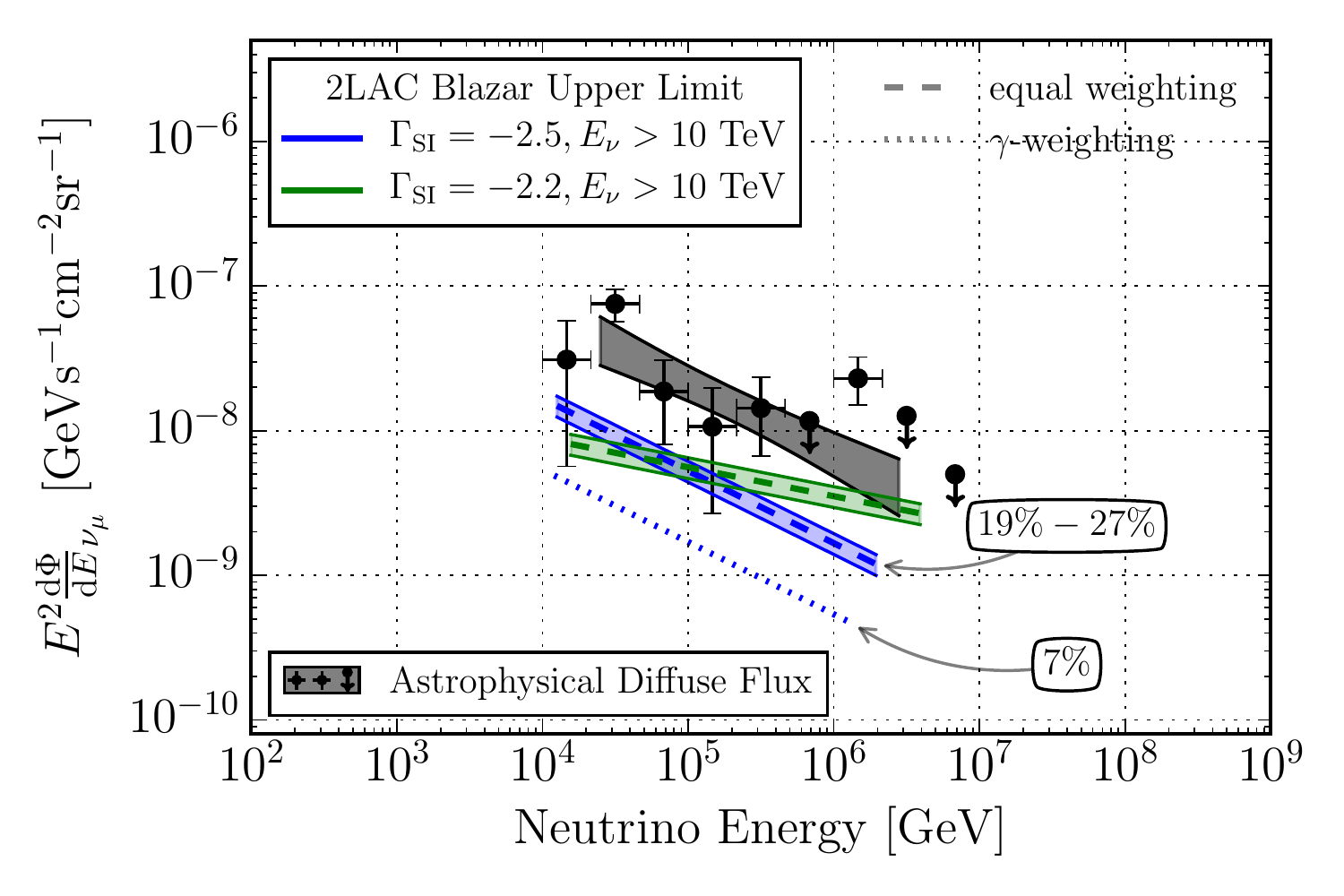}
\caption{Upper limit on the contribution of Fermi LAT observed blazars (2LAC catalog) to the cosmic neutrino flux, shown for two different power-law spectra for the neutrino flux with indices of 2.5 and 2.2, respectively~\cite{Aartsen:2016lir}. The width of the upper limit band indicates the dependence on the relative distribution of neutrino luminosities in the blazar sample if no strict proportionality is assumed between the $\gamma$-ray and neutrino luminosity of the source. The dotted line indicates the upper limit in case such proportionality is considered. }\label{fig:Blazar_Contribution}
\end{figure}

Presently, the sensitivity of IceCube to continuous point source emission in the Northern Hemisphere is at the level of $E^2\phi^{\rm PS}_{\nu_\mu+\bar\nu_\mu}\sim 10^{-12}$~TeV~cm$^{-2}$~s$^{-1}$ (see section \ref{sec:fluxlimits} and figure \ref{fig:PS_Limit}). This is already putting some tension on very rare source candidates like blazars ($\rho_0 \lesssim10^{-7}$~Mpc$^{-3}$). In fact, a dedicated IceCube analysis looking for the combined neutrino emission from Fermi~LAT identified blazars~\cite{Aartsen:2016lir} places an upper limit on their contribution that is at the level of about 25\% of the observed flux. Figure \ref{fig:Blazar_Contribution} shows the maximum contribution from blazars in the 2LAC catalog to the observed cosmic neutrino flux for two different spectral hypotheses. If additionally a strict proportionality is assumed between the emitted power at GeV energies and in TeV neutrinos, the 2FGL blazars can contribute less than 10\% to the observed flux (e.g. for sources for which the high-energy emission is dominated by pion-decay processes).

A similar line of argument can also be applied to transient sources~\cite{Ahlers:2014ioa}. Here, the experimental livetime does not increase the individual emission of transients, but the total size of the source sample with local burst density $\dot\rho_0$. For instance, the contribution of gamma-ray bursts ($\dot\rho_0 \simeq 10^{-9}$~Mpc$^{-3}$~yr$^{-1}$) to the diffuse emission is limited to less than 10\% due to IceCube's strong limit on the prompt neutrino emission of GRBs coincident with the gamma-ray signal~\cite{Abbasi:2012zw}.

\subsection{Neutrinos from the propagation of ultra-high energy CRs}

The CR spectrum extends to energies far above $10^{18}$~eV. These ultra-high energy (UHE) CRs are believed to be accelerated in extragalactic sources since Galactic magnetic fields are too weak to sufficiently confine the UHE CRs. Candidate sources include GRBs, active galactic nuclei or galaxy clusters (see Figure~\ref{fig:hillas}).


The propagation of UHE CRs over cosmic distances makes them susceptible to interactions with cosmic backgrounds. In particular, photo-pion production of CR nuclei on the cosmic microwave background (CMB) with a local density of about $410~{\rm cm}^{-3}$ becomes resonant at CR nucleon energies of about $7\times10^{11}$~GeV. This leads to a strong suppression of CR protons beyond an energy at about $E_{\rm GZK} \simeq 5\times10^{19}$~eV, which is known as the GZK suppression~\cite{Greisen:1966jv,Zatsepin:1966jv}. 

The neutrinos from the decaying pions are the so called {\it cosmogenic} or {\it GZK} neutrinos~\cite{Beresinsky:1969qj}. For proton-dominated UHE CR models, the expected flux peaks at EeV energies and is expected to be equally distributed between neutrino flavors after propagation~\cite{Yoshida:1993pt,Protheroe:1995ft,Engel:2001hd}. This flux is considered a {\it guaranteed} contribution to high-energy neutrino fluxes since it does not rely on specific neutrino production mechanisms in CR sources. However, even in the simplest case of proton-dominated models, the flux depends on the unknown UHE CR source redshift evolution function and maximal energy cutoff of the proton spectra. 

The largest contributions are predicted in proton models with a low energy cross-over between Galactic and extragalactic CRs, which typically require a strong redshift evolution of sources to fit the data~\cite{Berezinsky:2002nc,Fodor:2003ph,Yuksel:2006qb,Takami:2007pp}. However in this case the related production of high energy $\gamma$ rays, electrons and positrons predict a strong extragalactic diffuse $\gamma$-ray background~\cite{Berezinsky:2010xa,Ahlers:2010fw,Gelmini:2011kg,Decerprit:2011qe,Heinze:2015hhp,Supanitsky:2016gke} in excess of the observations with the Fermi LAT~\cite{Abdo:2010nz,Ackermann:2014usa}.

Proton-dominated UHE CR models generally 
are in reach of present neutrino observatories. However, the large experimental uncertainties on the relative contribution of heavier nuclei translates into large uncertainties in the cosmogenic neutrino predictions. The simple reason is that if the UHE CR spectrum is dominated by heavy nuclei with atomic mass number $A$, then the resonant interaction of CR nucleons with the CMB is shifted to higher CR energies, $(A/56) \times 4\times10^{13}$~GeV. For the extreme case of iron this would shift the required CR energies to a level beyond the observed CR spectrum. 

In the context of an increased threshold for GZK neutrino production by heavy nuclei, additional cosmic radiation backgrounds with higher photon energies can become a more important target. The extragalactic background light (EBL) in the infrared, optical and ultra-violet are included in most GZK neutrino predictions including heavy nuclei~\cite{Decerprit:2011qe,Hooper:2004jc,Ave:2004uj,Hooper:2006tn,Allard:2006mv,Anchordoqui:2007fi,Aloisio:2009sj,Kotera:2010yn,Ahlers:2011sd}. In general, these EBL neutrino predictions shift the peak neutrino production to the 1-10 PeV range but at an absolute level that is below present experimental sensitivities. As in the case of the proton dominated model, the cosmogenic neutrino prediction depends also on maximal energies and evolution of models. An estimate of a lower limit of these pessimistic models was given in Ref.~\cite{Ahlers:2012rz}.

The search for cosmogenic neutrinos is one of the standard analyses of IceCube.  At EeV energies, where the emission is expected to peak, there are practically no background events from atmospheric CR interactions. However, even after seven years of observation, no signal consistent with the expected emission spectra of various GZK models has been detected~\cite{Aartsen:2016ngq}. These limits now start to exclude some of the more optimistic scenarios of UHE CRs, dominated by light nuclei and/or strong source evolution (see also Ref.~\cite{Heinze:2015hhp}). In addition to constraining propagation, the non-observation of neutrinos with energies above a few PeV sets upper limits on the cumulative emission of neutrinos from astrophysical sources in the energy range between 10~PeV and 1~EeV. These limits are comparable to predictions from blazar \cite{Murase:2014,Padovani:2015mba} and Pulsar models \cite{Fang:2012rx}.

\section{Neutrino transients}
\label{S:Transients}

\subsection{Gamma-Ray Bursts}

GRBs are intense $\gamma$-ray flashes lasting from fractions of a second to tens of minutes. During their prompt emission phase they are the brightest explosions in the Universe reaching isotropic-equivalent energies of up to $10^{54}$\,ergs. They are likely powered by the core-collapse of a very massive star or the merger of two compact objects. Their locations are distributed isotropically and they have been measured up to a redshift $z=8$.
GRBs have been proposed as the sources of the highest-energy CRs~\cite{Waxman:1995vg}. The central engine produces highly relativistic collimated jets, which are predicted to host internal shocks, where particles are efficiently accelerated to high energies. In hadronic scenarios, accelerated protons interact with ambient synchrotron photons and produce high-energy neutrinos. The neutrino emission is expected to be collimated and in temporal coincidence with the prompt $\gamma$-ray emission. 

\begin{figure}[t]
\centering\includegraphics[width=0.7\textwidth]{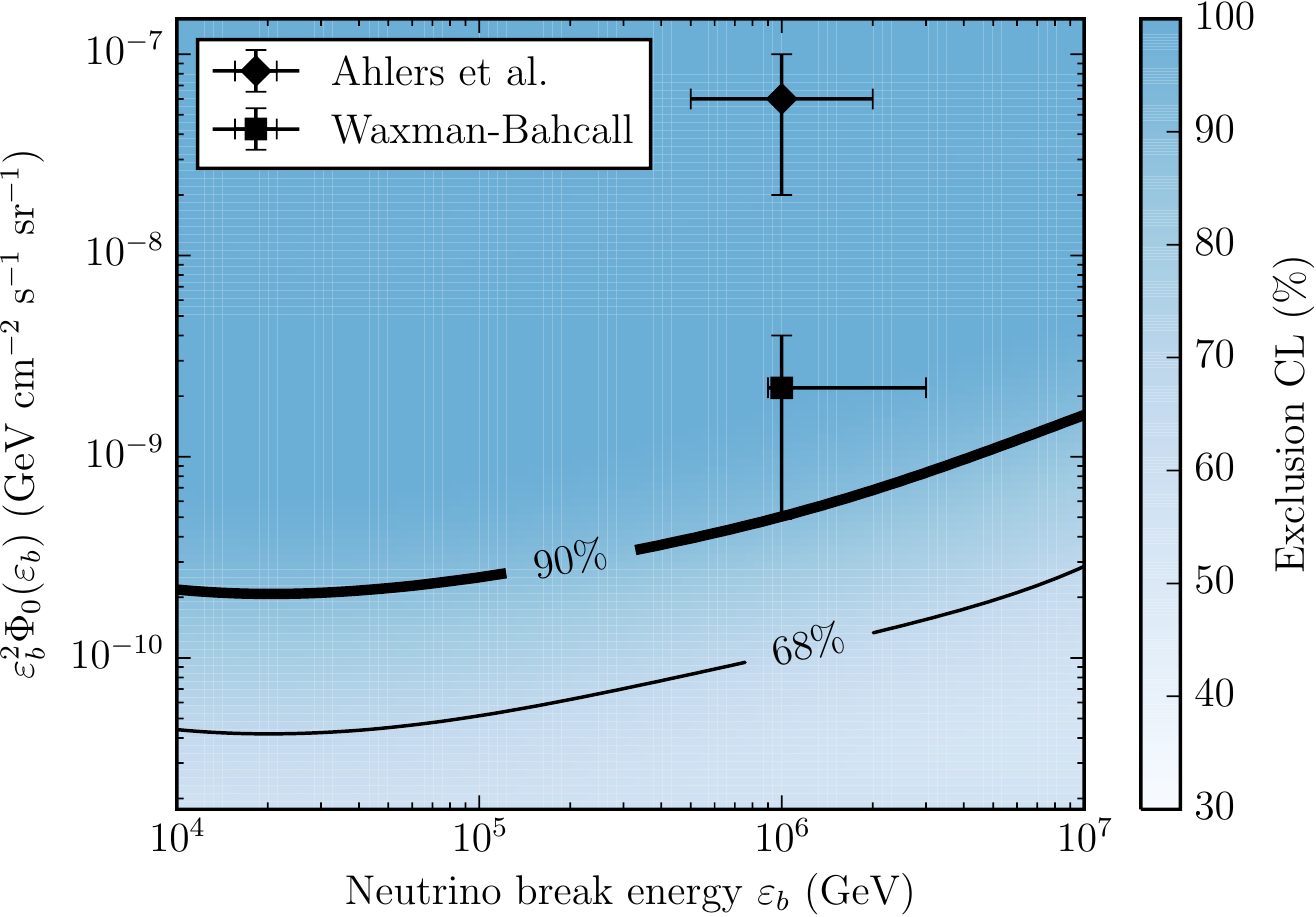}
  \caption{Limits on GRB model parameters adopted from~\citet{Aartsen:2016qcr}.}
  \label{fig:GRB}
\end{figure}

A search for high-energy neutrinos detected by IceCube from the locations of 807 GRBs in coincidence with their prompt $\gamma$-ray emission did not find a significant excess compared to background expectations~\citep{Aartsen:2016qcr}. This result provides tight constraints on models of neutrino and ultra-high-energy CR production in GRBs. Current models assuming acceleration of protons~\citep{Waxman:1997ti} and models assuming CR production through the decay of escaping neutrons~\citep{Ahlers:2011jj} are excluded at $90\%$ confidence (see Figure~\ref{fig:GRB}).
However, models assuming multiple emission regions predict a neutrino flux below our current
sensitivity~\citep{Bustamante:2014oka}.

Limits on the neutrino flux normalization allow us to constrain the contribution of $\gamma$-ray bright GRBs to less than $1\%$ of the observed cosmic neutrino flux~\cite{Aartsen:2016qcr}.
However, a possibly large population of choked-jet GRBs with low $\gamma$-ray luminosity might contribute a larger fraction of the astrophysical neutrino flux. 
Choked jets may explain trans-relativistic supernovae (SNe) and low-luminosity GRBs, giving a unified picture of GRBs and GRB-SNe~\citep{Senno:2015tsn}.
This scenario can be tested by correlating high-energy neutrinos with SNe.

\subsection{Supernovae}
Analogous to GRBs, high-energy neutrino production is predicted from SNe hosting mildy relativistic jets, which get choked in the envelope of the star~\citep{Razzaque:2005bh,Ando:2005xi,Tamborra:2015fzv}. Preferred candidates for choked-jet SNe are Type Ic SNe~\citep{2012grbu.book..169H}. The neutrino emission is expected at the time of the SN explosion and to last $\mathcal{O}$(10\,s), comparable to the typical GRB duration.
Other models predict neutrino emission from SNe exploding in a dense circum-stellar medium (CSM)~\citep{Murase:2013kda,Zirakashvili:2015mua}. Neutrinos are produced in the interactions of the SNe ejecta with the dense medium on time scales of months to years.

Supernovae are most easily discovered at optical wavelengths. However, current optical surveys cover only limited regions of the sky or do not go very deep. To overcome this limitation, the IceCube collaboration set up an optical follow-up program for neutrino events of interest~\citep{2012A&A...539A..60A} in 2008. The IceCube data are processed in real-time and the most interesting neutrino events are selected to trigger observations with optical telescopes aiming for the detection of an optical counterpart.

The Palomar Transient Factory (PTF)~\citep{2009PASP..121.1395L} found a Type IIn SN, triggered by two track-like IceCube events which arrived within 1.6\,s~\citep{Aartsen:2015trq}, at a location compatible with the direction of the neutrinos. Type IIn SNe are likely powered by interactions of the ejecta with a dense circumstellar medium, and are candidate neutrino sources with an expected duration of the neutrino emission of several months. 
Unfortunately, it turned out that the SN was already 160 days old at the time of the neutrino detection. 
It is very unlikely that two neutrinos arrive within 1.6\,s, so late after the SN explosion.
The observation can therefore be considered a chance coincidence. It nevertheless shows the potential of follow-up observations to reveal neutrino source candidates that would otherwise remain undetected.    
A dedicated search for neutrinos from an ensemble of Type IIn SNe observed independently of IceCube alerts is currently under development to test the possibility of long-term neutrino emission from Type IIn SNe.

Another supernova was found in a follow-up of a public IceCube alert (see Sec.~\ref{subsec:publicICAlerts}). The Pan-STARRS optical telescope found a SN -- possibly of type Ic -- with an explosion time consistent with the arrival time of a high-energy neutrino~\citep{PanSTARRS_GCN}. An analysis is currently in progress to investigate in detail the SN type classification, as well as the probability for a chance coincidence between such a SN and a high-energy neutrino.

\subsection{Blazar flares}
In addition to the optical follow-up program, IceCube has operated a $\gamma$-ray follow-up program~\citep{Aartsen:2016qbu} since March 2012, which alerts the Cherenkov telescopes MAGIC\footnote{\url{http://magic.mppmu.mpg.de}} and VERITAS\footnote{\url{http://veritas.sao.arizona.edu}}. This program is aiming for the detection of neutrinos in coincidence with $\gamma$-ray flares from blazars. A predefined list of known variable $\gamma$-ray sources is monitored by IceCube for an excess in neutrinos on time scales of up to three weeks. So far, no $\gamma$-ray flare was detected in coincidence with a neutrino excess.
A hint for neutrino production in blazar flares was claimed in~\citep{Kadler:2016ygj}, where a PeV neutrino shower event was found in spatial and temporal coincidence with a $\gamma$-ray outburst from the blazar PKS\,B1424-418.

\subsection{Public IceCube alerts}
\label{subsec:publicICAlerts}
Since the spring of 2016 a real-time selection for high-energy single track events with high probability of being of astrophysical origin is in place. An expected rate of four high-energy starting track events (HESE) and four extreme high-energy through-going track events (EHE) are selected per year and published in real-time through the Astrophysical Multi-messenger Observatory Network (AMON)~\citep{Smith:2012eu} via the Gamma-Ray Coordinate Network (GCN\footnote{\url{http://gcn.gsfc.nasa.gov}}). The first public neutrino alerts were followed up by various instruments in several wavelengths ranging from optical to $\gamma$-ray bands. 
A detailed overview of the different IceCube real-time channels can be found in~\cite{Aartsen:2016lmt}.

\subsection{Gravitational Wave Follow-Up}
\begin{figure}[t]
\centering\includegraphics[width=0.8\textwidth]{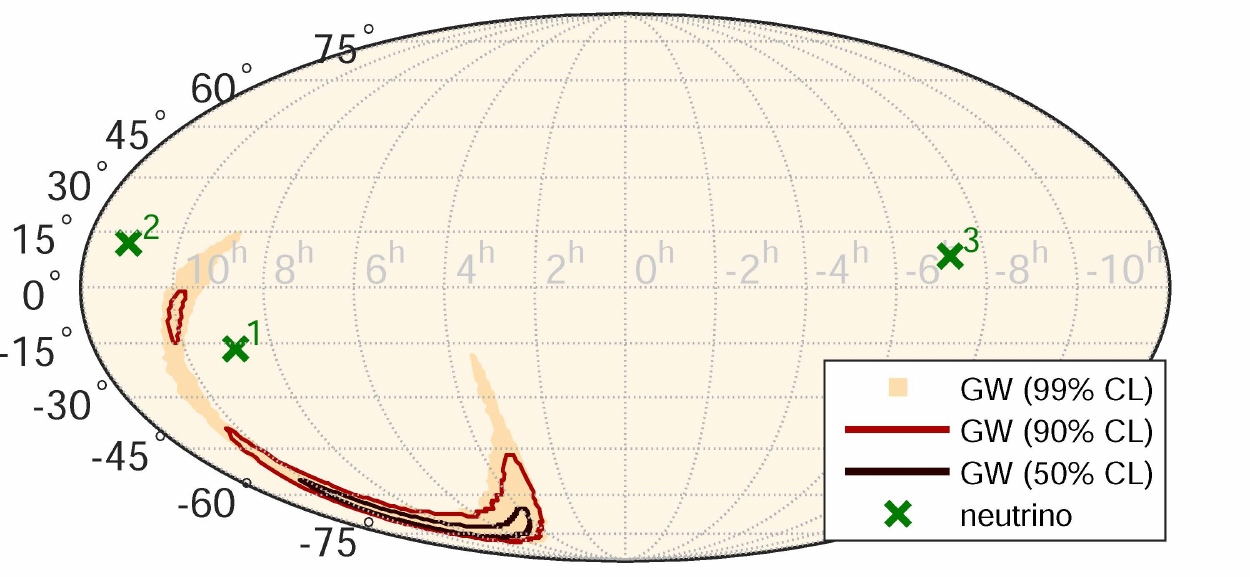}
  \caption{Skymap of the probability density contours of the GW event in equatorial coordinates together with the high-energy neutrino candidates detected by IceCube within a $\pm500$\,s time window centered on the GW event. Figure adopted from~\citet{Adrian-Martinez:2016xgn}.}
  \label{fig:GW}
\end{figure}

The detection of the first gravitational wave (GW) event by the advanced LIGO detectors\footnote{\url{http://ligo.org}} in September 2015 \cite{2016PhRvL.116f1102A} was accompanied by a broad multi-messenger follow-up program looking for the detection of a counterpart to the GW signal. 
IceCube and Antares searched their data in a $\pm500$\,s time window centered on the GW event for high-energy neutrinos~\citep{Adrian-Martinez:2016xgn}. No neutrino event was detected by Antares while IceCube found three events in  the time window, which is consistent with background expectations. Those events were not in spatial coincidence with the GW position as shown in Figure ~\ref{fig:GW}. Given the absence of a coincident signal, an upper limit on the total energy radiated in neutrinos of $5.4 \times 10^{51}$ -– $1.3 \times 10^{54}$~erg was derived assuming an energy spectrum following $dN/dE \sim E^{-2}$. Both of the distinct sky regions (see Figure~\ref{fig:GW}) are considered in the limit calculation to provide an inclusive range.

\section{Neutrinos from cosmic-ray interactions in the Galactic plane}
\label{S:GalacticPlane}

Cosmic rays up to a few PeV are believed to originate in Galactic sources. At this energy the CR spectrum shows a break, the so-called CR {\it knee}, which could indicate that the sources have reached their maximal acceleration energy for the lightest nuclei. It has long been speculated that Galactic core-collapse SNe, which occur at a rate of about 3 per century, could be responsible for the observed CR density~\cite{Baade1934}. These cataclysmic events produce ejecta with kinetic energy of the order of $10^{51}$~erg per SN explosion. Diffuse shocks that form as the ejecta run into the ambient medium could accelerate particles and transfer a significant fraction of this kinetic energy to a non-thermal population of cosmic rays.

Interactions of CRs with gas during their acceleration in the source or during their passage through nearby molecular clouds can lead to the production of charged and neutral pions. In the decay of neutral pions, $\pi^0\to\gamma\gamma$, these sources could be visible via their $\gamma$-ray emission. Indeed, recent observation of Fermi-LAT~\cite{Ackermann:2013wqa} indicate, that the $\gamma$-ray spectra of two Galactic supernova remnants, W44 and IC 443, show evidence for a characteristic rise in the spectra below $\sim200$~MeV resulting from pion decay. The corresponding decay of charged pions, e.g.\ $\pi^+\to\mu^+\nu_\mu$ followed by $\mu^+\to e^+\nu_e\bar\nu_\mu$, would be visible as high-energy neutrino emission. It has also been suggested that the injected $e^\pm$ from the production and decay of charged pions in the acceleration region would result in hard emission of $e^\pm$~\cite{Blasi:2009hv,Mertsch:2014poa} that could be responsible for the steep rise of the positron fraction in Galactic CRs above 10~GeV observed with PAMELA~\cite{Adriani:2008zr}, Fermi-LAT~\cite{FermiLAT:2011ab}, and AMS~\cite{Accardo:2014lma}. The associated neutrino emission could be observable in IceCube~\cite{Ahlers:2009ae}. 

After emission from their sources, CRs start to diffuse through the Galactic magnetic fields. This process has two effects. First, the arrival directions of the CRs become highly isotropized and obscure the position of the sources. Second, the diffusion process softens the spectra compared to the initial emission spectrum due to the enhanced loss of particles at higher energies. Diffusion also implies a rather smooth distribution of CRs throughout the Milky Way and therefore the local CR density, $n_{\rm CR} \simeq 4\pi\phi_{\rm CR}/c$, can be used as a proxy of the average density. The interaction of these CRs with gas in the vicinity of the Galactic plane then guarantees a diffuse Galactic emission of neutrinos and $\gamma$~rays ~\cite{Stecker:1978ah,Domokos:1991tt,Berezinsky:1992wr,Bertsch:1993,Ingelman:1996md,Evoli:2007iy,Gaggero:2015xza,Ahlers:2015moa}. 

\begin{figure}[t]
\begin{center}
\includegraphics[width=0.5\textwidth,clip=true,viewport= 10 0 610 400]{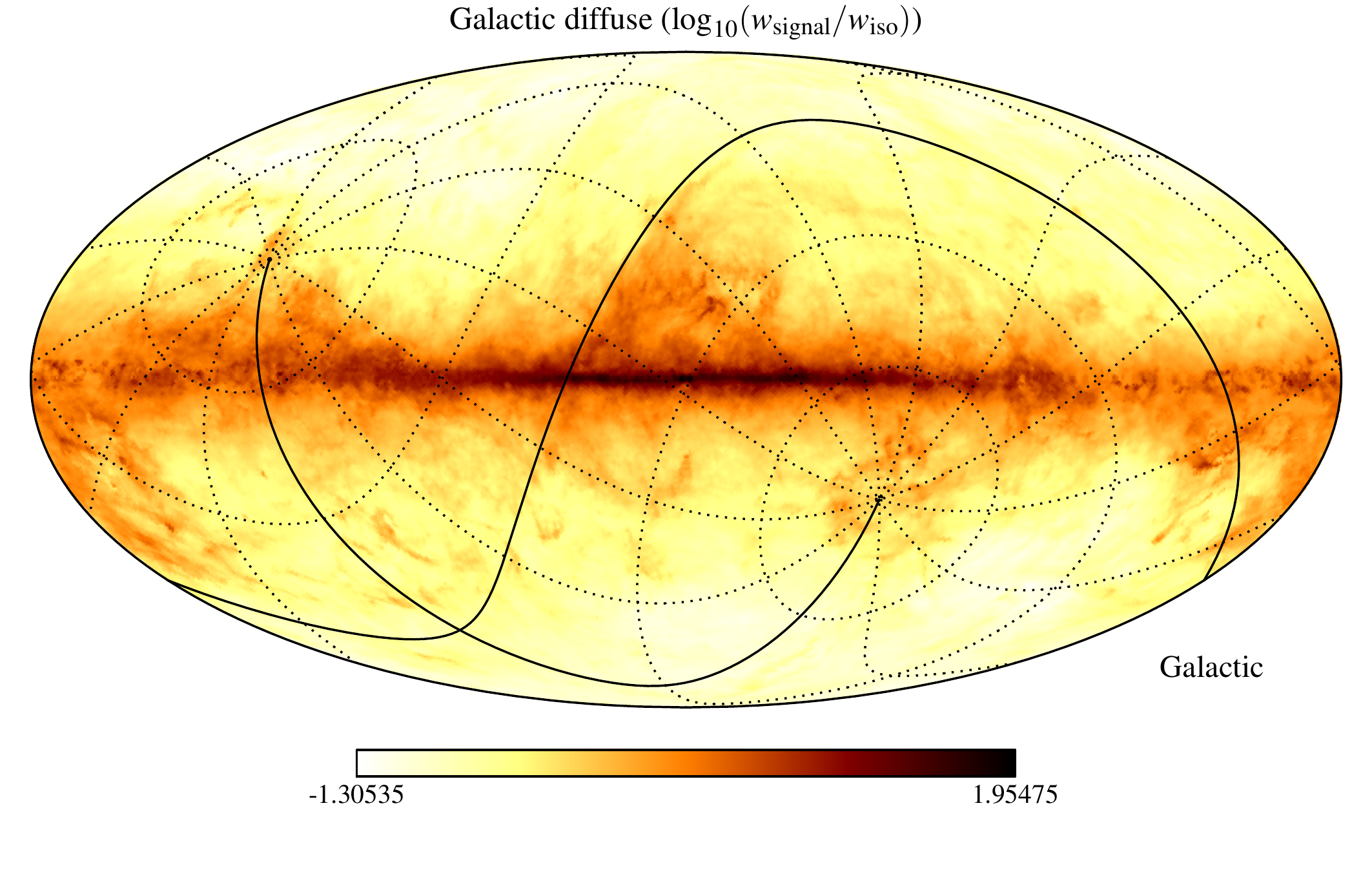}\includegraphics[width=0.5\textwidth,clip=true,viewport= 10 0 610 400]{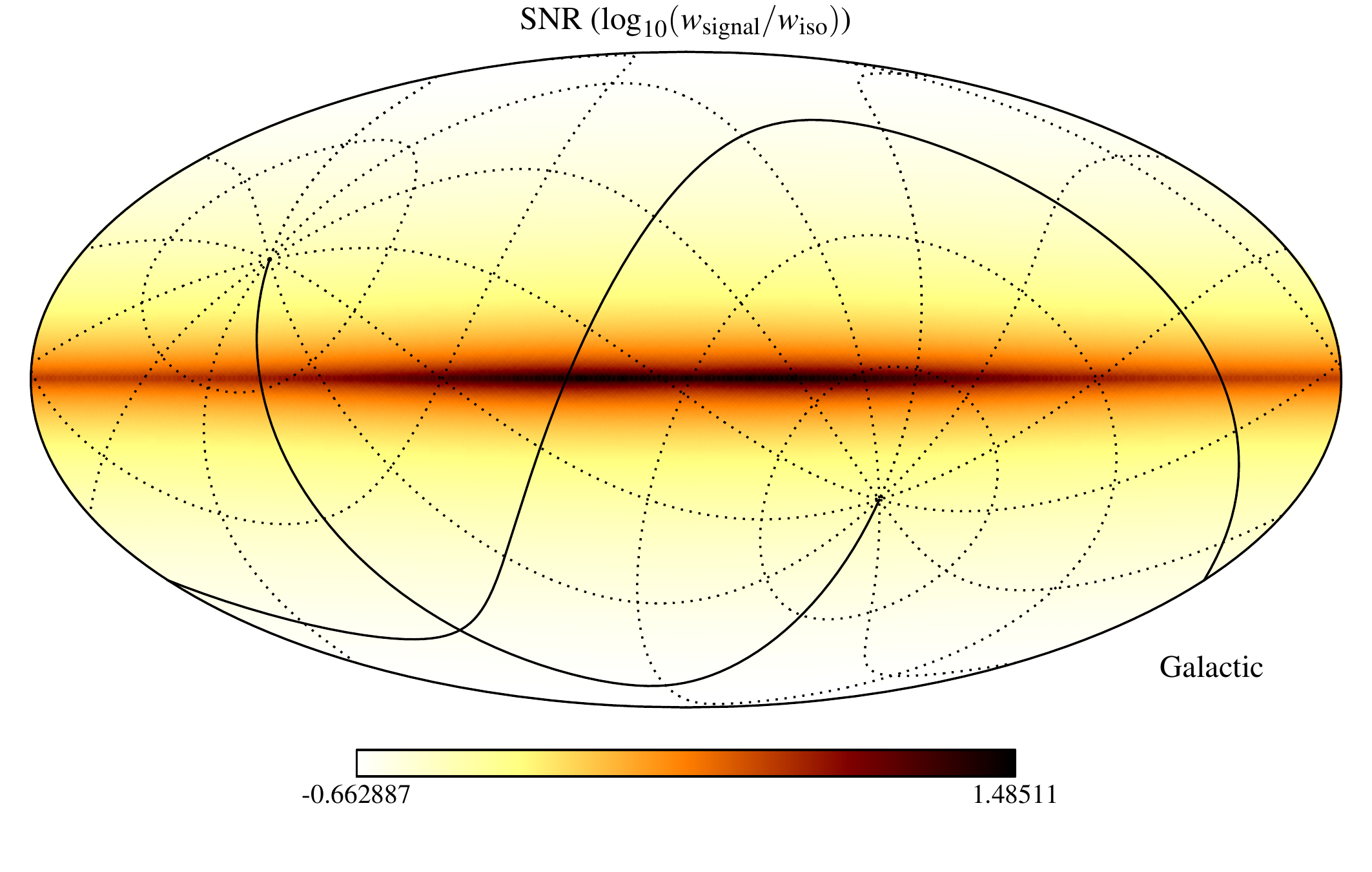}
\caption[]{Mollweide projections of expected diffuse Galactic neutrino emission~\cite{Ahlers:2015moa}. The left plot shows diffuse emission from CR propagation ($E_{\nu} = 10$~TeV) and the right plot the combined emission from supernovae remnants~\cite{Case:1998qg}. The mesh indicates the equatorial coordinate system with right ascension $\alpha=0^\circ$ and declination $\delta=0^\circ$ indicated as solid lines. The color reflects the logarithm of the intensity ratio between the Galactic and an isotropic signal.
}\label{fig:GalacticMap}
\end{center}
\end{figure}

The local emission rate of neutrinos (per flavor) from Galactic CR interactions can be estimated by the local nucleon density $n_N$ as 
\begin{equation}\label{eq:diffuse1}
E_\nu^2Q_\nu(E_\nu) \simeq \frac{1}{6}cn\kappa\sigma_{pp}\left[E_{\rm N}^2n_{\rm N}(E_{\rm N})\right]_{E_N = 20E_\nu}\,,
\end{equation}
where $\sigma_{pp}$ is the inelastic proton-proton cross section with inelasticity $\kappa\simeq0.5$~\cite{Kelner:2006tc,Block:2011vz}. The $1/6$ factor accounts for the per flavor emission ($\simeq1/3$), for the total neutrino energy fraction in the charged pion decay ($\simeq3/4$) and for the charged pion fraction in $pp$ collisions ($\simeq2/3$). The neutrino energy is related to the energy of CR nucleons ($N$) as $E_\nu \simeq E_N / 20$. The target gas density $n$ is mostly concentrated along the Galactic plane, but can also show high-latitude fluctuations from atomic and molecular gas clouds. The left plot of Figure~\ref{fig:GalacticMap} shows the predicted intensity of the diffuse emission from~\cite{Ahlers:2015moa} in terms of Galactic coordinates. Note that the map shows the intensity in logarithmic units. High-latitude intensity fluctuations are generally sub-dominant compared to the Galactic plane emission.

\begin{figure}[t]
\begin{center}
\includegraphics[width=0.75\textwidth]{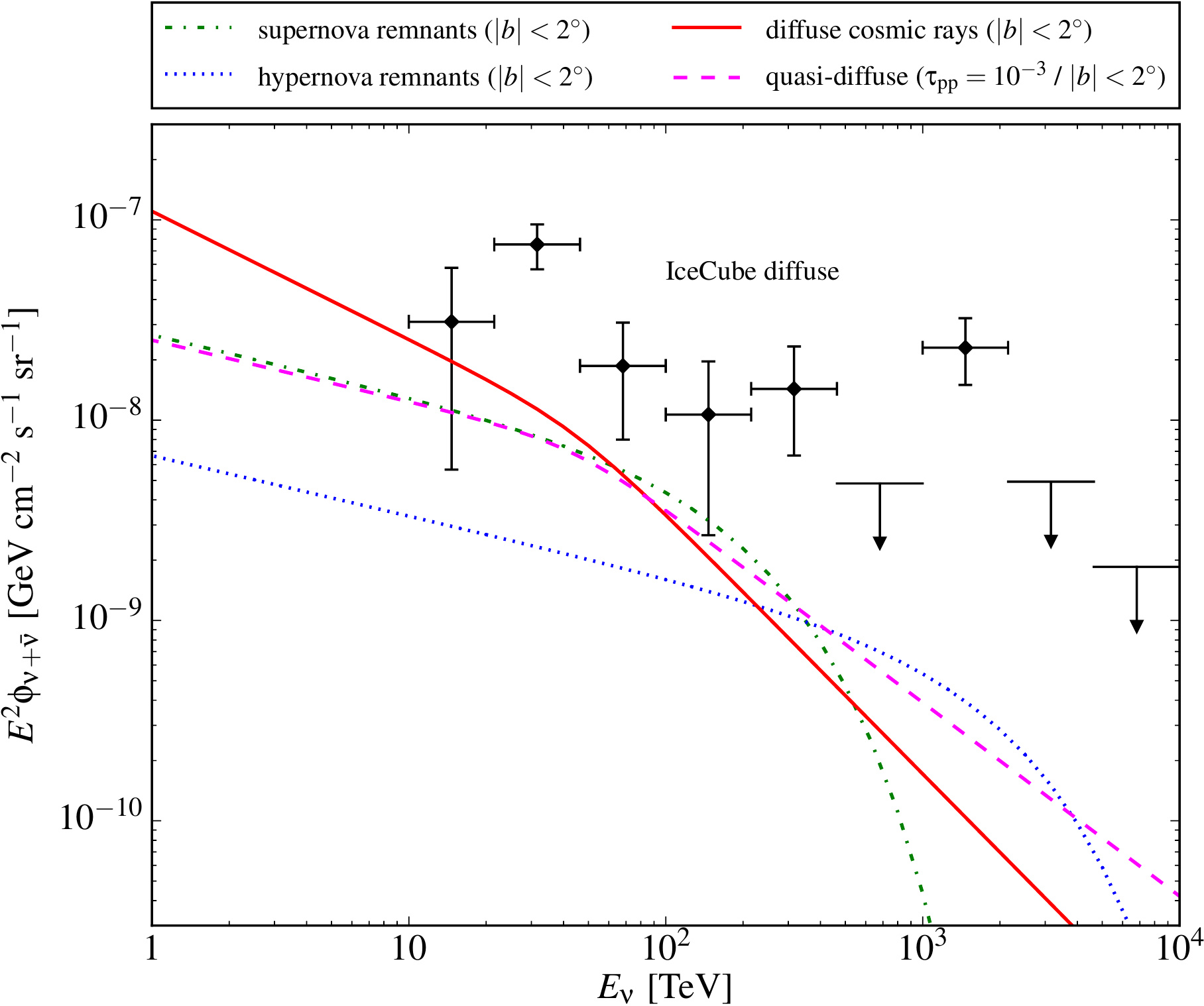}
\caption[]{Diffuse emission from the Galactic plane ($|b|\leq2^\circ$) in comparison to the isotropic diffuse neutrino flux (per flavor) observed by IceCube~\cite{Aartsen:2015knd}. We summarize here the Galactic diffuse flux (\ref{eq:diffuse1}), the quasi-diffuse flux (\ref{eq:diffuse3}) from weak Galactic sources (optical thickness $\tau_{\rm pp}\simeq 10^{-3}$ and diffusion index $\delta=1/3$), and the quasi-diffuse flux (\ref{eq:diffuse2}) of supernovae and hypernovae ($\Gamma=2.3$).}\label{fig:GalacticFlux}
\end{center}
\end{figure}

A simple estimate of the overall diffuse flux around the Galactic plane can be derived from a simple density scaling $n\simeq \exp(-|z|/0.1\,{\rm kpc})$~cm$^{-3}$ with distance $z$ from the Galactic plane and the corresponding integrated column density along the line-of-sight. The result is shown in Figure~\ref{fig:GalacticFlux} as a red solid line, where we averaged the diffuse emission over latitudes $|b|<2^\circ$. For the calculation we use Equation~(\ref{eq:diffuse1}) with the locally observed CR nucleon flux derived from the model of~\cite{Gaisser:2013ira}. This estimate agrees well with more elaborate studies using numerical CR propagation codes to evaluate the CR density across the Galaxy and using non-azimuthal target gas maps~\cite{Ahlers:2015moa}. 

Figure \ref{fig:GalacticFlux} also shows the diffuse flux of cosmic neutrinos observed by IceCube~\cite{Aartsen:2015knd}. This indicates that the diffuse flux close to the Galactic plane can dominate over the isotropic diffuse emission observed with IceCube for $E_{\nu}\leq10$~TeV. However, it is unlikely that this Galactic contribution has a strong impact on the interpretation of the IceCube data~\cite{Ahlers:2015moa,Ahlers:2013xia,Joshi:2013aua,Kachelriess:2014oma}.

Note that the previous estimate is based on the assumption that one can approximate the average Galactic CR density by the local CR flux. This is not necessarily the case with more general scenarios introducing spatial density fluctuation, e.g., by accounting for anisotropic diffusion~\cite{Effenberger:2012jc}, by inhomogeneous diffusion~\cite{Gaggero:2015xza}, or by strongly inhomogeneous source distributions~\cite{Gaggero:2013rya,Werner:2014sya}. Alternatively, a time-dependent local CR injection episode could be responsible for local CR spectra that are softer than the Galactic average~\cite{Neronov:2013lza} and could also lead to an increase of the overall Galactic diffuse emission.

At present, there are various dedicated IceCube analyses that are searching for Galactic diffuse neutrino emission, accounting for uncertainties of morphology and emission spectrum. The simplest test for a signal from the Galactic diffuse emission in the IceCube data is by checking for spatial correlations with the Galactic plane. No significant correlation of events with the Galactic plane was found in four years of high-energy starting event (HESE) data \cite{Aartsen:2015zva}. When letting the Galactic plane size float freely, the best fit returned a value of $|b|\leq7.5^\circ$ with a post-trial chance probability of 3.3\%. The recent analysis~\cite{Ahlers:2015moa} based on 3 years of HESE data~\cite{Aartsen:2014gkd} showed that even with the poor angular resolution of cascade events the anisotropy produced by a strong Galactic diffuse flux should be visible in data. The upper limit on the contribution to the high-energy data with deposited energy above 60~TeV is about 50\%. This is in contrast to the claim of~\cite{Neronov:2015osa} that the 4-year HESE update shows evidence of Galactic emission within latitudes $|b|\leq10^\circ$ above 100~TeV. In addition, the angular distribution of the muon neutrino data from the recent analysis~\cite{Aartsen:2016xlq} does not seem to support this claim. 

While individual Galactic neutrino sources have not been identified, the cumulative contribution of Galactic sources below the IceCube detection threshold~\cite{Casanova:2007cf} might be identified as extended emission concentrated along the Galactic plane. In general, if $N_{\rm N}$ is the (time-integrated) CR nucleon spectrum of a single source, we can define the Galactic neutrino emission from interactions of CRs with ambient gas as 
\begin{equation}\label{eq:diffuse2}
E^2_{\nu}{Q}_\nu(E_\nu) \simeq \frac{1}{6}cn\kappa\sigma_{pp}\rho_{\rm act}\left[E_N^2N_{\rm N}(E_N)\right]_{E_N = 20E_\nu}\,,
\end{equation}
where $n$ is the ambient gas density and $\rho_{\rm act}$ is the number density of {\it active} sources in the Galaxy. 

We consider now the case of neutrino emission from supernova remnants (SNR) in our Milky Way following the distribution of Ref.~\cite{Case:1998qg}. Similar to the diffuse emission from CR propagation, the intensity distribution of events is concentrated along the Galactic plane as shown in the right plot of Fig.~\ref{fig:GalacticMap}. The number of active SNRs can be estimated from the SN rate and the time-scale of the onset of the snow-plow phase which marks the end of the adiabatic Sedov--Taylor phase~\cite{Blondin1998}. From this, one can estimate that a few  thousand SNRs are CR emitters at any given time. The maximal energy can be estimated from the ambient gas density, ejecta mass, and velocity to reach $E_{\rm max,p}\simeq$5~PeV. An order of magnitude higher CR energies might be reached in very energetic SNe ($\simeq10^{52}$~erg), so-called {\it hypernovae}, but they are much less frequent than normal SNe with only 1\% - 2\% of the SNe rate~\cite{Fox:2013oza,Ahlers:2013xia}. 

Figure~\ref{fig:GalacticFlux} summarizes the estimated flux of SNRs (green dashed-dotted line) and hypernova remnants (blue dotted line) assuming a source spectral index $\Gamma\simeq2.3$ (see Ref.~\cite{Ahlers:2013xia} for details). While the source emission spectrum is subdominant at lower energies it is expected to become more important at higher energies because of its harder emission spectrum. In fact, for the choice of parameters in our example, the combined emission of sources becomes comparable to the diffuse emission at energies of 100~TeV, corresponding to neutrinos produced by CRs close to the knee region.

Note that the previous estimate applies more generally than to the case in which SNR are the main sources of Galactic CRs. If we focus on the sources of Galactic CRs, we can relate the (per flavor) neutrino emission rate to that of the CR nucleons as
\begin{equation}\label{eq:diffuse3}
E_\nu^2{Q}_\nu(E_\nu) \simeq \frac{1}{6}\kappa\tau_{pp} \left[E_N^2{Q}_{\rm N}(E_N)\right]_{E_N = 20E_\nu}\,,
\end{equation}
where $\tau_{pp}\ll1$ is the optical thickness of the source environment for CR-gas interactions, before CRs are released into the Galactic medium. The nucleon emission rate ${Q}_{\rm N}$ is now fixed to the observed CR spectrum by the steady-state solution of the CR diffusion equation. For an active emission period lasting over a time $t_{\rm act}$ and an ambient average gas density $n_{\rm gas}$, one can estimate the optical thickness as $\tau_{pp} \simeq ct_{\rm act}n_{\rm gas}\sigma_{pp}$. For the case of SNRs we can estimate $t_{\rm act}$ by the dynamical time-scale  $10^4$~yr (the end of the adiabatic Sedov--Taylor phase~\cite{Blondin1998}) and $n_{\rm gas}\simeq 1{\rm cm}^{-3}$ yielding $\tau_{pp} \simeq 3\times10^{-4}$. This flux is also shown in Figure~\ref{fig:GalacticFlux} as a magenta dashed line assuming $\tau_{\rm pp}\simeq 10^{-3}$ and a diffusion index $\delta = 1/3$. Not surprisingly, this is consistent with our previous estimates of the combined flux of supernova and hypernova remnants.

The combined neutrino emission of Galactic sources has been studied by various authors~\cite{Casanova:2007cf,Ahlers:2013xia,Mandelartz:2014sqa}. Analogous to the diffuse emission from CR propagation, the contribution of weak Galactic sources can be constrained in the simplest case by the absence of anisotropies. For instance, Ref.~\cite{Ahlers:2015moa} argues that a Galactic emission following the distributions of supernova remnants~\cite{Case:1998qg} or pulsars~\cite{Lorimer:2006qs} can not contribute more than 65\% to the HESE three-year data~\cite{Ahlers:2015moa}. Even stronger limits can be expected if also the emission spectrum is taken into account.

\section{Measurements of the local cosmic-ray spectrum and composition}
\label{S:IceTop}

From the point of view of CR physics, IceCube is a three-dimensional
air shower array.  The aperture for trajectories that pass through IceTop
and within the deep array at its mid-plane is $\approx 0.25$~km$^2$sr, which
corresponds to $\approx 1000$ events per year above $100$~PeV, but only a few per
year above one EeV.  Such coincident events provide information about the
primary composition from the ratio of the energy in the muon bundle in
the deep ice to the total shower size at the surface.  The measurement of
the primary spectrum can be extended to the EeV range by using events
over a larger angular range reconstructed with only the surface array~\cite{Aartsen:2013wda}.  Muon bundles reconstructed over a
large range of zenith angles with the deep array of IceCube extend
the acceptance into the EeV range and provide complementary
information to the surface array~\cite{Aartsen:2015nss}.  
IceCube can resolve muons in the deep array laterally separated from the main core by more than the string spacing.  The separation distribution 
measured out to 400 meters shows the concave shape expected from
the transition from an exponential to a power-law for the transverse momentum
distribution of the parent mesons~\cite{Abbasi:2012kza}.  Perturbative QCD can be used
to calculate the rate of high $p_T$ muons that reach the detector at large separation.
Since production of mesons at high $p_T$ depends on energy per nucleon, measurement of
laterally separated muons is in principle sensitive to primary composition.

\begin{figure}[ht]
\centering\includegraphics[width=0.9\textwidth]{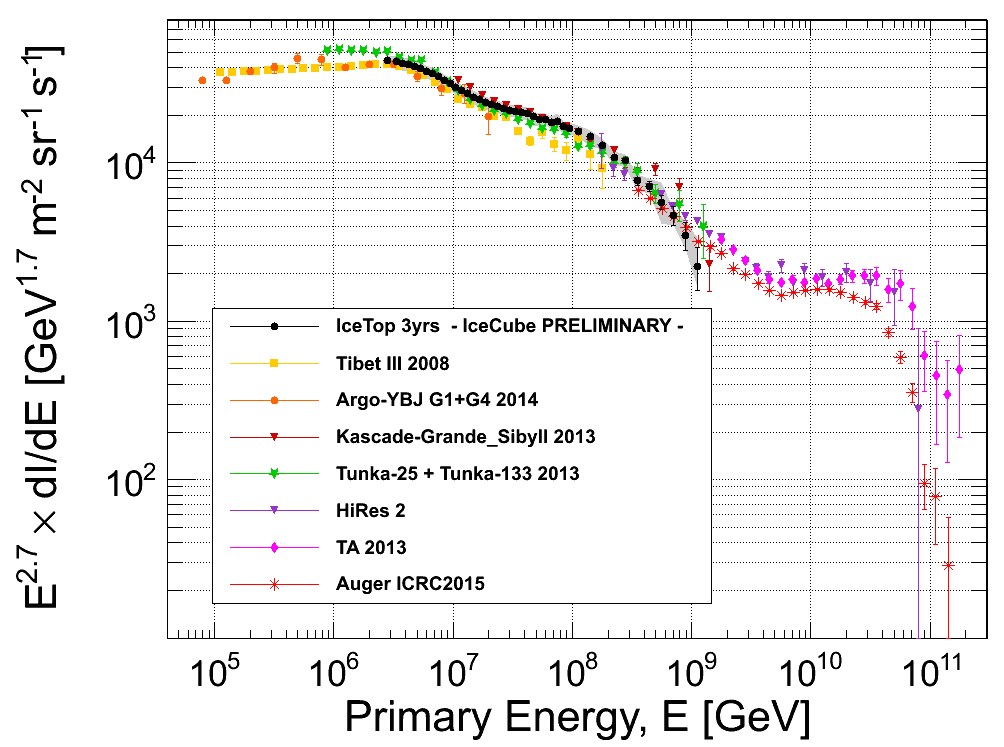}
\caption{A summary of the primary CR spectrum measured by selected air shower 
experiments.~\cite{Amenomori:2008aa,Sciascio:2014opa,Apel:2014uka,Prosin:2014dxa,Abbasi:2007sv,Abu-Zayyad:2013jra,Auger:2015abc}  
Measurement from three years of data in IceCube~\cite{Rawlins:2016bkc}
are shown by the black squares.
}\label{fig:CR1}
\end{figure}

Several aspects of IceTop lead to its good energy resolution and its
ability to distinguish features in the energy spectrum.  The array is
at a high altitude so that events are observed closer to shower
maximum.  As a consequence, fluctuations from event to event are less severe
than in an array near sea level.  The ice Cherenkov tanks are approximately
two radiation lengths deep so that the dominant photon component
of the surface shower is counted as well as the charged leptons.
In contrast, most photons pass through scintillators without converting.
The IceTop results are shown by the black
points in the compilation of air-shower data in Figure~\ref{fig:CR1}.
In addition to the knee above $3$ PeV, there is a significant
hardening of the spectrum around $20$~PeV, and the second knee
is visible above $200$~PeV.

IceCube is the only air shower array currently in operation that can
detect TeV muons in the shower core in coincidence with the
main shower at the surface.  It has much larger acceptance than its
predecessors, EASTOP-MACRO~\cite{Bellotti:1989pa} 
and SPASE-AMANDA~\cite{Ahrens:2004nn}.  Preliminary analysis of the
coincidence data in IceCube shows the composition becoming increasingly
heavy through the knee region to $100$~PeV and beyond~\cite{Rawlins:2016bkc}, 
although the results become statistically limited at the highest energies.

The increasing fraction of heavy primaries is expected if the knee
is the result of Galactic CR accelerators reaching their upper
limit.  Air shower experiments make calorimetric measurements of
the total energy per particle.
Since acceleration and propagation of CRs are
both determined by magnetic fields, features in the spectrum should instead
depend on magnetic rigidity~\cite{Peters:1961}.  Thus, for example,
if the characteristic maximum energy for protons is 4 PeV, there should be
a corresponding steepening for iron nuclei around 100 PeV total energy.
Several air-shower measurements, as reviewed in~\cite{Kampert:2012mx}, 
show the composition changing back toward a lighter composition above $100$~PeV as
might be expected with the onset of an extra-galactic component at higher energy.
The IceCube coincidence analysis gives composition results that agree well with
$\langle\ln(A)\rangle$ measurements summarized in Ref.~\cite{Kampert:2012mx} up to $100$~PeV, but
the mass value remains high above that energy in some tension 
with the other data (see Figure~8 in~\cite{Gaisser:2016ytm}).

\begin{figure}[ht]
\centering\includegraphics[width=0.6\textwidth]{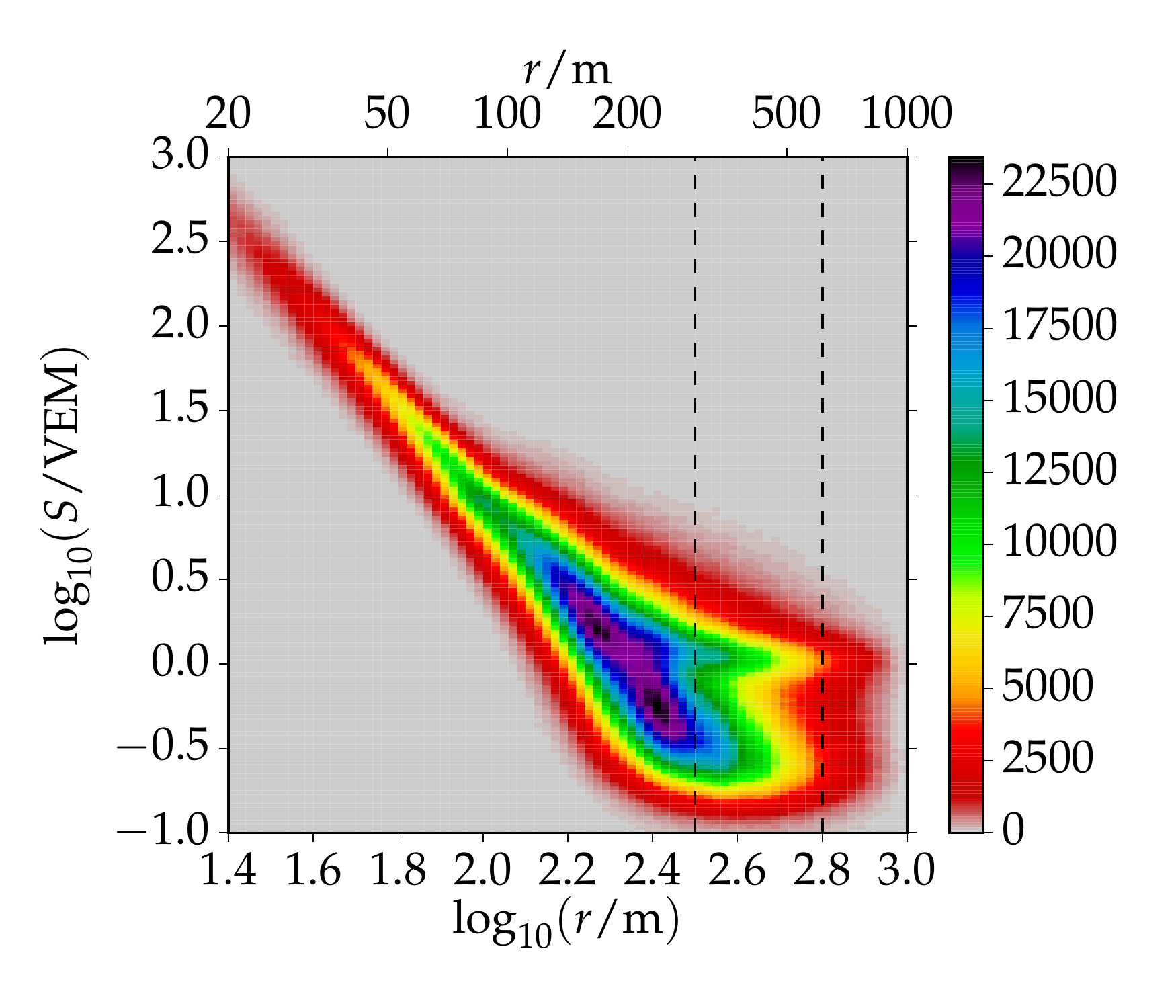}
\caption{Two-dimensional distribution of signals in showers
with primary energies of $\approx 3$~PeV and zenith angles around $13^\circ$
as a function of signal in VEM and reconstructed core distance.  The horizontal
dashed lines indicate the distances at which muon densities as a
function of primary energy have been reported~\cite{Dembinski:2015xtn}.
}\label{fig:CR2}
\end{figure}

Muons produce a characteristic signal in IceTop tanks because they generate
a charge proportional to the length of their tracks.  In addition, as the
main electromagnetic part of the signal falls off at large distance from the
shower core, muons become increasingly prominent, as indicated by the ``thumb''
centered near one Vertical Equivalent Muon (VEM) in Figure~\ref{fig:CR2}.  This leads to
the possibility of measuring the contribution of $\sim$~GeV muons to the
showers at the surface~\cite{Dembinski:2015xtn}.  Such a measure of the fraction of muons
at the surface opens the possibility of utilizing a different quantity that is sensitive
to primary composition.  Information from the low energy muons at the surface
is complementary to the TeV muons in the shower cores in the coincident event
analysis.  The comparison, which is ongoing, is of particular interest in light
of the fact that different hadronic interaction models show a different behavior
for the ratio of GeV to TeV muons. In addition, there are indications that
all the standard event generators for $>$EeV air showers produce fewer muons
at the surface than observed~\cite{Aab:2014pza,Aab:2016hkv}.

Because muons are rare in cascades initiated by photons, the muon content
can also be used to reduce the CR background in a search for
$\sim$~PeV $\gamma$-rays.  A shower reconstructed
at the surface with a trajectory that passes through the deep array of IceCube
without leaving a signal is a $\gamma$-ray candidate.  Because of energy losses in
the CMB, only Galactic sources would be visible in PeV photons.  Using one year of data
taken when IceCube was partially complete with 40 strings, a limit on $\gamma$~rays
of several PeV from the Galactic plane was set~\cite{Aartsen:2012gka}.  
Because of the small zenith angle
required for events to pass through both components of IceCube, the search was limited
to Southern declinations $<-60^\circ$.  Therefore the analysis covers a limited region of the
Galactic plane, $-80^\circ<\ell <-30^\circ$ in longitude and $-10^\circ < b < 5^\circ$
in latitude.  The sensitivity with five years of data from the full IceCube
detector is estimated in this region to be comparable to expectations from some known TeV
$\gamma$-ray sources if their spectra continue to PeV energies without steepening.
An analysis with the completed IceCube detector is underway.  Including muon information
from the surface detector will allow a larger region of the sky to be explored.

Another search for Galactic CR sources looks for neutrons~\cite{Aartsen:2016asr},
which would show up as point sources of air showers above the
smooth background of charged CRs.  No such excesses are identified
in 4 years of IceTop data.
Limits are placed on potential accelerators of CR protons and nuclei, 
including millisecond
pulsars and high-mass x-ray binaries, by using events with energy $>100$~PeV
for which the mean distance a neutron would travel before decaying is $100$~kpc.
Assuming an $E^{-2}$ spectrum, the limits are of the same order of magnitude
in energy flux as might be expected for sources that produce photons in association
with acceleration of nuclei that fragment in or near their sources to produce neutrons.

\section{Anisotropy of local cosmic rays}
\label{S:Anisotropy}

Through measurement of the energy spectrum and composition of the CR 
flux, we hope to gain a better understanding of CR sources and acceleration 
mechanisms.  Another quantity accessible to experimental measurement is 
the arrival direction of the CR particles.  In principle, the sky map of 
CR arrival directions should give us the most direct indication of where 
the sources might be located.  Below several PeV, the sources of CRs are 
Galactic and the arrival direction distribution should show a correlation with
the Galactic plane.  However, unlike $\gamma$~rays and neutrinos, CR particles 
are charged and therefore repeatedly scattered in the chaotic interstellar 
magnetic fields.  Their arrival direction distribution at Earth is highly isotropic, 
although a small residual dipole anisotropy is expected from diffusion theory.

Observations made over the last few decades with various surface and underground 
detectors, together covering an energy range from tens of GeV to tens of PeV, have 
indeed provided statistically significant evidence for a faint anisotropy in the 
CR arrival direction distribution~\cite{Tibet:2005jun, 
Tibet:2006oct, SuperK:2007mar, Milagro:2009jun, MINOS:2011icrc, Bartoli:2013, 
Nagashima:1998aug, Hall:1999apr, ARGO:2013oct, 
IceCube:2010aug, IceCube:2011oct, IceCube:2013mar, Aartsen:2016ivj}.  The anisotropy 
is small, with an amplitude on the order of $10^{-3}$, and it shows a strong dependence 
on energy~\cite{Bartoli:2013, Aartsen:2016ivj, Aglietta:2009feb, IceCube:2012feb}.  
It is, however, not well described by a simple dipole.  A quantitative description of 
the anisotropy as a superposition of spherical harmonics~\cite{Aartsen:2016ivj,
HAWC:2014dec} shows that while 
most of the power is in the low-multipole ($\ell \leq 4$) terms, i.e. in the dipole,
quadrupole, and octupole terms, features with smaller angular scale down to sizes of a
few degrees are also present.  These small-scale features have been observed in the
TeV range by several experiments~\cite{ARGO:2013oct,IceCube:2011oct,Aartsen:2016ivj, 
HAWC:2014dec,Tibet:2007aug, Milagro:2008nov}, and their relative intensity is on the 
order of $10^{-5}$~--~$10^{-4}$. Given the complex nature of the anisotropy, its range 
from large to small angular scales, and its strong dependence on energy, it has become 
clear that there is no single process that can account for all observations.  Rather, 
multiple phenomena likely contribute to the anisotropy.

Before IceCube, high-statistics measurements of the CR anisotropy in the TeV energy range 
were only available from experiments in the Northern Hemisphere.  Over the last few 
years, IceCube has accumulated one of the largest CR data sets at TeV to PeV energies, 
and a detailed study of the morphology, energy dependence, and stability of the anisotropy 
over time is possible for the southern sky.

CRs can be studied with IceCube in two independent ways.  The in-ice component
of IceCube detects downward-going muons created in extensive air showers caused by
CR entering the atmosphere above the detector.  Simulations show that the 
detected muon events are generated by primary CR particles with median energy
of about 20\,TeV.  The trigger rate ranges between 2.5~kHz and 2.9~kHz, with the modulation 
caused by seasonal variations of the stratospheric temperature and 
density~\cite{Tilav:2010jan, Desiati:2011aug,Aartsen:2013llab}.
 
The anisotropy can also be studied using the CR air showers detected by IceTop.
Its surface location near the shower maximum makes it sensitive to the full electromagnetic 
component of the shower, not just the muonic component.  The detection rate 
is approximately 30\,Hz and the minimum primary particle energy threshold is about 300\,TeV.
Requiring a minimum of eight IceTop stations leads to a median energy of 1.6\,PeV.  
The IceTop data set therefore provides an independent measurement at PeV energies, close 
to the knee of the CR spectrum.

\begin{figure}[t]
\centering\includegraphics[width=\textwidth]{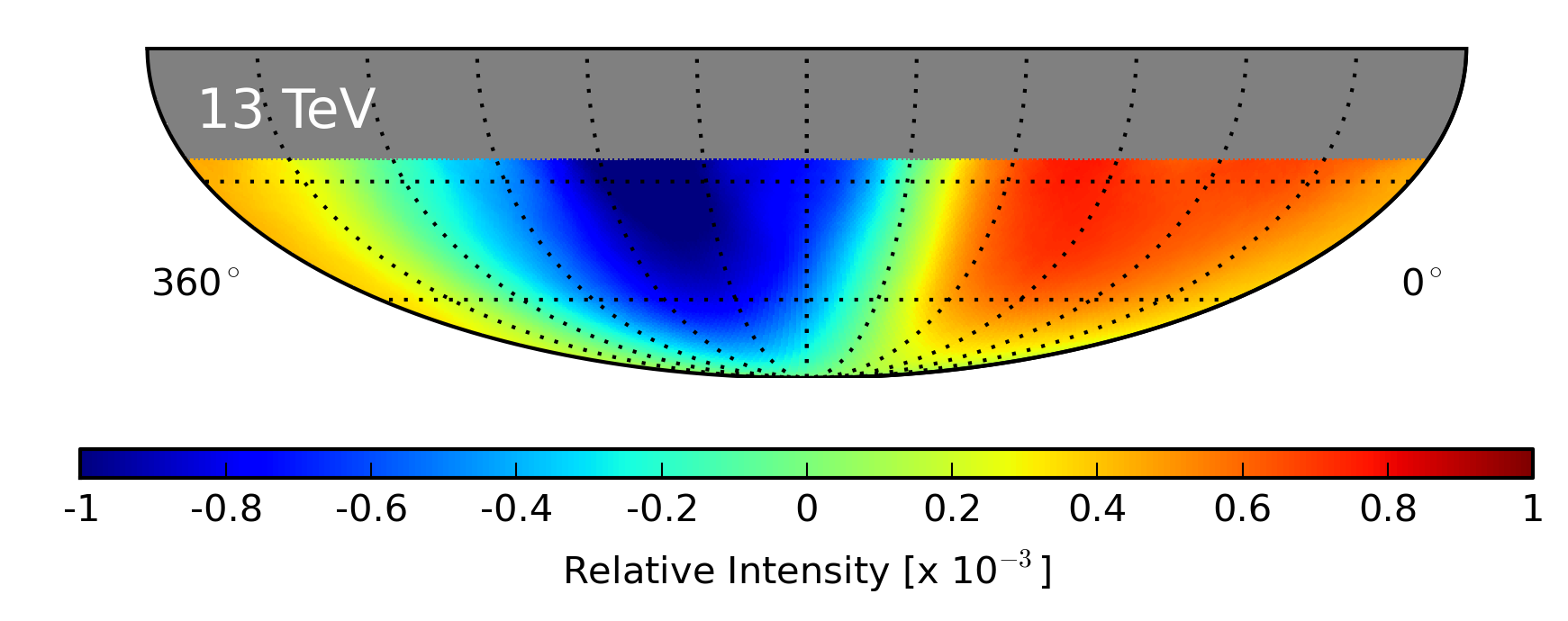}
\centering\includegraphics[width=\textwidth]{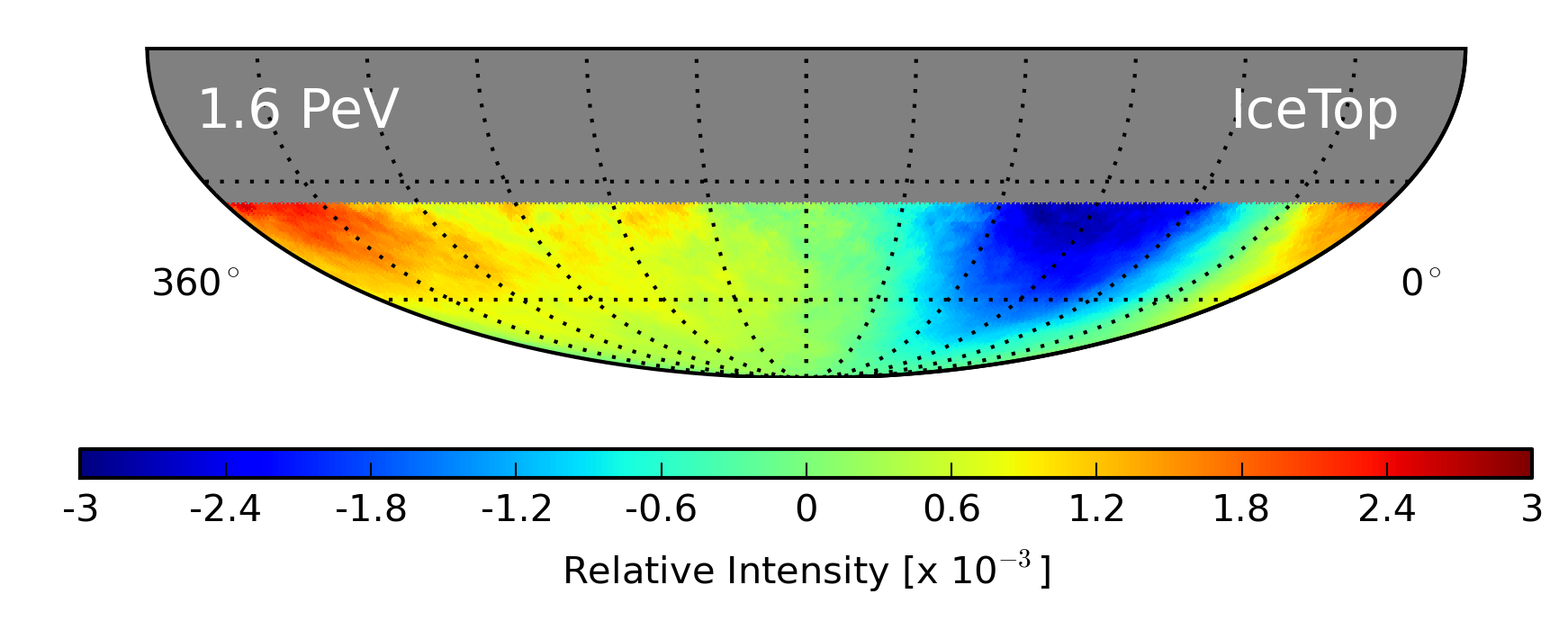}
  \caption{Maps of the relative intensity of CRs in equatorial coordinates for a 
  median energy of 13\,TeV (top) and 1.6\,PeV (bottom)~\cite{Aartsen:2016ivj}.  The low-energy map is based 
  on IceCube data, the high-energy map on IceTop data.  Maps have been smoothed with 
  a $20^{\circ}$ smoothing radius.}
  \label{skymaps}
\end{figure}

A recent study of the CR anisotropy in IceCube and IceTop~\cite{Aartsen:2016ivj} 
is based on six years of data taken between May 2009 and May 2015.
The data set contains 318 billion CR events observed by IceCube and 172 million
events observed with IceTop at higher energies.  In order to study the energy dependence of
the anisotropy, the IceCube data set is split into nine bins of increasing median energy, 
ranging from TeV to PeV.  The resolution of this energy assignment depends on the detector
configuration and energy band but is on the order of 0.5 in $\log_{10}(E/\mathrm{GeV})$.
It is primarily limited by the relatively large fluctuations in the fraction
of the total shower energy that is transferred to the muon bundle.

The most prominent anisotropy observed in the IceCube data at energies below 50\,TeV
is characterized by a large excess from $30^{\circ}$ to $120^{\circ}$ in right ascension
and a deficit from $150^{\circ}$ to $250^{\circ}$.  The relative intensity of the anisotropy
is at the $10^{-3}$ level.  This large-scale structure that dominates the sky map at lower 
energies gradually disappears above 50\,TeV.  Above 100\,TeV, a change in the morphology is
observed.  At higher energies, the anisotropy is characterized by a wide relative deficit
from $30^{\circ}$ to $120^{\circ}$, with an amplitude increasing with energy up to at least 
5\,PeV, the highest energies currently accessible to IceCube.  The IceTop map at 1.6\,PeV
shows the same morphology as the IceCube maps at comparable energies.  To illustrate this
change of the phase of the large-scale anisotropy between TeV and PeV energies, 
Figure~\ref{skymaps} shows the IceCube map at a median energy of 13\,TeV (top) to the IceTop 
map at 1.6\,PeV (bottom).

\begin{figure}[t]
\centering\includegraphics[width=0.8\textwidth]{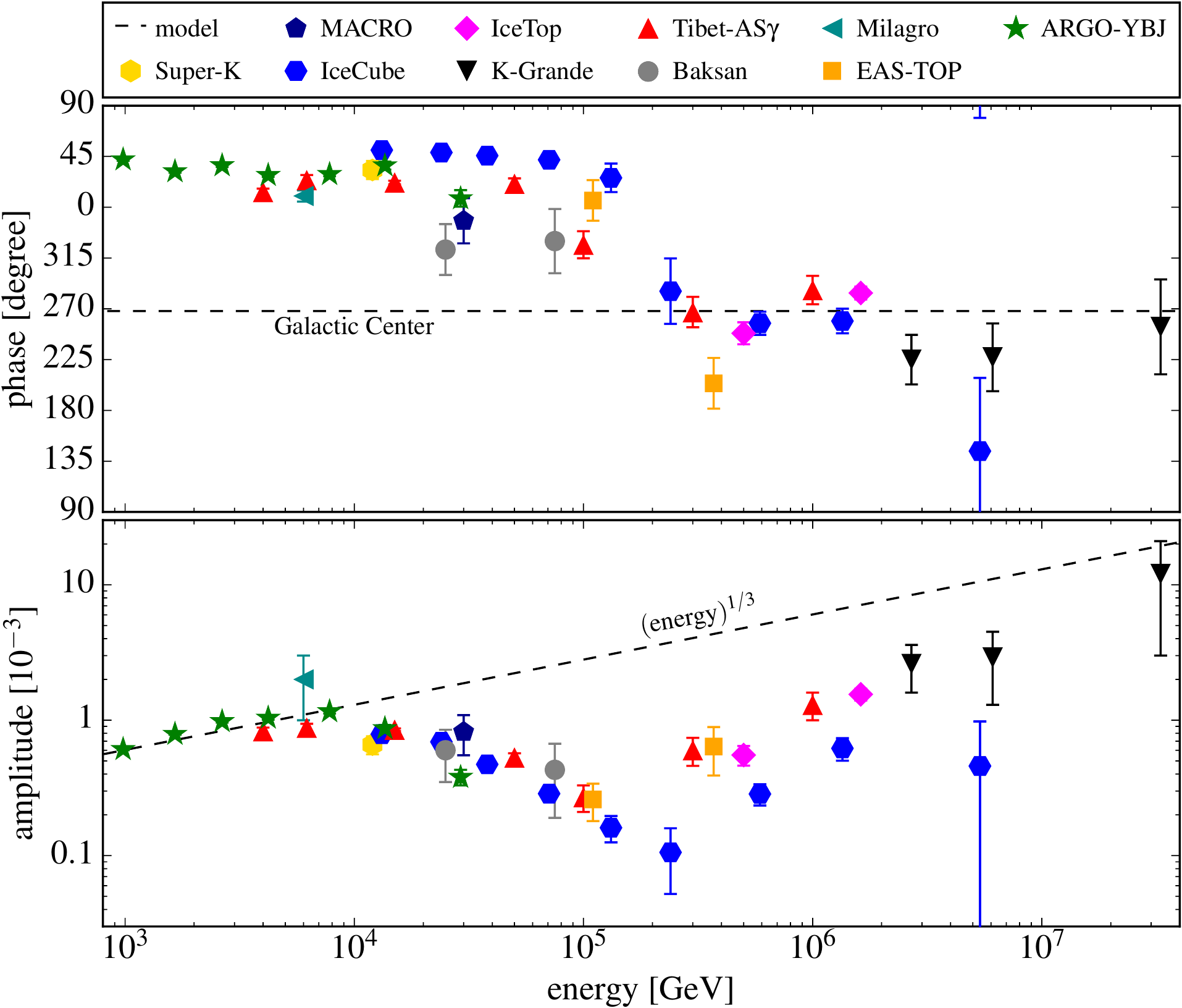}
  \caption{Phase (top) and amplitude (bottom) of the dipole moment of the
  CR relative intensity map as a function of energy for IceCube (blue), IceTop (pink),
  and other experiments.  Taken from~\cite{ahlers_review}.}
  \label{dipole} 
\end{figure}

Figure~\ref{dipole} shows the phase (top) and amplitude (bottom) of the dipole component 
as a function of energy.  Since the data are not well described by a dipole, the actual 
fit is performed including higher-order multipoles, but only the amplitude and phase of 
the dipole are reported here.  The phase shift in the dipole component of the large-scale 
anisotropy occurs rather rapidly between 100\,TeV and 200\,TeV.  The amplitude of the 
dipole component rises with energy up to about 10\,TeV.  Above this energy, it slowly
decreases until it has essentially dropped by an order of magnitude at around 200\,TeV.
It then increases again, with a different phase, up to the highest detected energies.
The figure also shows the results from several other experiments in the 
Northern Hemisphere.  The results are generally in good agreement.  The difference in
the amplitude measured by IceCube and IceTop above 1\,PeV is likely due to a 
difference in the chemical composition of the two data sets.  At this energy, the IceTop
data set has on average a lighter composition than the IceCube data set because IceTop 
is not yet fully sensitive to heavier nuclei.

Measurements of a dipole amplitude and phase of the CR flux have also been performed
at even higher energies, although the small event rate makes these measurements increasingly
difficult.  Nevertheless, the Pierre Auger Observatory found that a shift in the phase of 
the anisotropy occurs again at EeV energies~\cite{Auger:2011mar}.  Below 1\,EeV, the dipole 
phase is consistent with the phase observed by IceCube at PeV energies.  Around 4\,EeV, the
phase changes and the relative excess moves towards the range in right
ascension that includes the Galactic anti-center direction.  In between the IceCube and
Pierre Auger measurements, KASCADE-Grande data shows a dipole phase between median
energies of 2.7\,PeV and 33\,PeV~\cite{Chiavassa:2015jbg}, which is consistent with the IceCube results at PeV
energies.

\begin{figure}[ht]
\centering\includegraphics[width=\textwidth]{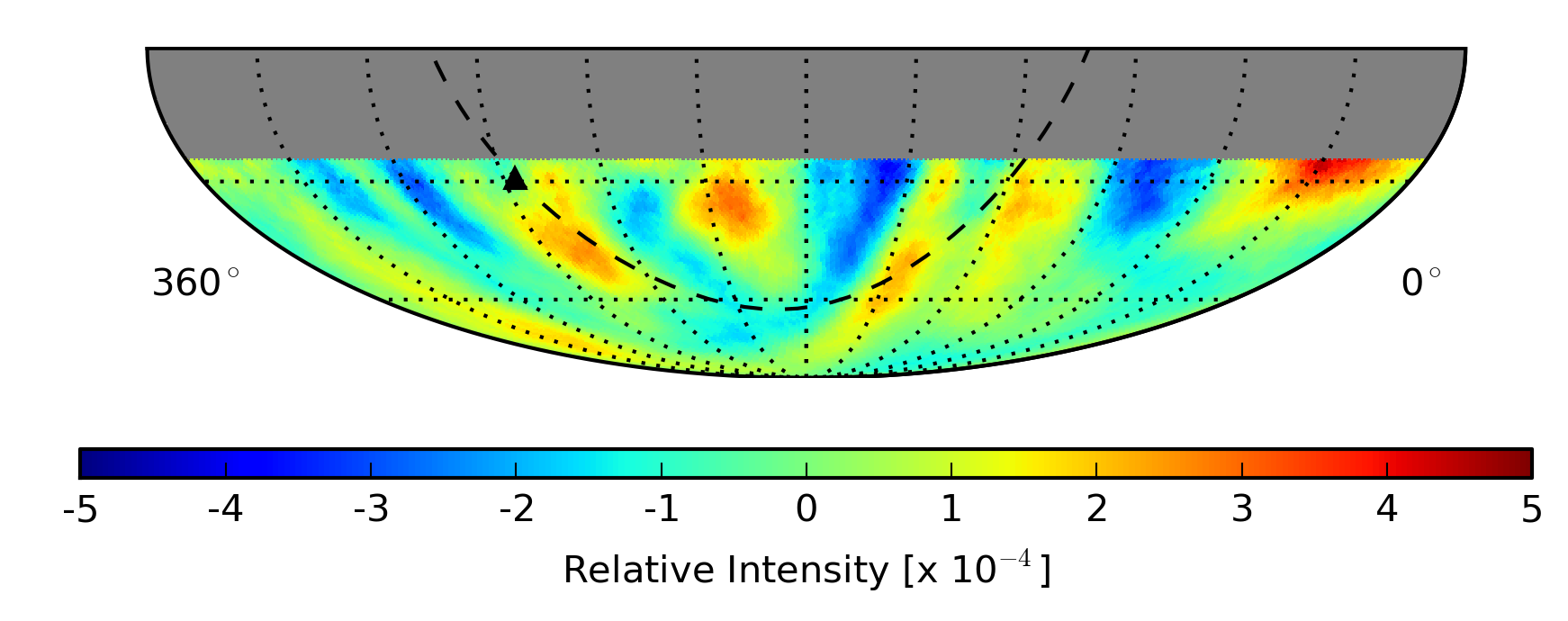}
  \caption{Relative intensity map of the full 6-year IceCube data set~\cite{Aartsen:2016ivj} for all energies
  (median energy 20\,TeV) after dipole- and quadrupole-subtraction. The subtraction of
  the dominant low-order multipoles reveals the small-scale structure with a relative 
  intensity of order $10^{-4}$.  The dashed line indicates the Galactic plane and the 
  triangle indicates the Galactic center.}
  \label{smallscale}
\end{figure}

While the large-scale structure dominates the anisotropy, there is also anisotropy on 
smaller scales.  The small-scale structure, with a relative intensity on the order of 
$10^{-4}$, and therefore roughly one order of magnitude weaker, becomes visible after the 
best-fit dipole and quadrupole are subtracted from the sky map.  Figure~\ref{smallscale} 
shows the relative intensity of the residual map.  Several excess and deficit regions
are visible at angular scales approaching the angular resolution of IceCube 
for CR primaries.  The strongest of these regions have statistical
significances exceeding $10\,\sigma$.

A study of the time dependence of the large- and small-scale structure
over the six-year period covered by this analysis reveals no significant change with
time.  An analysis of data taken with the AMANDA detector between 2000 and 2006 also
did not find any significant time variation of the observed large-scale
anisotropy~\cite{Aartsen:2013llaa}.

The source of the CR anisotropy remains unknown.  Homogeneous and isotropic 
diffusive propagation of CRs in the Galaxy from discrete sources leads to a density 
gradient of CRs, which produces a dipole.  While a small residual dipole anisotropy 
is therefore expected from diffusion theory, the observed anisotropy has a considerably 
more complex morphology than simple diffusion models suggest.  To explain the formation 
of the non-dipolar structures, additional processes like non-diffusive propagation of 
CRs in perturbed magnetic fields need to be considered.  

To complicate matters further, the observed amplitude of the anisotropy is smaller 
than predicted from diffusion theory.  Numerical studies show that particular 
realizations of Galactic source distributions reproduce the observed energy dependence 
of the anisotropy~\cite{Sveshnikova:2013dec}, but simulations based on plausible source 
distributions typically predict a larger amplitude for the anisotropy than what is 
observed~\cite{Sveshnikova:2013dec, Blasi:2012jan, Ptuskin:2012dec, Pohl:2013mar, 
Erlykin:2006apr}.  The  misalignment between the direction of the CR density 
gradient and the interstellar magnetic field lines may explain the smaller observed 
amplitude component (see~\cite{Kumar:2014apr, Mertsch:2015jan}). 

Non-dipolar structures may be produced by the interactions of CRs
with an isotropically turbulent interstellar magnetic field.  Scattering
processes with stochastic magnetic instabilities produce perturbations in the
arrival directions within the scattering mean free path.  Such perturbations may be
observed as stochastic localized excess or deficit regions~\cite{Giacinti:2012aug,
Biermann:2013may, Ahlers:2014jan, Ahlers:2015dwa, Lopez-Barquero:2015qpa, 
Harding:2015pna, Scherer:2016uyr}.

It has been shown recently that one or more local sources at Galactic longitudes 
between $120^{\circ}$ and $300^{\circ}$ in the presence of a strong ordered magnetic 
field in our local environment can explain the observations~\cite{Ahlers:2016njd}. 
The Vela SNR, created about 12,000 years ago, is identified as a candidate local source. 
The discrepancy between the predicted and observed amplitude could, at least in part, be 
a result of the limited capabilities of ground-based detectors to reconstruct
the true underlying anisotropy.  New analysis methods to correct for some of these 
observational biases have recently been developed~\cite{Ahlers:2016njl}.

There are other sources of magnetic perturbations on smaller scales, for example
the heliosphere, formed by the interaction between the solar wind and the
interstellar flow.  The heliosphere constitutes a perturbation in the 3\,$\mu$G 
local interstellar magnetic field.  The local magnetic field draping around the 
heliosphere might be a significant source of resonant scattering, capable of 
redistributing the arrival directions of TeV CR particles~\cite{Desiati:2013jan, 
IBEX:2014feb, Zhang:2014jul, Lopez-Barquero:2016wnt}.  CR acceleration 
from magnetic reconnection in the heliotail has also been proposed as an explanation 
of the localized small-scale excess regions and their harder 
spectrum~\cite{Lazarian:2010oct, Desiati:2012jun}.

\section{Conclusions}
\label{S:Conclusions}

We have reviewed how observations of neutrinos and cosmic rays with the IceCube neutrino telescope and its surface array IceTop have impacted our knowledge about the high-energy non-thermal universe. Only three years after their first detection, we know the spectrum and flavor composition of cosmic neutrinos in the energy interval between 10~TeV and several PeV with encouraging precision. The distribution of the neutrinos on the sky is compatible with an isotropic distribution, excluding a purely Galactic origin.

Surprisingly, no individual neutrino sources or transients have been observed so far that would pinpoint the origin of the cosmic neutrinos. However, putting all the information together, we can already make important statements about their origin. Blazar jets and GRBs can only be responsible for small fractions of the observed cosmic neutrinos. Less luminous sources with higher number densities are needed to explain the observed level of astrophysical neutrinos and the absence of detectable point sources at the same time. Coincidences of neutrino events with transient phenomena, a SN explosion and a Blazar flare, have been observed, however the circumstances make it impossible to exclude a chance occurrence. 
No neutrinos have been observed thus far that could be attributed to the GZK effect. Again, the non-observation of associated neutrinos starts to constrain evolution scenarios for UHE CR sources.

Direct observations of the spectrum and the anisotropy of CRs at TeV and PeV energies with IceCube and IceTop have provided accurate measurements of the shape of the CR spectrum from few~PeV to above one~EeV. Searches for point sources of photons or neutrons among the CR air showers recorded have been negative so far. Additionally, the large statistics of CR air showers collected by IceCube has allowed the most precise measurement of the CR anisotropy in the Southern hemisphere, confirming and extending the measurements from the Northern hemisphere. Both large-scale and small-scale components have been detected, however their origin is still not well understood.

Both IceCube and IceTop continue to collect data, likely for at least another decade. As the statistics of cosmic neutrinos and CRs increases, and the understanding of systematic effects improves, we can expect significant advances in understanding the neutrino sky, the origin of CRs, and their propagation and arrival at Earth.
However, the prospects for what IceCube can achieve within a reasonable time span of a few decades are limited by its size. Based on the experience and success of IceCube, efforts are underway to develop a next generation instrument, IceCube-Gen2 \cite{Aartsen:2014njl}. With its five times better sensitivity for sources than IceCube, ten times the statistics for cosmic neutrinos and at least ten times larger area for a surface array, it will truly mark the next big step towards understanding the origin and propagation of cosmic rays.

\section{Acknowledgements}

We acknowledge the support from the following agencies: U.S. National Science Foundation - Office of Polar Programs, U.S. National Science Foundation - Physics Division, University of Wisconsin Alumni Research Foundation, the Grid Laboratory Of Wisconsin (GLOW) grid infrastructure at the University of Wisconsin - Madison, the Open Science Grid (OSG) grid infrastructure; U.S. Department of Energy, and National Energy Research Scientific Computing Center, the Louisiana Optical Network Initiative (LONI) grid computing resources; Natural Sciences and Engineering Research Council of Canada, WestGrid and Compute/Calcul Canada; Swedish Research Council, Swedish Polar Research Secretariat, Swedish National Infrastructure for Computing (SNIC), and Knut and Alice Wallenberg Foundation, Sweden; German Ministry for Education and Research (BMBF), Deutsche Forschungsgemeinschaft (DFG), Helmholtz Alliance for Astroparticle Physics (HAP), Research Department of Plasmas with Complex Interactions (Bo\-chum), Germany; Fund for Scientific Research (FNRS-FWO), FWO Odysseus programme, Flanders Institute to encourage scientific and technological research in industry (IWT), Belgian Federal Science Policy Office (Belspo); University of Oxford, United Kingdom; Marsden Fund, New Zealand; Australian Research Council; Japan Society for Promotion of Science (JSPS); the Swiss National Science Foundation (SNSF), Switzerland; National Research Foundation of Korea (NRF); Villum Fonden, Danish National Research Foundation (DNRF), Denmark




\newpage
\bibliographystyle{model1-num-names}
\bibliography{review_article.bib}

\begin{thebibliography}{209}
\expandafter\ifx\csname natexlab\endcsname\relax\def\natexlab#1{#1}\fi
\providecommand{\bibinfo}[2]{#2}
\ifx\xfnm\relax \def\xfnm[#1]{\unskip,\space#1}\fi
\bibitem[{Aartsen et~al.(2013)}]{Aartsen:2013jdh}
\bibinfo{author}{M.~G. Aartsen}, et~al.,
\newblock \bibinfo{title}{{Evidence for High-Energy Extraterrestrial Neutrinos
  at the IceCube Detector}},
\newblock \bibinfo{journal}{Science} \bibinfo{volume}{342}
  (\bibinfo{year}{2013}) \bibinfo{pages}{1242856}.
\bibitem[{{Hillas}(1984)}]{Hillas:1984}
\bibinfo{author}{A.~M. {Hillas}},
\newblock \bibinfo{title}{{The Origin of Ultra-High-Energy Cosmic Rays}},
\newblock \bibinfo{journal}{Ann. Rev. Astron. Astrophys.} \bibinfo{volume}{22}
  (\bibinfo{year}{1984}) \bibinfo{pages}{425--444}.
\bibitem[{Waxman and Bahcall(1997)}]{Waxman:1997ti}
\bibinfo{author}{E.~Waxman}, \bibinfo{author}{J.~N. Bahcall},
\newblock \bibinfo{title}{{High-energy neutrinos from cosmological gamma-ray
  burst fireballs}},
\newblock \bibinfo{journal}{Phys. Rev. Lett.} \bibinfo{volume}{78}
  (\bibinfo{year}{1997}) \bibinfo{pages}{2292--2295}.
\bibitem[{Fang et~al.(2012)Fang, Kotera, and Olinto}]{Fang:2012rx}
\bibinfo{author}{K.~Fang}, \bibinfo{author}{K.~Kotera}, \bibinfo{author}{A.~V.
  Olinto},
\newblock \bibinfo{title}{{Newly-born pulsars as sources of ultrahigh energy
  cosmic rays}},
\newblock \bibinfo{journal}{Astrophys. J.} \bibinfo{volume}{750}
  (\bibinfo{year}{2012}) \bibinfo{pages}{118}.
\bibitem[{{Mannheim}(1993)}]{Mannheim:1993}
\bibinfo{author}{K.~{Mannheim}},
\newblock \bibinfo{title}{{The proton blazar}},
\newblock \bibinfo{journal}{Astronomy and Astrophysics} \bibinfo{volume}{269}
  (\bibinfo{year}{1993}) \bibinfo{pages}{67--76}.
\bibitem[{{Stecker} et~al.(1991){Stecker}, {Done}, {Salamon}, and
  {Sommers}}]{Stecker:1991}
\bibinfo{author}{F.~W. {Stecker}}, \bibinfo{author}{C.~{Done}},
  \bibinfo{author}{M.~H. {Salamon}}, \bibinfo{author}{P.~{Sommers}},
\newblock \bibinfo{title}{{High-energy neutrinos from active galactic nuclei}},
\newblock \bibinfo{journal}{Phys. Rev. Lett.} \bibinfo{volume}{66}
  (\bibinfo{year}{1991}) \bibinfo{pages}{2697--2700}.
\bibitem[{{Kashiyama} and {M{\'e}sz{\'a}ros}(2014)}]{Kashiyama:2014}
\bibinfo{author}{K.~{Kashiyama}}, \bibinfo{author}{P.~{M{\'e}sz{\'a}ros}},
\newblock \bibinfo{title}{{Galaxy Mergers as a Source of Cosmic Rays,
  Neutrinos, and Gamma Rays}},
\newblock \bibinfo{journal}{Astrophys. J. Lett.} \bibinfo{volume}{790}
  (\bibinfo{year}{2014}) \bibinfo{pages}{L14}.
\bibitem[{Ahlers et~al.(2010)}]{Ahlers:Fermilab}
\bibinfo{author}{M.~Ahlers}, et~al.  (\bibinfo{year}{2010}).
  \bibinfo{note}{FERMILAB-FN-0847-A, YITP-SB-10-01.}
\bibitem[{Greisen(1966)}]{Greisen:1966jv}
\bibinfo{author}{K.~Greisen},
\newblock \bibinfo{title}{{End to the cosmic ray spectrum?}},
\newblock \bibinfo{journal}{Phys. Rev. Lett.} \bibinfo{volume}{16}
  (\bibinfo{year}{1966}) \bibinfo{pages}{748--750}.
\bibitem[{Zatsepin and Kuzmin(1966)}]{Zatsepin:1966jv}
\bibinfo{author}{G.~T. Zatsepin}, \bibinfo{author}{V.~A. Kuzmin},
\newblock \bibinfo{title}{{Upper limit of the spectrum of cosmic rays}},
\newblock \bibinfo{journal}{JETP Lett.} \bibinfo{volume}{4}
  (\bibinfo{year}{1966}) \bibinfo{pages}{78--80}. \bibinfo{note}{[Pisma Zh.
  Eksp. Teor. Fiz.4,114(1966)]}.
\bibitem[{Olive(2016)}]{Olive:2016xmw}
\bibinfo{author}{K.~A. Olive},
\newblock \bibinfo{title}{{Review of Particle Physics}},
\newblock \bibinfo{journal}{Chin. Phys.} \bibinfo{volume}{C40}
  (\bibinfo{year}{2016}) \bibinfo{pages}{100001}.
\bibitem[{Amenomori et~al.(2005)}]{Tibet:2005jun}
\bibinfo{author}{M.~Amenomori}, et~al.,
\newblock \bibinfo{title}{{Large-Scale Sidereal Anisotropy of Galactic
  Cosmic-Ray Intensity Observed by the Tibet Air Shower Array}},
\newblock \bibinfo{journal}{Astrophys. J. Lett.} \bibinfo{volume}{626}
  (\bibinfo{year}{2005}) \bibinfo{pages}{L29--L32}.
\bibitem[{Amenomori et~al.(2006)}]{Tibet:2006oct}
\bibinfo{author}{M.~Amenomori}, et~al.,
\newblock \bibinfo{title}{{Anisotropy and Corotation of Galactic Cosmic Rays}},
\newblock \bibinfo{journal}{Science} \bibinfo{volume}{314}
  (\bibinfo{year}{2006}) \bibinfo{pages}{439--443}.
\bibitem[{Guillian et~al.(2007)}]{SuperK:2007mar}
\bibinfo{author}{G.~Guillian}, et~al.,
\newblock \bibinfo{title}{{Observation of the anisotropy of 10 TeV primary
  cosmic ray nuclei flux with the Super-Kamiokande-I detector}},
\newblock \bibinfo{journal}{Phys. Rev. D} \bibinfo{volume}{75}
  (\bibinfo{year}{2007}) \bibinfo{pages}{062003}.
\bibitem[{Abdo et~al.(2009)}]{Milagro:2009jun}
\bibinfo{author}{A.~Abdo}, et~al.,
\newblock \bibinfo{title}{{The Large-Scale Cosmic-Ray Anisotropy as Observed
  with Milagro}},
\newblock \bibinfo{journal}{Astrophys. J.} \bibinfo{volume}{698}
  (\bibinfo{year}{2009}) \bibinfo{pages}{2121--2130}.
\bibitem[{De~Jong et~al.(2011)}]{MINOS:2011icrc}
\bibinfo{author}{J.~De~Jong}, et~al.,
\newblock \bibinfo{title}{{Observations of Large Scale Sidereal Anisotropy in 1
  and 11 TeV cosmic rays from the MINOS experiment. (arXiv:1201.2621)}},
\newblock \bibinfo{journal}{Proc. 32nd Int. Cosmic Ray Conference, Beijing,
  China} \bibinfo{volume}{4} (\bibinfo{year}{2011}) \bibinfo{pages}{46}.
\bibitem[{Bartoli et~al.(2015)}]{Bartoli:2013}
\bibinfo{author}{B.~Bartoli}, et~al.,
\newblock \bibinfo{title}{{ARGO-YBJ Observation of the Large-scale Cosmic Ray
  Anisotropy During the Solar Minimum between Cycles 23 and 24}},
\newblock \bibinfo{journal}{Astrophys. J.} \bibinfo{volume}{809}
  (\bibinfo{year}{2015}) \bibinfo{pages}{90}.
\bibitem[{Belolaptikov et~al.(1997)}]{Belolaptikov:1997ry}
\bibinfo{author}{I.~A. Belolaptikov}, et~al.,
\newblock \bibinfo{title}{{The Baikal underwater neutrino telescope: Design,
  performance and first results}},
\newblock \bibinfo{journal}{Astropart. Phys.} \bibinfo{volume}{7}
  (\bibinfo{year}{1997}) \bibinfo{pages}{263--282}.
\bibitem[{Ageron et~al.(2011)}]{Ageron:2011nsa}
\bibinfo{author}{M.~Ageron}, et~al.,
\newblock \bibinfo{title}{{ANTARES: the first undersea neutrino telescope}},
\newblock \bibinfo{journal}{Nucl. Instrum. Meth.} \bibinfo{volume}{A656}
  (\bibinfo{year}{2011}) \bibinfo{pages}{11--38}.
\bibitem[{Adrian-Martinez et~al.(2016)}]{Adrian-Martinez:2016fdl}
\bibinfo{author}{S.~Adrian-Martinez}, et~al.,
\newblock \bibinfo{title}{{Letter of intent for KM3NeT 2.0}},
\newblock \bibinfo{journal}{J. Phys.} \bibinfo{volume}{G43}
  (\bibinfo{year}{2016}) \bibinfo{pages}{084001}.
\bibitem[{Avrorin et~al.(2015)}]{Avrorin:2014vca}
\bibinfo{author}{A.~D. Avrorin}, et~al.,
\newblock \bibinfo{title}{{Sensitivity of the Baikal-GVD neutrino telescope to
  neutrino emission toward the center of the galactic dark matter halo}},
\newblock \bibinfo{journal}{JETP Lett.} \bibinfo{volume}{101}
  (\bibinfo{year}{2015}) \bibinfo{pages}{289--294}.
\bibitem[{Achterberg et~al.(2006)}]{Achterberg:2006md}
\bibinfo{author}{A.~Achterberg}, et~al.,
\newblock \bibinfo{title}{{First Year Performance of The IceCube Neutrino
  Telescope}},
\newblock \bibinfo{journal}{Astropart. Phys.} \bibinfo{volume}{26}
  (\bibinfo{year}{2006}) \bibinfo{pages}{155--173}.
\bibitem[{Abbasi et~al.(2009)}]{Abbasi:2008aa}
\bibinfo{author}{R.~Abbasi}, et~al.,
\newblock \bibinfo{title}{{The IceCube Data Acquisition System: Signal Capture,
  Digitization, and Timestamping}},
\newblock \bibinfo{journal}{Nucl. Instrum. Meth.} \bibinfo{volume}{A601}
  (\bibinfo{year}{2009}) \bibinfo{pages}{294--316}.
\bibitem[{Aartsen et~al.(2016)}]{Aartsen:2016nxy}
\bibinfo{author}{M.~G. Aartsen}, et~al.,
\newblock \bibinfo{title}{{The IceCube Neutrino Observatory: Instrumentation
  and Online Systems}},
\newblock \bibinfo{journal}{arXiv eprints}  (\bibinfo{year}{2016})
  \bibinfo{pages}{astro--ph/1612.05093}.
\bibitem[{Abbasi et~al.(2013)}]{IceCube:2012nn}
\bibinfo{author}{R.~Abbasi}, et~al.,
\newblock \bibinfo{title}{{IceTop: The surface component of IceCube}},
\newblock \bibinfo{journal}{Nucl. Instrum. Meth.} \bibinfo{volume}{A700}
  (\bibinfo{year}{2013}) \bibinfo{pages}{188--220}.
\bibitem[{Abbasi et~al.(2012)}]{Abbasi:2011ym}
\bibinfo{author}{R.~Abbasi}, et~al.,
\newblock \bibinfo{title}{{The Design and Performance of IceCube DeepCore}},
\newblock \bibinfo{journal}{Astropart. Phys.} \bibinfo{volume}{35}
  (\bibinfo{year}{2012}) \bibinfo{pages}{615--624}.
\bibitem[{Learned and Pakvasa(1995)}]{Learned:1994wg}
\bibinfo{author}{J.~G. Learned}, \bibinfo{author}{S.~Pakvasa},
\newblock \bibinfo{title}{{Detecting tau-neutrino oscillations at PeV
  energies}},
\newblock \bibinfo{journal}{Astropart. Phys.} \bibinfo{volume}{3}
  (\bibinfo{year}{1995}) \bibinfo{pages}{267--274}.
\bibitem[{Aartsen et~al.(2013)}]{SpicePaper}
\bibinfo{author}{M.~G. Aartsen}, et~al.,
\newblock \bibinfo{title}{{Measurement of South Pole ice transparency with the
  IceCube LED calibration system}},
\newblock \bibinfo{journal}{Nucl. Instrum. Meth.} \bibinfo{volume}{A711}
  (\bibinfo{year}{2013}) \bibinfo{pages}{73--89}.
\bibitem[{{Aartsen} et~al.(2014)}]{Aartsen:ereco}
\bibinfo{author}{M.~G. {Aartsen}}, et~al.,
\newblock \bibinfo{title}{{Energy reconstruction methods in the IceCube
  neutrino telescope}},
\newblock \bibinfo{journal}{Journal of Instrumentation} \bibinfo{volume}{9}
  (\bibinfo{year}{2014}) \bibinfo{pages}{P03009}.
\bibitem[{Abbasi et~al.(2011)}]{Abbasi:2011jx}
\bibinfo{author}{R.~Abbasi}, et~al.,
\newblock \bibinfo{title}{{A Search for a Diffuse Flux of Astrophysical Muon
  Neutrinos with the IceCube 40-String Detector}},
\newblock \bibinfo{journal}{Phys. Rev.} \bibinfo{volume}{D84}
  (\bibinfo{year}{2011}) \bibinfo{pages}{082001}.
\bibitem[{Aartsen et~al.(2015)}]{Aartsen:2014qna}
\bibinfo{author}{M.~G. Aartsen}, et~al.,
\newblock \bibinfo{title}{{Development of a General Analysis and Unfolding
  Scheme and its Application to Measure the Energy Spectrum of Atmospheric
  Neutrinos with IceCube}},
\newblock \bibinfo{journal}{Eur. Phys. J.} \bibinfo{volume}{C75}
  (\bibinfo{year}{2015}) \bibinfo{pages}{116}.
\bibitem[{Aartsen et~al.(2013)}]{Aartsen:2012uu}
\bibinfo{author}{M.~G. Aartsen}, et~al.,
\newblock \bibinfo{title}{{Measurement of the Atmospheric $\nu_e$ flux in
  IceCube}},
\newblock \bibinfo{journal}{Phys. Rev. Lett.} \bibinfo{volume}{110}
  (\bibinfo{year}{2013}) \bibinfo{pages}{151105}.
\bibitem[{Aartsen et~al.(2015)}]{Aartsen:2015xup}
\bibinfo{author}{M.~G. Aartsen}, et~al.,
\newblock \bibinfo{title}{{Measurement of the Atmospheric $\nu_e$ Spectrum with
  IceCube}},
\newblock \bibinfo{journal}{Phys. Rev.} \bibinfo{volume}{D91}
  (\bibinfo{year}{2015}) \bibinfo{pages}{122004}.
\bibitem[{Enberg et~al.(2009)Enberg, Reno, and Sarcevic}]{Enberg:2008jm}
\bibinfo{author}{R.~Enberg}, \bibinfo{author}{M.~H. Reno},
  \bibinfo{author}{I.~Sarcevic},
\newblock \bibinfo{title}{{High energy neutrinos from charm in astrophysical
  sources}},
\newblock \bibinfo{journal}{Phys. Rev.} \bibinfo{volume}{D79}
  (\bibinfo{year}{2009}) \bibinfo{pages}{053006}.
\bibitem[{Aartsen et~al.(2013)}]{Aartsen:2013bka}
\bibinfo{author}{M.~G. Aartsen}, et~al.,
\newblock \bibinfo{title}{{First observation of PeV-energy neutrinos with
  IceCube}},
\newblock \bibinfo{journal}{Phys. Rev. Lett.} \bibinfo{volume}{111}
  (\bibinfo{year}{2013}) \bibinfo{pages}{021103}.
\bibitem[{Aartsen et~al.(2014)}]{Aartsen:2014gkd}
\bibinfo{author}{M.~G. Aartsen}, et~al.,
\newblock \bibinfo{title}{{Observation of High-Energy Astrophysical Neutrinos
  in Three Years of IceCube Data}},
\newblock \bibinfo{journal}{Phys. Rev. Lett.} \bibinfo{volume}{113}
  (\bibinfo{year}{2014}) \bibinfo{pages}{101101}.
\bibitem[{Aartsen et~al.(2015{\natexlab{a}})}]{Aartsen:2015zva}
\bibinfo{author}{M.~G. Aartsen}, et~al.,
\newblock \bibinfo{title}{{The IceCube Neutrino Observatory - Contributions to
  ICRC 2015 Part II: Atmospheric and Astrophysical Diffuse Neutrino Searches of
  All Flavors, 6: Observation of Astrophysical Neutrinos in four Years of
  IceCube Data}},
\newblock in: \bibinfo{booktitle}{{Proceedings, 34th International Cosmic Ray
  Conference (ICRC 2015): The Hague, The Netherlands, July 30-August 6, 2015
  (arXiv:1510.03223)}}.
\bibitem[{Aartsen et~al.(2015{\natexlab{b}})}]{Aartsen:2015ivb}
\bibinfo{author}{M.~G. Aartsen}, et~al.,
\newblock \bibinfo{title}{{Flavor Ratio of Astrophysical Neutrinos above 35 TeV
  in IceCube}},
\newblock \bibinfo{journal}{Phys. Rev. Lett.} \bibinfo{volume}{114}
  (\bibinfo{year}{2015}{\natexlab{b}}) \bibinfo{pages}{171102}.
\bibitem[{Aartsen et~al.(2015{\natexlab{c}})}]{Aartsen:2014muf}
\bibinfo{author}{M.~G. Aartsen}, et~al.,
\newblock \bibinfo{title}{{Atmospheric and astrophysical neutrinos above 1 TeV
  interacting in IceCube}},
\newblock \bibinfo{journal}{Phys. Rev.} \bibinfo{volume}{D91}
  (\bibinfo{year}{2015}{\natexlab{c}}) \bibinfo{pages}{022001}.
\bibitem[{Aartsen et~al.(2015{\natexlab{d}})}]{Aartsen:2015zvb}
\bibinfo{author}{M.~G. Aartsen}, et~al.,
\newblock \bibinfo{title}{{The IceCube Neutrino Observatory - Contributions to
  ICRC 2015 Part II: Atmospheric and Astrophysical Diffuse Neutrino Searches of
  All Flavors, 9: High Energy Astrophysical Neutrino Flux Characteristics for
  Neutrino-induced Cascades Using IC79 and IC86-String IceCube
  Configurations}},
\newblock in: \bibinfo{booktitle}{{Proceedings, 34th International Cosmic Ray
  Conference (ICRC 2015): The Hague, The Netherlands, July 30-August 6, 2015
  (arXiv:1510.05223)}}.
\bibitem[{Aartsen et~al.(2015{\natexlab{e}})}]{Aartsen:2015rwa}
\bibinfo{author}{M.~G. Aartsen}, et~al.,
\newblock \bibinfo{title}{{Evidence for Astrophysical Muon Neutrinos from the
  Northern Sky with IceCube}},
\newblock \bibinfo{journal}{Phys. Rev. Lett.} \bibinfo{volume}{115}
  (\bibinfo{year}{2015}{\natexlab{e}}) \bibinfo{pages}{081102}.
\bibitem[{Aartsen et~al.(2016{\natexlab{a}})}]{Aartsen:2016xlq}
\bibinfo{author}{M.~G. Aartsen}, et~al.,
\newblock \bibinfo{title}{{Observation and Characterization of a Cosmic Muon
  Neutrino Flux from the Northern Hemisphere using six years of IceCube data}},
\newblock \bibinfo{journal}{Astrophys. J.} \bibinfo{volume}{833}
  (\bibinfo{year}{2016}{\natexlab{a}}) \bibinfo{pages}{3}.
\bibitem[{Aartsen et~al.(2016{\natexlab{b}})}]{Aartsen:2015dlt}
\bibinfo{author}{M.~G. Aartsen}, et~al.,
\newblock \bibinfo{title}{{Search for Astrophysical Tau Neutrinos in Three
  Years of IceCube Data}},
\newblock \bibinfo{journal}{Phys. Rev.} \bibinfo{volume}{D93}
  (\bibinfo{year}{2016}{\natexlab{b}}) \bibinfo{pages}{022001}.
\bibitem[{Aartsen et~al.(2015{\natexlab{a}})}]{Aartsen:2015knd}
\bibinfo{author}{M.~G. Aartsen}, et~al.,
\newblock \bibinfo{title}{{A combined maximum-likelihood analysis of the
  high-energy astrophysical neutrino flux measured with IceCube}},
\newblock \bibinfo{journal}{Astrophys. J.} \bibinfo{volume}{809}
  (\bibinfo{year}{2015}{\natexlab{a}}) \bibinfo{pages}{98}.
\bibitem[{Aartsen et~al.(2015{\natexlab{b}})}]{Aartsen:2015zva3}
\bibinfo{author}{M.~G. Aartsen}, et~al.,
\newblock \bibinfo{title}{{The IceCube Neutrino Observatory - Contributions to
  ICRC 2015 Part II: Atmospheric and Astrophysical Diffuse Neutrino Searches of
  All Flavors, 3: Combined Analysis of the High-Energy Cosmic Neutrino Flux at
  the IceCube Detector}},
\newblock in: \bibinfo{booktitle}{{Proceedings, 34th International Cosmic Ray
  Conference (ICRC 2015): The Hague, The Netherlands, July 30-August 6, 2015
  (arXiv:1510.05223)}}.
\bibitem[{Klein et~al.(2013)Klein, Mikkelsen, and Becker~Tjus}]{Klein:2012ug}
\bibinfo{author}{S.~R. Klein}, \bibinfo{author}{R.~E. Mikkelsen},
  \bibinfo{author}{J.~Becker~Tjus},
\newblock \bibinfo{title}{{Muon Acceleration in Cosmic-ray Sources}},
\newblock \bibinfo{journal}{Astrophys. J.} \bibinfo{volume}{779}
  (\bibinfo{year}{2013}) \bibinfo{pages}{106}.
\bibitem[{Aartsen et~al.(2016)}]{Aartsen:2016oji}
\bibinfo{author}{M.~G. Aartsen}, et~al.,
\newblock \bibinfo{title}{{All-sky search for time-integrated neutrino emission
  from astrophysical sources with 7 years of IceCube data}},
\newblock \bibinfo{journal}{arXiv eprints}  (\bibinfo{year}{2016})
  \bibinfo{pages}{astro--ph/1609.04981}.
\bibitem[{Adrian-Martinez et~al.(2014)}]{Adrian-Martinez:2014wzf}
\bibinfo{author}{S.~Adrian-Martinez}, et~al.,
\newblock \bibinfo{title}{{Searches for Point-like and extended neutrino
  sources close to the Galactic Centre using the ANTARES neutrino Telescope}},
\newblock \bibinfo{journal}{Astrophys. J.} \bibinfo{volume}{786}
  (\bibinfo{year}{2014}) \bibinfo{pages}{L5}.
\bibitem[{Reimer(2015)}]{Reimer:2015abc}
\bibinfo{author}{A.~Reimer},
\newblock \bibinfo{title}{Photon-neutrino flux correlations from hadronic
  models of {AGN}?},
\newblock \bibinfo{journal}{{Proceedings, 34th International Cosmic Ray
  Conference (ICRC 2015)}}  (\bibinfo{year}{2015}).
\bibitem[{{Petropoulou} et~al.(2015){Petropoulou}, {Dimitrakoudis}, {Padovani},
  {Mastichiadis}, and {Resconi}}]{Petropoulou:2015}
\bibinfo{author}{M.~{Petropoulou}}, \bibinfo{author}{S.~{Dimitrakoudis}},
  \bibinfo{author}{P.~{Padovani}}, \bibinfo{author}{A.~{Mastichiadis}},
  \bibinfo{author}{E.~{Resconi}},
\newblock \bibinfo{title}{{Photohadronic origin of {$\gamma$} -ray BL Lac
  emission: implications for IceCube neutrinos}},
\newblock \bibinfo{journal}{\mnras} \bibinfo{volume}{448}
  (\bibinfo{year}{2015}) \bibinfo{pages}{2412--2429}.
\bibitem[{Aartsen et~al.(2015)}]{Aartsen:2015wto}
\bibinfo{author}{M.~G. Aartsen}, et~al.,
\newblock \bibinfo{title}{{Searches for Time Dependent Neutrino Sources with
  IceCube Data from 2008 to 2012}},
\newblock \bibinfo{journal}{Astrophys. J.} \bibinfo{volume}{807}
  (\bibinfo{year}{2015}) \bibinfo{pages}{46}.
\bibitem[{Aartsen et~al.(2014)}]{Aartsen:2014cva}
\bibinfo{author}{M.~G. Aartsen}, et~al.,
\newblock \bibinfo{title}{{Searches for Extended and Point-like Neutrino
  Sources with Four Years of IceCube Data}},
\newblock \bibinfo{journal}{Astrophys. J.} \bibinfo{volume}{796}
  (\bibinfo{year}{2014}) \bibinfo{pages}{109}.
\bibitem[{Aartsen et~al.(2013)}]{Aartsen:2013uuv}
\bibinfo{author}{M.~G. Aartsen}, et~al.,
\newblock \bibinfo{title}{{Search for Time-independent Neutrino Emission from
  Astrophysical Sources with 3 yr of IceCube Data}},
\newblock \bibinfo{journal}{Astrophys. J.} \bibinfo{volume}{779}
  (\bibinfo{year}{2013}) \bibinfo{pages}{132}.
\bibitem[{Abbasi et~al.(2012)}]{IceCube:2011ai}
\bibinfo{author}{R.~Abbasi}, et~al.,
\newblock \bibinfo{title}{{Time-Dependent Searches for Point Sources of
  Neutrinos with the 40-String and 22-String Configurations of IceCube}},
\newblock \bibinfo{journal}{Astrophys. J.} \bibinfo{volume}{744}
  (\bibinfo{year}{2012}) \bibinfo{pages}{1}.
\bibitem[{Adrian-Martinez et~al.(2012)}]{AdrianMartinez:2012rp}
\bibinfo{author}{S.~Adrian-Martinez}, et~al.,
\newblock \bibinfo{title}{{Search for Cosmic Neutrino Point Sources with Four
  Year Data of the ANTARES Telescope}},
\newblock \bibinfo{journal}{Astrophys. J.} \bibinfo{volume}{760}
  (\bibinfo{year}{2012}) \bibinfo{pages}{53}.
\bibitem[{Abbasi et~al.(2012)}]{Abbasi:2012zw}
\bibinfo{author}{R.~Abbasi}, et~al.,
\newblock \bibinfo{title}{{An absence of neutrinos associated with cosmic-ray
  acceleration in $\gamma$-ray bursts}},
\newblock \bibinfo{journal}{Nature} \bibinfo{volume}{484}
  (\bibinfo{year}{2012}) \bibinfo{pages}{351--353}.
\bibitem[{Adrian-Martinez et~al.(2013)}]{Adrian-Martinez:2013dsk}
\bibinfo{author}{S.~Adrian-Martinez}, et~al.,
\newblock \bibinfo{title}{{Search for muon neutrinos from gamma-ray bursts with
  the ANTARES neutrino telescope using 2008 to 2011 data}},
\newblock \bibinfo{journal}{Astron. Astrophys.} \bibinfo{volume}{559}
  (\bibinfo{year}{2013}) \bibinfo{pages}{A9}.
\bibitem[{Lipari(2006)}]{Lipari:2006uw}
\bibinfo{author}{P.~Lipari},
\newblock \bibinfo{title}{{Perspectives of High Energy Neutrino Astronomy}},
\newblock \bibinfo{journal}{Nucl. Instrum. Meth.} \bibinfo{volume}{A567}
  (\bibinfo{year}{2006}) \bibinfo{pages}{405--417}.
\bibitem[{Becker et~al.(2008)Becker, Rhode, Biermann, and
  Muenich}]{Becker:2006gi}
\bibinfo{author}{J.~K. Becker}, \bibinfo{author}{W.~Rhode},
  \bibinfo{author}{P.~L. Biermann}, \bibinfo{author}{K.~Muenich},
\newblock \bibinfo{title}{{Astrophysical implications of high energy neutrino
  limits. 1. Overall diffuse limits}},
\newblock \bibinfo{journal}{Astropart. Phys.} \bibinfo{volume}{28}
  (\bibinfo{year}{2008}) \bibinfo{pages}{98--118}.
\bibitem[{Silvestri and Barwick(2010)}]{Silvestri:2009xb}
\bibinfo{author}{A.~Silvestri}, \bibinfo{author}{S.~W. Barwick},
\newblock \bibinfo{title}{{Constraints on Extragalactic Point Source Flux from
  Diffuse Neutrino Limits}},
\newblock \bibinfo{journal}{Phys. Rev.} \bibinfo{volume}{D81}
  (\bibinfo{year}{2010}) \bibinfo{pages}{023001}.
\bibitem[{Ahlers and Halzen(2014)}]{Ahlers:2014ioa}
\bibinfo{author}{M.~Ahlers}, \bibinfo{author}{F.~Halzen},
\newblock \bibinfo{title}{{Pinpointing Extragalactic Neutrino Sources in Light
  of Recent IceCube Observations}},
\newblock \bibinfo{journal}{Phys. Rev.} \bibinfo{volume}{D90}
  (\bibinfo{year}{2014}) \bibinfo{pages}{043005}.
\bibitem[{Hopkins and Beacom(2006)}]{Hopkins:2006bw}
\bibinfo{author}{A.~M. Hopkins}, \bibinfo{author}{J.~F. Beacom},
\newblock \bibinfo{title}{{On the normalisation of the cosmic star formation
  history}},
\newblock \bibinfo{journal}{Astrophys. J.} \bibinfo{volume}{651}
  (\bibinfo{year}{2006}) \bibinfo{pages}{142--154}.
\bibitem[{Yuksel et~al.(2008)Yuksel, Kistler, Beacom, and
  Hopkins}]{Yuksel:2008cu}
\bibinfo{author}{H.~Yuksel}, \bibinfo{author}{M.~D. Kistler},
  \bibinfo{author}{J.~F. Beacom}, \bibinfo{author}{A.~M. Hopkins},
\newblock \bibinfo{title}{{Revealing the High-Redshift Star Formation Rate with
  Gamma-Ray Bursts}},
\newblock \bibinfo{journal}{Astrophys. J.} \bibinfo{volume}{683}
  (\bibinfo{year}{2008}) \bibinfo{pages}{L5--L8}.
\bibitem[{Aartsen et~al.(2016)}]{Aartsen:2016lir}
\bibinfo{author}{M.~G. Aartsen}, et~al.,
\newblock \bibinfo{title}{{The contribution of Fermi-2LAC blazars to the
  diffuse TeV-PeV neutrino flux}},
\newblock \bibinfo{journal}{arXiv eprints}  (\bibinfo{year}{2016})
  \bibinfo{pages}{astro--ph/1611.03874}.
\bibitem[{Berezinsky and Zatsepin(1969)}]{Beresinsky:1969qj}
\bibinfo{author}{V.~S. Berezinsky}, \bibinfo{author}{G.~T. Zatsepin},
\newblock \bibinfo{title}{{Cosmic rays at ultrahigh-energies (neutrino?)}},
\newblock \bibinfo{journal}{Phys. Lett.} \bibinfo{volume}{B28}
  (\bibinfo{year}{1969}) \bibinfo{pages}{423--424}.
\bibitem[{Yoshida and Teshima(1993)}]{Yoshida:1993pt}
\bibinfo{author}{S.~Yoshida}, \bibinfo{author}{M.~Teshima},
\newblock \bibinfo{title}{{Energy spectrum of ultrahigh-energy cosmic rays with
  extragalactic origin}},
\newblock \bibinfo{journal}{Prog. Theor. Phys.} \bibinfo{volume}{89}
  (\bibinfo{year}{1993}) \bibinfo{pages}{833--845}.
\bibitem[{Protheroe and Johnson(1996)}]{Protheroe:1995ft}
\bibinfo{author}{R.~J. Protheroe}, \bibinfo{author}{P.~A. Johnson},
\newblock \bibinfo{title}{{Propagation of ultrahigh-energy protons over
  cosmological distances and implications for topological defect models}},
\newblock \bibinfo{journal}{Astropart. Phys.} \bibinfo{volume}{4}
  (\bibinfo{year}{1996}) \bibinfo{pages}{253}.
\bibitem[{Engel et~al.(2001)Engel, Seckel, and Stanev}]{Engel:2001hd}
\bibinfo{author}{R.~Engel}, \bibinfo{author}{D.~Seckel},
  \bibinfo{author}{T.~Stanev},
\newblock \bibinfo{title}{{Neutrinos from propagation of ultrahigh-energy
  protons}},
\newblock \bibinfo{journal}{Phys. Rev.} \bibinfo{volume}{D64}
  (\bibinfo{year}{2001}) \bibinfo{pages}{093010}.
\bibitem[{Berezinsky et~al.(2006)Berezinsky, Gazizov, and
  Grigorieva}]{Berezinsky:2002nc}
\bibinfo{author}{V.~Berezinsky}, \bibinfo{author}{A.~Z. Gazizov},
  \bibinfo{author}{S.~I. Grigorieva},
\newblock \bibinfo{title}{{On astrophysical solution to ultrahigh-energy cosmic
  rays}},
\newblock \bibinfo{journal}{Phys. Rev.} \bibinfo{volume}{D74}
  (\bibinfo{year}{2006}) \bibinfo{pages}{043005}.
\bibitem[{Fodor et~al.(2003)Fodor, Katz, Ringwald, and Tu}]{Fodor:2003ph}
\bibinfo{author}{Z.~Fodor}, \bibinfo{author}{S.~D. Katz},
  \bibinfo{author}{A.~Ringwald}, \bibinfo{author}{H.~Tu},
\newblock \bibinfo{title}{{Bounds on the cosmogenic neutrino flux}},
\newblock \bibinfo{journal}{JCAP} \bibinfo{volume}{0311} (\bibinfo{year}{2003})
  \bibinfo{pages}{015}.
\bibitem[{Yuksel and Kistler(2007)}]{Yuksel:2006qb}
\bibinfo{author}{H.~Yuksel}, \bibinfo{author}{M.~D. Kistler},
\newblock \bibinfo{title}{{Enhanced cosmological GRB rates and implications for
  cosmogenic neutrinos}},
\newblock \bibinfo{journal}{Phys. Rev.} \bibinfo{volume}{D75}
  (\bibinfo{year}{2007}) \bibinfo{pages}{083004}.
\bibitem[{Takami et~al.(2009)Takami, Murase, Nagataki, and
  Sato}]{Takami:2007pp}
\bibinfo{author}{H.~Takami}, \bibinfo{author}{K.~Murase},
  \bibinfo{author}{S.~Nagataki}, \bibinfo{author}{K.~Sato},
\newblock \bibinfo{title}{{Cosmogenic neutrinos as a probe of the transition
  from Galactic to extragalactic cosmic rays}},
\newblock \bibinfo{journal}{Astropart. Phys.} \bibinfo{volume}{31}
  (\bibinfo{year}{2009}) \bibinfo{pages}{201--211}.
\bibitem[{Berezinsky et~al.(2011)Berezinsky, Gazizov, Kachelriess, and
  Ostapchenko}]{Berezinsky:2010xa}
\bibinfo{author}{V.~Berezinsky}, \bibinfo{author}{A.~Gazizov},
  \bibinfo{author}{M.~Kachelriess}, \bibinfo{author}{S.~Ostapchenko},
\newblock \bibinfo{title}{{Restricting UHECRs and cosmogenic neutrinos with
  Fermi-LAT}},
\newblock \bibinfo{journal}{Phys. Lett.} \bibinfo{volume}{B695}
  (\bibinfo{year}{2011}) \bibinfo{pages}{13--18}.
\bibitem[{Ahlers et~al.(2010)Ahlers, Anchordoqui, Gonzalez-Garcia, Halzen, and
  Sarkar}]{Ahlers:2010fw}
\bibinfo{author}{M.~Ahlers}, \bibinfo{author}{L.~A. Anchordoqui},
  \bibinfo{author}{M.~C. Gonzalez-Garcia}, \bibinfo{author}{F.~Halzen},
  \bibinfo{author}{S.~Sarkar},
\newblock \bibinfo{title}{{GZK Neutrinos after the Fermi-LAT Diffuse Photon
  Flux Measurement}},
\newblock \bibinfo{journal}{Astropart. Phys.} \bibinfo{volume}{34}
  (\bibinfo{year}{2010}) \bibinfo{pages}{106--115}.
\bibitem[{Gelmini et~al.(2012)Gelmini, Kalashev, and Semikoz}]{Gelmini:2011kg}
\bibinfo{author}{G.~B. Gelmini}, \bibinfo{author}{O.~Kalashev},
  \bibinfo{author}{D.~V. Semikoz},
\newblock \bibinfo{title}{{Gamma-Ray Constraints on Maximum Cosmogenic Neutrino
  Fluxes and UHECR Source Evolution Models}},
\newblock \bibinfo{journal}{JCAP} \bibinfo{volume}{1201} (\bibinfo{year}{2012})
  \bibinfo{pages}{044}.
\bibitem[{Decerprit and Allard(2011)}]{Decerprit:2011qe}
\bibinfo{author}{G.~Decerprit}, \bibinfo{author}{D.~Allard},
\newblock \bibinfo{title}{{Constraints on the origin of ultrahigh energy cosmic
  rays from cosmogenic neutrinos and photons}},
\newblock \bibinfo{journal}{Astron. Astrophys.} \bibinfo{volume}{535}
  (\bibinfo{year}{2011}) \bibinfo{pages}{A66}.
\bibitem[{Heinze et~al.(2016)Heinze, Boncioli, Bustamante, and
  Winter}]{Heinze:2015hhp}
\bibinfo{author}{J.~Heinze}, \bibinfo{author}{D.~Boncioli},
  \bibinfo{author}{M.~Bustamante}, \bibinfo{author}{W.~Winter},
\newblock \bibinfo{title}{{Cosmogenic Neutrinos Challenge the Cosmic Ray Proton
  Dip Model}},
\newblock \bibinfo{journal}{Astrophys. J.} \bibinfo{volume}{825}
  (\bibinfo{year}{2016}) \bibinfo{pages}{122}.
\bibitem[{Supanitsky(2016)}]{Supanitsky:2016gke}
\bibinfo{author}{A.~D. Supanitsky},
\newblock \bibinfo{title}{{Implications of gamma-ray observations on proton
  models of ultrahigh energy cosmic rays}},
\newblock \bibinfo{journal}{Phys. Rev.} \bibinfo{volume}{D94}
  (\bibinfo{year}{2016}) \bibinfo{pages}{063002}.
\bibitem[{Abdo et~al.(2010)}]{Abdo:2010nz}
\bibinfo{author}{A.~A. Abdo}, et~al.,
\newblock \bibinfo{title}{{The Spectrum of the Isotropic Diffuse Gamma-Ray
  Emission Derived From First-Year Fermi Large Area Telescope Data}},
\newblock \bibinfo{journal}{Phys. Rev. Lett.} \bibinfo{volume}{104}
  (\bibinfo{year}{2010}) \bibinfo{pages}{101101}.
\bibitem[{Ackermann et~al.(2015)}]{Ackermann:2014usa}
\bibinfo{author}{M.~Ackermann}, et~al.,
\newblock \bibinfo{title}{{The spectrum of isotropic diffuse gamma-ray emission
  between 100 MeV and 820 GeV}},
\newblock \bibinfo{journal}{Astrophys. J.} \bibinfo{volume}{799}
  (\bibinfo{year}{2015}) \bibinfo{pages}{86}.
\bibitem[{Hooper et~al.(2005)Hooper, Taylor, and Sarkar}]{Hooper:2004jc}
\bibinfo{author}{D.~Hooper}, \bibinfo{author}{A.~Taylor},
  \bibinfo{author}{S.~Sarkar},
\newblock \bibinfo{title}{{The Impact of heavy nuclei on the cosmogenic
  neutrino flux}},
\newblock \bibinfo{journal}{Astropart. Phys.} \bibinfo{volume}{23}
  (\bibinfo{year}{2005}) \bibinfo{pages}{11--17}.
\bibitem[{Ave et~al.(2005)Ave, Busca, Olinto, Watson, and
  Yamamoto}]{Ave:2004uj}
\bibinfo{author}{M.~Ave}, \bibinfo{author}{N.~Busca}, \bibinfo{author}{A.~V.
  Olinto}, \bibinfo{author}{A.~A. Watson}, \bibinfo{author}{T.~Yamamoto},
\newblock \bibinfo{title}{{Cosmogenic neutrinos from ultra-high energy
  nuclei}},
\newblock \bibinfo{journal}{Astropart. Phys.} \bibinfo{volume}{23}
  (\bibinfo{year}{2005}) \bibinfo{pages}{19--29}.
\bibitem[{Hooper et~al.(2007)Hooper, Sarkar, and Taylor}]{Hooper:2006tn}
\bibinfo{author}{D.~Hooper}, \bibinfo{author}{S.~Sarkar},
  \bibinfo{author}{A.~M. Taylor},
\newblock \bibinfo{title}{{The intergalactic propagation of ultrahigh energy
  cosmic ray nuclei}},
\newblock \bibinfo{journal}{Astropart. Phys.} \bibinfo{volume}{27}
  (\bibinfo{year}{2007}) \bibinfo{pages}{199--212}.
\bibitem[{Allard et~al.(2006)}]{Allard:2006mv}
\bibinfo{author}{D.~Allard}, et~al.,
\newblock \bibinfo{title}{{Cosmogenic Neutrinos from the propagation of
  Ultrahigh Energy Nuclei}},
\newblock \bibinfo{journal}{JCAP} \bibinfo{volume}{0609} (\bibinfo{year}{2006})
  \bibinfo{pages}{005}.
\bibitem[{Anchordoqui et~al.(2007)Anchordoqui, Goldberg, Hooper, Sarkar, and
  Taylor}]{Anchordoqui:2007fi}
\bibinfo{author}{L.~A. Anchordoqui}, \bibinfo{author}{H.~Goldberg},
  \bibinfo{author}{D.~Hooper}, \bibinfo{author}{S.~Sarkar},
  \bibinfo{author}{A.~M. Taylor},
\newblock \bibinfo{title}{{Predictions for the Cosmogenic Neutrino Flux in
  Light of New Data from the Pierre Auger Observatory}},
\newblock \bibinfo{journal}{Phys. Rev.} \bibinfo{volume}{D76}
  (\bibinfo{year}{2007}) \bibinfo{pages}{123008}.
\bibitem[{Aloisio et~al.(2011)Aloisio, Berezinsky, and
  Gazizov}]{Aloisio:2009sj}
\bibinfo{author}{R.~Aloisio}, \bibinfo{author}{V.~Berezinsky},
  \bibinfo{author}{A.~Gazizov},
\newblock \bibinfo{title}{{Ultra High Energy Cosmic Rays: The disappointing
  model}},
\newblock \bibinfo{journal}{Astropart. Phys.} \bibinfo{volume}{34}
  (\bibinfo{year}{2011}) \bibinfo{pages}{620--626}.
\bibitem[{Kotera et~al.(2010)Kotera, Allard, and Olinto}]{Kotera:2010yn}
\bibinfo{author}{K.~Kotera}, \bibinfo{author}{D.~Allard},
  \bibinfo{author}{A.~V. Olinto},
\newblock \bibinfo{title}{{Cosmogenic Neutrinos: parameter space and
  detectabilty from PeV to ZeV}},
\newblock \bibinfo{journal}{JCAP} \bibinfo{volume}{1010} (\bibinfo{year}{2010})
  \bibinfo{pages}{013}.
\bibitem[{Ahlers and Salvado(2011)}]{Ahlers:2011sd}
\bibinfo{author}{M.~Ahlers}, \bibinfo{author}{J.~Salvado},
\newblock \bibinfo{title}{{Cosmogenic gamma-rays and the composition of cosmic
  rays}},
\newblock \bibinfo{journal}{Phys. Rev.} \bibinfo{volume}{D84}
  (\bibinfo{year}{2011}) \bibinfo{pages}{085019}.
\bibitem[{Ahlers and Halzen(2012)}]{Ahlers:2012rz}
\bibinfo{author}{M.~Ahlers}, \bibinfo{author}{F.~Halzen},
\newblock \bibinfo{title}{{Minimal Cosmogenic Neutrinos}},
\newblock \bibinfo{journal}{Phys. Rev.} \bibinfo{volume}{D86}
  (\bibinfo{year}{2012}) \bibinfo{pages}{083010}.
\bibitem[{Aartsen et~al.(2016)}]{Aartsen:2016ngq}
\bibinfo{author}{M.~G. Aartsen}, et~al.,
\newblock \bibinfo{title}{{Constraints on ultra-high-energy cosmic ray sources
  from a search for neutrinos above 10 PeV with IceCube}},
\newblock \bibinfo{journal}{Phys. Rev. Lett.} \bibinfo{volume}{117}
  (\bibinfo{year}{2016}) \bibinfo{pages}{241101}.
\bibitem[{{Murase} et~al.(2014){Murase}, {Inoue}, and {Dermer}}]{Murase:2014}
\bibinfo{author}{K.~{Murase}}, \bibinfo{author}{Y.~{Inoue}},
  \bibinfo{author}{C.~D. {Dermer}},
\newblock \bibinfo{title}{{Diffuse neutrino intensity from the inner jets of
  active galactic nuclei: Impacts of external photon fields and the blazar
  sequence}},
\newblock \bibinfo{journal}{\prd} \bibinfo{volume}{90} (\bibinfo{year}{2014})
  \bibinfo{pages}{023007}.
\bibitem[{Padovani et~al.(2015)Padovani, Petropoulou, Giommi, and
  Resconi}]{Padovani:2015mba}
\bibinfo{author}{P.~Padovani}, \bibinfo{author}{M.~Petropoulou},
  \bibinfo{author}{P.~Giommi}, \bibinfo{author}{E.~Resconi},
\newblock \bibinfo{title}{{A simplified view of blazars: the neutrino
  background}},
\newblock \bibinfo{journal}{Mon. Not. Roy. Astron. Soc.} \bibinfo{volume}{452}
  (\bibinfo{year}{2015}) \bibinfo{pages}{1877--1887}.
\bibitem[{Waxman(1995)}]{Waxman:1995vg}
\bibinfo{author}{E.~Waxman},
\newblock \bibinfo{title}{{Cosmological gamma-ray bursts and the highest energy
  cosmic rays}},
\newblock \bibinfo{journal}{Phys. Rev. Lett.} \bibinfo{volume}{75}
  (\bibinfo{year}{1995}) \bibinfo{pages}{386--389}.
\bibitem[{Aartsen et~al.(2016)}]{Aartsen:2016qcr}
\bibinfo{author}{M.~G. Aartsen}, et~al.,
\newblock \bibinfo{title}{{An All-Sky Search for Three Flavors of Neutrinos
  from Gamma-Ray Bursts with the IceCube Neutrino Observatory}},
\newblock \bibinfo{journal}{Astrophys. J.} \bibinfo{volume}{824}
  (\bibinfo{year}{2016}) \bibinfo{pages}{115}.
\bibitem[{Ahlers et~al.(2011)Ahlers, Gonzalez-Garcia, and
  Halzen}]{Ahlers:2011jj}
\bibinfo{author}{M.~Ahlers}, \bibinfo{author}{M.~C. Gonzalez-Garcia},
  \bibinfo{author}{F.~Halzen},
\newblock \bibinfo{title}{{GRBs on probation: testing the UHE CR paradigm with
  IceCube}},
\newblock \bibinfo{journal}{Astropart. Phys.} \bibinfo{volume}{35}
  (\bibinfo{year}{2011}) \bibinfo{pages}{87--94}.
\bibitem[{Bustamante et~al.(2015)Bustamante, Baerwald, Murase, and
  Winter}]{Bustamante:2014oka}
\bibinfo{author}{M.~Bustamante}, \bibinfo{author}{P.~Baerwald},
  \bibinfo{author}{K.~Murase}, \bibinfo{author}{W.~Winter},
\newblock \bibinfo{title}{{Neutrino and cosmic-ray emission from multiple
  internal shocks in gamma-ray bursts}},
\newblock \bibinfo{journal}{Nature Commun.} \bibinfo{volume}{6}
  (\bibinfo{year}{2015}) \bibinfo{pages}{6783}.
\bibitem[{Senno et~al.(2016)Senno, Murase, and Meszaros}]{Senno:2015tsn}
\bibinfo{author}{N.~Senno}, \bibinfo{author}{K.~Murase},
  \bibinfo{author}{P.~Meszaros},
\newblock \bibinfo{title}{{Choked Jets and Low-Luminosity Gamma-Ray Bursts as
  Hidden Neutrino Sources}},
\newblock \bibinfo{journal}{Phys. Rev.} \bibinfo{volume}{D93}
  (\bibinfo{year}{2016}) \bibinfo{pages}{083003}.
\bibitem[{Razzaque et~al.(2005)Razzaque, Meszaros, and
  Waxman}]{Razzaque:2005bh}
\bibinfo{author}{S.~Razzaque}, \bibinfo{author}{P.~Meszaros},
  \bibinfo{author}{E.~Waxman},
\newblock \bibinfo{title}{{High energy neutrinos from a slow jet model of core
  collapse supernovae}},
\newblock \bibinfo{journal}{Mod. Phys. Lett.} \bibinfo{volume}{A20}
  (\bibinfo{year}{2005}) \bibinfo{pages}{2351--2368}.
\bibitem[{Ando and Beacom(2005)}]{Ando:2005xi}
\bibinfo{author}{S.~Ando}, \bibinfo{author}{J.~F. Beacom},
\newblock \bibinfo{title}{{Revealing the supernova-gamma-ray burst connection
  with TeV neutrinos}},
\newblock \bibinfo{journal}{Phys. Rev. Lett.} \bibinfo{volume}{95}
  (\bibinfo{year}{2005}) \bibinfo{pages}{061103}.
\bibitem[{Tamborra and Ando(2016)}]{Tamborra:2015fzv}
\bibinfo{author}{I.~Tamborra}, \bibinfo{author}{S.~Ando},
\newblock \bibinfo{title}{{Inspecting the supernova -- gamma-ray-burst
  connection with high-energy neutrinos}},
\newblock \bibinfo{journal}{Phys. Rev.} \bibinfo{volume}{D93}
  (\bibinfo{year}{2016}) \bibinfo{pages}{053010}.
\bibitem[{{Hjorth} and {Bloom}(2012)}]{2012grbu.book..169H}
\bibinfo{author}{J.~{Hjorth}}, \bibinfo{author}{J.~S. {Bloom}},
  \bibinfo{title}{{The Gamma-Ray Burst - Supernova Connection}},
  volume~\bibinfo{volume}{51} of \textit{\bibinfo{series}{Cambridge
  Astrophysics Series}}, \bibinfo{publisher}{Cambridge University Press},
  \bibinfo{address}{Cambridge}, pp. \bibinfo{pages}{169--190}.
\bibitem[{Murase et~al.(2014)Murase, Thompson, and Ofek}]{Murase:2013kda}
\bibinfo{author}{K.~Murase}, \bibinfo{author}{T.~A. Thompson},
  \bibinfo{author}{E.~O. Ofek},
\newblock \bibinfo{title}{{Probing Cosmic-Ray Ion Acceleration with Radio-Submm
  and Gamma-Ray Emission from Interaction-Powered Supernovae}},
\newblock \bibinfo{journal}{Mon. Not. Roy. Astron. Soc.} \bibinfo{volume}{440}
  (\bibinfo{year}{2014}) \bibinfo{pages}{2528--2543}.
\bibitem[{Zirakashvili and Ptuskin(2016)}]{Zirakashvili:2015mua}
\bibinfo{author}{V.~N. Zirakashvili}, \bibinfo{author}{V.~S. Ptuskin},
\newblock \bibinfo{title}{{Type IIn supernovae as sources of high energy
  astrophysical neutrinos}},
\newblock \bibinfo{journal}{Astropart. Phys.} \bibinfo{volume}{78}
  (\bibinfo{year}{2016}) \bibinfo{pages}{28--34}.
\bibitem[{{Abbasi} et~al.(2012)}]{2012A&A...539A..60A}
\bibinfo{author}{R.~{Abbasi}}, et~al.,
\newblock \bibinfo{title}{{Searching for soft relativistic jets in
  core-collapse supernovae with the IceCube optical follow-up program}},
\newblock \bibinfo{journal}{Astron. Astrophys.} \bibinfo{volume}{539}
  (\bibinfo{year}{2012}) \bibinfo{pages}{A60}.
\bibitem[{{Law} et~al.(2009)}]{2009PASP..121.1395L}
\bibinfo{author}{N.~M. {Law}}, et~al.,
\newblock \bibinfo{title}{{The Palomar Transient Factory: System Overview,
  Performance, and First Results}},
\newblock \bibinfo{journal}{Publications of the Astronomical Society of the
  Pacific} \bibinfo{volume}{121} (\bibinfo{year}{2009})
  \bibinfo{pages}{1395--1408}.
\bibitem[{Aartsen et~al.(2015)}]{Aartsen:2015trq}
\bibinfo{author}{M.~G. Aartsen}, et~al.,
\newblock \bibinfo{title}{{The Detection of a SN IIn in Optical Follow-up
  Observations of IceCube Neutrino Events}},
\newblock \bibinfo{journal}{Astrophys. J.} \bibinfo{volume}{811}
  (\bibinfo{year}{2015}) \bibinfo{pages}{52}.
\bibitem[{{Smartt} et~al.(2016)}]{PanSTARRS_GCN}
\bibinfo{author}{S.~{Smartt}}, et~al.,
\newblock \bibinfo{journal}{GCN Circular} \bibinfo{volume}{19381}
  (\bibinfo{year}{2016}).
\bibitem[{Aartsen et~al.(2016)}]{Aartsen:2016qbu}
\bibinfo{author}{M.~G. Aartsen}, et~al.,
\newblock \bibinfo{title}{{Very High-Energy Gamma-Ray Follow-Up Program Using
  Neutrino Triggers from IceCube}},
\newblock \bibinfo{journal}{arXiv eprints}  (\bibinfo{year}{2016})
  \bibinfo{pages}{astro--ph/1610.01814}.
\bibitem[{Kadler et~al.(2016)}]{Kadler:2016ygj}
\bibinfo{author}{M.~Kadler}, et~al.,
\newblock \bibinfo{title}{{Coincidence of a high-fluence blazar outburst with a
  PeV-energy neutrino event}},
\newblock \bibinfo{journal}{Nature Phys.} \bibinfo{volume}{12}
  (\bibinfo{year}{2016}) \bibinfo{pages}{807--814}.
\bibitem[{Smith et~al.(2013)}]{Smith:2012eu}
\bibinfo{author}{M.~W.~E. Smith}, et~al.,
\newblock \bibinfo{title}{{The Astrophysical Multimessenger Observatory Network
  (AMON)}},
\newblock \bibinfo{journal}{Astropart. Phys.} \bibinfo{volume}{45}
  (\bibinfo{year}{2013}) \bibinfo{pages}{56--70}.
\bibitem[{Aartsen et~al.(2016)}]{Aartsen:2016lmt}
\bibinfo{author}{M.~G. Aartsen}, et~al.,
\newblock \bibinfo{title}{{The IceCube Realtime Alert System}},
\newblock \bibinfo{journal}{arXiv eprints}  (\bibinfo{year}{2016})
  \bibinfo{pages}{astro--ph/1612.06028}.
\bibitem[{{Abbott} and et~al.(2016)}]{2016PhRvL.116f1102A}
\bibinfo{author}{B.~P. {Abbott}}, \bibinfo{author}{et~al.},
\newblock \bibinfo{title}{{Observation of Gravitational Waves from a Binary
  Black Hole Merger}},
\newblock \bibinfo{journal}{Physical Review Letters} \bibinfo{volume}{116}
  (\bibinfo{year}{2016}) \bibinfo{pages}{061102}.
\bibitem[{Adrian-Martinez et~al.(2016)}]{Adrian-Martinez:2016xgn}
\bibinfo{author}{S.~Adrian-Martinez}, et~al.,
\newblock \bibinfo{title}{{High-energy Neutrino follow-up search of
  Gravitational Wave Event GW150914 with ANTARES and IceCube}},
\newblock \bibinfo{journal}{Phys. Rev.} \bibinfo{volume}{D93}
  (\bibinfo{year}{2016}) \bibinfo{pages}{122010}.
\bibitem[{{Baade} and {Zwicky}(1934)}]{Baade1934}
\bibinfo{author}{W.~{Baade}}, \bibinfo{author}{F.~{Zwicky}},
\newblock \bibinfo{title}{{Cosmic Rays from Super-novae}},
\newblock \bibinfo{journal}{Proceedings of the National Academy of Science}
  \bibinfo{volume}{20} (\bibinfo{year}{1934}) \bibinfo{pages}{259--263}.
\bibitem[{Ackermann et~al.(2013)}]{Ackermann:2013wqa}
\bibinfo{author}{M.~Ackermann}, et~al.,
\newblock \bibinfo{title}{{Detection of the Characteristic Pion-Decay Signature
  in Supernova Remnants}},
\newblock \bibinfo{journal}{Science} \bibinfo{volume}{339}
  (\bibinfo{year}{2013}) \bibinfo{pages}{807}.
\bibitem[{Blasi(2009)}]{Blasi:2009hv}
\bibinfo{author}{P.~Blasi},
\newblock \bibinfo{title}{{The origin of the positron excess in cosmic rays}},
\newblock \bibinfo{journal}{Phys. Rev. Lett.} \bibinfo{volume}{103}
  (\bibinfo{year}{2009}) \bibinfo{pages}{051104}.
\bibitem[{Mertsch and Sarkar(2014)}]{Mertsch:2014poa}
\bibinfo{author}{P.~Mertsch}, \bibinfo{author}{S.~Sarkar},
\newblock \bibinfo{title}{{AMS-02 data confront acceleration of cosmic ray
  secondaries in nearby sources}},
\newblock \bibinfo{journal}{Phys. Rev.} \bibinfo{volume}{D90}
  (\bibinfo{year}{2014}) \bibinfo{pages}{061301}.
\bibitem[{Adriani et~al.(2009)}]{Adriani:2008zr}
\bibinfo{author}{O.~Adriani}, et~al.,
\newblock \bibinfo{title}{{An anomalous positron abundance in cosmic rays with
  energies 1.5-100 GeV}},
\newblock \bibinfo{journal}{Nature} \bibinfo{volume}{458}
  (\bibinfo{year}{2009}) \bibinfo{pages}{607--609}.
\bibitem[{Ackermann et~al.(2012)}]{FermiLAT:2011ab}
\bibinfo{author}{M.~Ackermann}, et~al.,
\newblock \bibinfo{title}{{Measurement of separate cosmic-ray electron and
  positron spectra with the Fermi Large Area Telescope}},
\newblock \bibinfo{journal}{Phys. Rev. Lett.} \bibinfo{volume}{108}
  (\bibinfo{year}{2012}) \bibinfo{pages}{011103}.
\bibitem[{Accardo et~al.(2014)}]{Accardo:2014lma}
\bibinfo{author}{L.~Accardo}, et~al.,
\newblock \bibinfo{title}{{High Statistics Measurement of the Positron Fraction
  in Primary Cosmic Rays of 0.5Ð500 GeV with the Alpha Magnetic Spectrometer
  on the International Space Station}},
\newblock \bibinfo{journal}{Phys. Rev. Lett.} \bibinfo{volume}{113}
  (\bibinfo{year}{2014}) \bibinfo{pages}{121101}.
\bibitem[{Ahlers et~al.(2009)Ahlers, Mertsch, and Sarkar}]{Ahlers:2009ae}
\bibinfo{author}{M.~Ahlers}, \bibinfo{author}{P.~Mertsch},
  \bibinfo{author}{S.~Sarkar},
\newblock \bibinfo{title}{{On cosmic ray acceleration in supernova remnants and
  the FERMI/PAMELA data}},
\newblock \bibinfo{journal}{Phys. Rev.} \bibinfo{volume}{D80}
  (\bibinfo{year}{2009}) \bibinfo{pages}{123017}.
\bibitem[{Stecker(1979)}]{Stecker:1978ah}
\bibinfo{author}{F.~W. Stecker},
\newblock \bibinfo{title}{{Diffuse Fluxes of Cosmic High-Energy Neutrinos}},
\newblock \bibinfo{journal}{Astrophys. J.} \bibinfo{volume}{228}
  (\bibinfo{year}{1979}) \bibinfo{pages}{919--927}.
\bibitem[{Domokos et~al.(1993)Domokos, Elliott, and
  Kovesi-Domokos}]{Domokos:1991tt}
\bibinfo{author}{G.~Domokos}, \bibinfo{author}{B.~Elliott},
  \bibinfo{author}{S.~Kovesi-Domokos},
\newblock \bibinfo{title}{{Cosmic neutrino production in the Milky Way}},
\newblock \bibinfo{journal}{J. Phys.} \bibinfo{volume}{G19}
  (\bibinfo{year}{1993}) \bibinfo{pages}{899--912}.
\bibitem[{Berezinsky et~al.(1993)Berezinsky, Gaisser, Halzen, and
  Stanev}]{Berezinsky:1992wr}
\bibinfo{author}{V.~S. Berezinsky}, \bibinfo{author}{T.~K. Gaisser},
  \bibinfo{author}{F.~Halzen}, \bibinfo{author}{T.~Stanev},
\newblock \bibinfo{title}{{Diffuse radiation from cosmic ray interactions in
  the galaxy}},
\newblock \bibinfo{journal}{Astropart. Phys.} \bibinfo{volume}{1}
  (\bibinfo{year}{1993}) \bibinfo{pages}{281--288}.
\bibitem[{{Bertsch} et~al.(1993)}]{Bertsch:1993}
\bibinfo{author}{D.~L. {Bertsch}}, et~al.,
\newblock \bibinfo{title}{{Diffuse Gamma-Ray Emission in the Galactic Plane
  from Cosmic-Ray, Matter, and Photon Interactions}},
\newblock \bibinfo{journal}{Astrophys. J.} \bibinfo{volume}{416}
  (\bibinfo{year}{1993}) \bibinfo{pages}{587}.
\bibitem[{Ingelman and Thunman(1996)}]{Ingelman:1996md}
\bibinfo{author}{G.~Ingelman}, \bibinfo{author}{M.~Thunman},
\newblock \bibinfo{title}{{Particle production in the interstellar medium}},
\newblock \bibinfo{journal}{arXiv eprints}  (\bibinfo{year}{1996})
  \bibinfo{pages}{hep--ph/9604286}.
\bibitem[{Evoli et~al.(2007)Evoli, Grasso, and Maccione}]{Evoli:2007iy}
\bibinfo{author}{C.~Evoli}, \bibinfo{author}{D.~Grasso},
  \bibinfo{author}{L.~Maccione},
\newblock \bibinfo{title}{{Diffuse Neutrino and Gamma-ray Emissions of the
  Galaxy above the TeV}},
\newblock \bibinfo{journal}{JCAP} \bibinfo{volume}{0706} (\bibinfo{year}{2007})
  \bibinfo{pages}{003}.
\bibitem[{Gaggero et~al.(2015)Gaggero, Grasso, Marinelli, Urbano, and
  Valli}]{Gaggero:2015xza}
\bibinfo{author}{D.~Gaggero}, \bibinfo{author}{D.~Grasso},
  \bibinfo{author}{A.~Marinelli}, \bibinfo{author}{A.~Urbano},
  \bibinfo{author}{M.~Valli},
\newblock \bibinfo{title}{{The gamma-ray and neutrino sky: A consistent picture
  of Fermi-LAT, Milagro, and IceCube results}},
\newblock \bibinfo{journal}{Astrophys. J.} \bibinfo{volume}{815}
  (\bibinfo{year}{2015}) \bibinfo{pages}{L25}.
\bibitem[{Ahlers et~al.(2016)Ahlers, Bai, Barger, and Lu}]{Ahlers:2015moa}
\bibinfo{author}{M.~Ahlers}, \bibinfo{author}{Y.~Bai},
  \bibinfo{author}{V.~Barger}, \bibinfo{author}{R.~Lu},
\newblock \bibinfo{title}{{Galactic neutrinos in the TeV to PeV range}},
\newblock \bibinfo{journal}{Phys. Rev.} \bibinfo{volume}{D93}
  (\bibinfo{year}{2016}) \bibinfo{pages}{013009}.
\bibitem[{Kelner et~al.(2006)Kelner, Aharonian, and Bugayov}]{Kelner:2006tc}
\bibinfo{author}{S.~R. Kelner}, \bibinfo{author}{F.~A. Aharonian},
  \bibinfo{author}{V.~V. Bugayov},
\newblock \bibinfo{title}{{Energy spectra of gamma-rays, electrons and
  neutrinos produced at proton-proton interactions in the very high energy
  regime}},
\newblock \bibinfo{journal}{Phys. Rev.} \bibinfo{volume}{D74}
  (\bibinfo{year}{2006}) \bibinfo{pages}{034018}. \bibinfo{note}{[Erratum:
  Phys. Rev.D79,039901(2009)]}.
\bibitem[{Block and Halzen(2011)}]{Block:2011vz}
\bibinfo{author}{M.~M. Block}, \bibinfo{author}{F.~Halzen},
\newblock \bibinfo{title}{{Experimental Confirmation that the Proton is
  Asymptotically a Black Disk}},
\newblock \bibinfo{journal}{Phys. Rev. Lett.} \bibinfo{volume}{107}
  (\bibinfo{year}{2011}) \bibinfo{pages}{212002}.
\bibitem[{Gaisser(2013)}]{Gaisser:2013ira}
\bibinfo{author}{T.~K. Gaisser},
\newblock \bibinfo{title}{{Atmospheric leptons}},
\newblock \bibinfo{journal}{EPJ Web Conf.} \bibinfo{volume}{52}
  (\bibinfo{year}{2013}) \bibinfo{pages}{09004}.
\bibitem[{Case and Bhattacharya(1998)}]{Case:1998qg}
\bibinfo{author}{G.~L. Case}, \bibinfo{author}{D.~Bhattacharya},
\newblock \bibinfo{title}{{A new sigma-d relation and its application to the
  galactic supernova remnant distribution}},
\newblock \bibinfo{journal}{Astrophys. J.} \bibinfo{volume}{504}
  (\bibinfo{year}{1998}) \bibinfo{pages}{761}.
\bibitem[{Ahlers and Murase(2014)}]{Ahlers:2013xia}
\bibinfo{author}{M.~Ahlers}, \bibinfo{author}{K.~Murase},
\newblock \bibinfo{title}{{Probing the Galactic Origin of the IceCube Excess
  with Gamma-Rays}},
\newblock \bibinfo{journal}{Phys. Rev.} \bibinfo{volume}{D90}
  (\bibinfo{year}{2014}) \bibinfo{pages}{023010}.
\bibitem[{Joshi et~al.(2014)Joshi, Winter, and Gupta}]{Joshi:2013aua}
\bibinfo{author}{J.~C. Joshi}, \bibinfo{author}{W.~Winter},
  \bibinfo{author}{N.~Gupta},
\newblock \bibinfo{title}{{How Many of the Observed Neutrino Events Can Be
  Described by Cosmic Ray Interactions in the Milky Way?}},
\newblock \bibinfo{journal}{Mon. Not. Roy. Astron. Soc.} \bibinfo{volume}{439}
  (\bibinfo{year}{2014}) \bibinfo{pages}{3414--3419}. \bibinfo{note}{[Erratum:
  Mon. Not. Roy. Astron. Soc.446,no.1,892(2014)]}.
\bibitem[{Kachelrie{\ss} and Ostapchenko(2014)}]{Kachelriess:2014oma}
\bibinfo{author}{M.~Kachelrie{\ss}}, \bibinfo{author}{S.~Ostapchenko},
\newblock \bibinfo{title}{{Neutrino yield from Galactic cosmic rays}},
\newblock \bibinfo{journal}{Phys. Rev.} \bibinfo{volume}{D90}
  (\bibinfo{year}{2014}) \bibinfo{pages}{083002}.
\bibitem[{Effenberger et~al.(2012)Effenberger, Fichtner, Scherer, and
  Busching}]{Effenberger:2012jc}
\bibinfo{author}{F.~Effenberger}, \bibinfo{author}{H.~Fichtner},
  \bibinfo{author}{K.~Scherer}, \bibinfo{author}{I.~Busching},
\newblock \bibinfo{title}{{Anisotropic diffusion of galactic cosmic ray protons
  and their steady-state azimuthal distribution}},
\newblock \bibinfo{journal}{Astron. Astrophys.} \bibinfo{volume}{547}
  (\bibinfo{year}{2012}) \bibinfo{pages}{A120}.
\bibitem[{Gaggero et~al.(2013)Gaggero, Maccione, Di~Bernardo, Evoli, and
  Grasso}]{Gaggero:2013rya}
\bibinfo{author}{D.~Gaggero}, \bibinfo{author}{L.~Maccione},
  \bibinfo{author}{G.~Di~Bernardo}, \bibinfo{author}{C.~Evoli},
  \bibinfo{author}{D.~Grasso},
\newblock \bibinfo{title}{{Three-Dimensional Model of Cosmic-Ray Lepton
  Propagation Reproduces Data from the Alpha Magnetic Spectrometer on the
  International Space Station}},
\newblock \bibinfo{journal}{Phys. Rev. Lett.} \bibinfo{volume}{111}
  (\bibinfo{year}{2013}) \bibinfo{pages}{021102}.
\bibitem[{Werner et~al.(2015)Werner, Kissmann, Strong, and
  Reimer}]{Werner:2014sya}
\bibinfo{author}{M.~Werner}, \bibinfo{author}{R.~Kissmann},
  \bibinfo{author}{A.~W. Strong}, \bibinfo{author}{O.~Reimer},
\newblock \bibinfo{title}{{Spiral Arms as Cosmic Ray Source Distributions}},
\newblock \bibinfo{journal}{Astropart. Phys.} \bibinfo{volume}{64}
  (\bibinfo{year}{2015}) \bibinfo{pages}{18--33}.
\bibitem[{Neronov et~al.(2014)Neronov, Semikoz, and Tchernin}]{Neronov:2013lza}
\bibinfo{author}{A.~Neronov}, \bibinfo{author}{D.~V. Semikoz},
  \bibinfo{author}{C.~Tchernin},
\newblock \bibinfo{title}{{PeV neutrinos from interactions of cosmic rays with
  the interstellar medium in the Galaxy}},
\newblock \bibinfo{journal}{Phys. Rev.} \bibinfo{volume}{D89}
  (\bibinfo{year}{2014}) \bibinfo{pages}{103002}.
\bibitem[{Neronov and Semikoz(2016)}]{Neronov:2015osa}
\bibinfo{author}{A.~Neronov}, \bibinfo{author}{D.~V. Semikoz},
\newblock \bibinfo{title}{{Evidence the Galactic contribution to the IceCube
  astrophysical neutrino flux}},
\newblock \bibinfo{journal}{Astropart. Phys.} \bibinfo{volume}{75}
  (\bibinfo{year}{2016}) \bibinfo{pages}{60--63}.
\bibitem[{Casanova and Dingus(2008)}]{Casanova:2007cf}
\bibinfo{author}{S.~Casanova}, \bibinfo{author}{B.~L. Dingus},
\newblock \bibinfo{title}{{Constraints on the TeV source population and its
  contribution to the galactic diffuse TeV emission}},
\newblock \bibinfo{journal}{Astropart. Phys.} \bibinfo{volume}{29}
  (\bibinfo{year}{2008}) \bibinfo{pages}{63--69}.
\bibitem[{{Blondin} et~al.(1998){Blondin}, {Wright}, {Borkowski}, and
  {Reynolds}}]{Blondin1998}
\bibinfo{author}{J.~M. {Blondin}}, \bibinfo{author}{E.~B. {Wright}},
  \bibinfo{author}{K.~J. {Borkowski}}, \bibinfo{author}{S.~P. {Reynolds}},
\newblock \bibinfo{title}{{Transition to the Radiative Phase in Supernova
  Remnants}},
\newblock \bibinfo{journal}{Astrophys. J.} \bibinfo{volume}{500}
  (\bibinfo{year}{1998}) \bibinfo{pages}{342--354}.
\bibitem[{Fox et~al.(2013)Fox, Kashiyama, and {M{\'e}sz{\'a}ros}}]{Fox:2013oza}
\bibinfo{author}{D.~B. Fox}, \bibinfo{author}{K.~Kashiyama},
  \bibinfo{author}{P.~{M{\'e}sz{\'a}ros}},
\newblock \bibinfo{title}{{Sub-PeV Neutrinos from TeV Unidentified Sources in
  the Galaxy}},
\newblock \bibinfo{journal}{Astrophys. J.} \bibinfo{volume}{774}
  (\bibinfo{year}{2013}) \bibinfo{pages}{74}.
\bibitem[{Mandelartz and Becker~Tjus(2015)}]{Mandelartz:2014sqa}
\bibinfo{author}{M.~Mandelartz}, \bibinfo{author}{J.~Becker~Tjus},
\newblock \bibinfo{title}{{Prediction of the diffuse neutrino flux from cosmic
  ray interactions near supernova remnants}},
\newblock \bibinfo{journal}{Astropart. Phys.} \bibinfo{volume}{65}
  (\bibinfo{year}{2015}) \bibinfo{pages}{80--100}.
\bibitem[{Lorimer et~al.(2006)}]{Lorimer:2006qs}
\bibinfo{author}{D.~R. Lorimer}, et~al.,
\newblock \bibinfo{title}{{The Parkes multibeam pulsar survey: VI. Discovery
  and timing of 142 pulsars and a Galactic population analysis}},
\newblock \bibinfo{journal}{Mon. Not. Roy. Astron. Soc.} \bibinfo{volume}{372}
  (\bibinfo{year}{2006}) \bibinfo{pages}{777--800}.
\bibitem[{Aartsen et~al.(2013)}]{Aartsen:2013wda}
\bibinfo{author}{M.~G. Aartsen}, et~al.,
\newblock \bibinfo{title}{{Measurement of the cosmic ray energy spectrum with
  IceTop-73}},
\newblock \bibinfo{journal}{Phys. Rev.} \bibinfo{volume}{D88}
  (\bibinfo{year}{2013}) \bibinfo{pages}{042004}.
\bibitem[{Aartsen et~al.(2016)}]{Aartsen:2015nss}
\bibinfo{author}{M.~G. Aartsen}, et~al.,
\newblock \bibinfo{title}{{Characterization of the Atmospheric Muon Flux in
  IceCube}},
\newblock \bibinfo{journal}{Astropart. Phys.} \bibinfo{volume}{78}
  (\bibinfo{year}{2016}) \bibinfo{pages}{1--27}.
\bibitem[{Abbasi et~al.(2013)}]{Abbasi:2012kza}
\bibinfo{author}{R.~Abbasi}, et~al.,
\newblock \bibinfo{title}{{Lateral Distribution of Muons in IceCube Cosmic Ray
  Events}},
\newblock \bibinfo{journal}{Phys. Rev.} \bibinfo{volume}{D87}
  (\bibinfo{year}{2013}) \bibinfo{pages}{012005}.
\bibitem[{Amenomori et~al.(2008)}]{Amenomori:2008aa}
\bibinfo{author}{M.~Amenomori}, et~al.,
\newblock \bibinfo{title}{{The All-particle spectrum of primary cosmic rays in
  the wide energy range from 10**14 eV to 10**17 eV observed with the Tibet-III
  air-shower array}},
\newblock \bibinfo{journal}{Astrophys. J.} \bibinfo{volume}{678}
  (\bibinfo{year}{2008}) \bibinfo{pages}{1165--1179}.
\bibitem[{Di~Sciascio(2014)}]{Sciascio:2014opa}
\bibinfo{author}{G.~Di~Sciascio},
\newblock \bibinfo{title}{{Measurement of the Cosmic Ray Energy Spectrum with
  ARGO-YBJ}},
\newblock \bibinfo{journal}{Frascati Phys. Ser.} \bibinfo{volume}{58}
  (\bibinfo{year}{2014}) \bibinfo{pages}{215 (arXiv:1408.6739)}.
\bibitem[{Apel et~al.(2014)}]{Apel:2014uka}
\bibinfo{author}{W.~D. Apel}, et~al.,
\newblock \bibinfo{title}{{The KASCADE-Grande energy spectrum of cosmic rays
  and the role of hadronic interaction models}},
\newblock \bibinfo{journal}{Adv. Space Res.} \bibinfo{volume}{53}
  (\bibinfo{year}{2014}) \bibinfo{pages}{1456--1469}.
\bibitem[{Prosin et~al.(2014)}]{Prosin:2014dxa}
\bibinfo{author}{V.~V. Prosin}, et~al.,
\newblock \bibinfo{title}{{Tunka-133: Results of 3 year operation}},
\newblock \bibinfo{journal}{Nucl. Instrum. Meth.} \bibinfo{volume}{A756}
  (\bibinfo{year}{2014}) \bibinfo{pages}{94--101}.
\bibitem[{Abbasi et~al.(2008)}]{Abbasi:2007sv}
\bibinfo{author}{R.~U. Abbasi}, et~al.,
\newblock \bibinfo{title}{{First observation of the Greisen-Zatsepin-Kuzmin
  suppression}},
\newblock \bibinfo{journal}{Phys. Rev. Lett.} \bibinfo{volume}{100}
  (\bibinfo{year}{2008}) \bibinfo{pages}{101101}.
\bibitem[{Abu-Zayyad et~al.(2013)}]{Abu-Zayyad:2013jra}
\bibinfo{author}{T.~Abu-Zayyad}, et~al.,
\newblock \bibinfo{title}{{The Energy Spectrum of Ultra-High-Energy Cosmic Rays
  Measured by the Telescope Array FADC Fluorescence Detectors in Monocular
  Mode}},
\newblock \bibinfo{journal}{Astropart. Phys.} \bibinfo{volume}{48}
  (\bibinfo{year}{2013}) \bibinfo{pages}{16--24}.
\bibitem[{Aug(????)}]{Auger:2015abc}
\bibinfo{title}{{The Pierre Auger Observatory: Contributions to the 34th
  International Cosmic Ray Conference (ICRC 2015), arXiv:1509.03732}},
\newblock in: \bibinfo{booktitle}{{Proceedings, 34th International Cosmic Ray
  Conference (ICRC 2015): The Hague, The Netherlands, July 30-August 6, 2015}}.
\bibitem[{Rawlins(2016)}]{Rawlins:2016bkc}
\bibinfo{author}{K.~Rawlins},
\newblock \bibinfo{title}{{Cosmic ray spectrum and composition from three years
  of IceTop and IceCube}},
\newblock \bibinfo{journal}{J. Phys. Conf. Ser.} \bibinfo{volume}{718}
  (\bibinfo{year}{2016}) \bibinfo{pages}{052033}.
\bibitem[{Bellotti et~al.(1990)}]{Bellotti:1989pa}
\bibinfo{author}{R.~Bellotti}, et~al.,
\newblock \bibinfo{title}{{Simultaneous Observation of Extensive Air Showers
  and Deep Underground Muons at the Gran Sasso Laboratory}},
\newblock \bibinfo{journal}{Phys. Rev.} \bibinfo{volume}{D42}
  (\bibinfo{year}{1990}) \bibinfo{pages}{1396--1403}.
\bibitem[{Ahrens et~al.(2004)}]{Ahrens:2004nn}
\bibinfo{author}{J.~Ahrens}, et~al.,
\newblock \bibinfo{title}{{Measurement of the cosmic ray composition at the
  knee with the SPASE-2/AMANDA-B10 detectors}},
\newblock \bibinfo{journal}{Astropart. Phys.} \bibinfo{volume}{21}
  (\bibinfo{year}{2004}) \bibinfo{pages}{565--581}.
\bibitem[{Peters(1961)}]{Peters:1961}
\bibinfo{author}{B.~Peters},
\newblock \bibinfo{title}{Primary cosmic radiation and extensive air showers},
\newblock \bibinfo{journal}{Nuovo Cimento} \bibinfo{volume}{XXII}
  (\bibinfo{year}{1961}) \bibinfo{pages}{800--819}.
\bibitem[{Kampert and Unger(2012)}]{Kampert:2012mx}
\bibinfo{author}{K.-H. Kampert}, \bibinfo{author}{M.~Unger},
\newblock \bibinfo{title}{{Measurements of the Cosmic Ray Composition with Air
  Shower Experiments}},
\newblock \bibinfo{journal}{Astropart. Phys.} \bibinfo{volume}{35}
  (\bibinfo{year}{2012}) \bibinfo{pages}{660--678}.
\bibitem[{Gaisser(2016)}]{Gaisser:2016ytm}
\bibinfo{author}{T.~K. Gaisser},
\newblock \bibinfo{title}{{Primary spectrum and composition with
  IceCube/IceTop}},
\newblock in: \bibinfo{booktitle}{{Cosmic Ray International Seminar: The status
  and the future of the UHE Cosmic Ray Physics in the post LHC era (CRIS 2015)
  Gallipoli, Italy, September 14-16, 2015} (arXiv:1601.06670)}.
\bibitem[{Dembinski and Gonzalez(2016)}]{Dembinski:2015xtn}
\bibinfo{author}{H.~P. Dembinski}, \bibinfo{author}{J.~Gonzalez},
\newblock \bibinfo{title}{{Surface muons in IceTop}},
\newblock \bibinfo{journal}{PoS} \bibinfo{volume}{ICRC2015}
  (\bibinfo{year}{2016}) \bibinfo{pages}{267}.
\bibitem[{Aab et~al.(2015)}]{Aab:2014pza}
\bibinfo{author}{A.~Aab}, et~al.,
\newblock \bibinfo{title}{{Muons in air showers at the Pierre Auger
  Observatory: Mean number in highly inclined events}},
\newblock \bibinfo{journal}{Phys. Rev.} \bibinfo{volume}{D91}
  (\bibinfo{year}{2015}) \bibinfo{pages}{032003}. \bibinfo{note}{[Erratum:
  Phys. Rev.D91,no.5,059901(2015)]}.
\bibitem[{Aab et~al.(2016)}]{Aab:2016hkv}
\bibinfo{author}{A.~Aab}, et~al.,
\newblock \bibinfo{title}{{Testing Hadronic Interactions at Ultrahigh Energies
  with Air Showers Measured by the Pierre Auger Observatory}},
\newblock \bibinfo{journal}{Phys. Rev. Lett.} \bibinfo{volume}{117}
  (\bibinfo{year}{2016}) \bibinfo{pages}{192001}.
\bibitem[{Aartsen et~al.(2013)}]{Aartsen:2012gka}
\bibinfo{author}{M.~G. Aartsen}, et~al.,
\newblock \bibinfo{title}{{Search for Galactic PeV Gamma Rays with the IceCube
  Neutrino Observatory}},
\newblock \bibinfo{journal}{Phys. Rev.} \bibinfo{volume}{D87}
  (\bibinfo{year}{2013}) \bibinfo{pages}{062002}.
\bibitem[{Aartsen et~al.(2016)}]{Aartsen:2016asr}
\bibinfo{author}{M.~G. Aartsen}, et~al.,
\newblock \bibinfo{title}{{Search for Sources of High Energy Neutrons with Four
  Years of Data from the IceTop Detector}},
\newblock \bibinfo{journal}{Astrophys. J.} \bibinfo{volume}{830}
  (\bibinfo{year}{2016}) \bibinfo{pages}{129}.
\bibitem[{{Nagashima} et~al.(1998){Nagashima}, {Fujimoto}, and
  {Jacklyn}}]{Nagashima:1998aug}
\bibinfo{author}{K.~{Nagashima}}, \bibinfo{author}{K.~{Fujimoto}},
  \bibinfo{author}{R.~M. {Jacklyn}},
\newblock \bibinfo{title}{{Galactic and heliotail-in anisotropies of cosmic
  rays as the origin of sidereal daily variation in the energy region
  $<$10$^{4}$~GeV}},
\newblock \bibinfo{journal}{J. Geophys. Res.} \bibinfo{volume}{103}
  (\bibinfo{year}{1998}) \bibinfo{pages}{17429--17440}.
\bibitem[{{Hall} et~al.(1999)}]{Hall:1999apr}
\bibinfo{author}{D.~L. {Hall}}, et~al.,
\newblock \bibinfo{title}{{Gaussian analysis of two hemisphere observations of
  galactic cosmic ray sidereal anisotropies}},
\newblock \bibinfo{journal}{J. Geophys. Res.} \bibinfo{volume}{104}
  (\bibinfo{year}{1999}) \bibinfo{pages}{6737--6750}.
\bibitem[{Bartoli et~al.(2013)}]{ARGO:2013oct}
\bibinfo{author}{B.~Bartoli}, et~al.,
\newblock \bibinfo{title}{{Medium scale anisotropy in the TeV cosmic ray flux
  observed by ARGO-YBJ}},
\newblock \bibinfo{journal}{Phys. Rev. D} \bibinfo{volume}{88}
  (\bibinfo{year}{2013}) \bibinfo{pages}{082001}.
\bibitem[{Abbasi et~al.(2010)}]{IceCube:2010aug}
\bibinfo{author}{R.~Abbasi}, et~al.,
\newblock \bibinfo{title}{{Measurement of the Anisotropy of Cosmic-ray Arrival
  Directions with IceCube}},
\newblock \bibinfo{journal}{Astrophys. J. Lett.} \bibinfo{volume}{718}
  (\bibinfo{year}{2010}) \bibinfo{pages}{L194--L198}.
\bibitem[{Abbasi et~al.(2011)}]{IceCube:2011oct}
\bibinfo{author}{R.~Abbasi}, et~al.,
\newblock \bibinfo{title}{{Observation of Anisotropy in the Arrival Directions
  of Galactic Cosmic Rays at Multiple Angular Scales with IceCube}},
\newblock \bibinfo{journal}{Astrophys. J.} \bibinfo{volume}{740}
  (\bibinfo{year}{2011}) \bibinfo{pages}{16}.
\bibitem[{Aartsen et~al.(2013)}]{IceCube:2013mar}
\bibinfo{author}{M.~Aartsen}, et~al.,
\newblock \bibinfo{title}{{Observation of Cosmic-Ray Anisotropy with the IceTop
  Air Shower Array}},
\newblock \bibinfo{journal}{Astrophys. J.} \bibinfo{volume}{765}
  (\bibinfo{year}{2013}) \bibinfo{pages}{55}.
\bibitem[{Aartsen et~al.(2016)}]{Aartsen:2016ivj}
\bibinfo{author}{M.~G. Aartsen}, et~al.,
\newblock \bibinfo{title}{{Anisotropy in Cosmic-ray Arrival Directions in the
  Southern Hemisphere Based on Six Years of Data From the IceCube Detector}},
\newblock \bibinfo{journal}{Astrophys. J.} \bibinfo{volume}{826}
  (\bibinfo{year}{2016}) \bibinfo{pages}{220}.
\bibitem[{Aglietta et~al.(2009)}]{Aglietta:2009feb}
\bibinfo{author}{M.~Aglietta}, et~al.,
\newblock \bibinfo{title}{{Evolution of the Cosmic-Ray Anisotropy Above
  10$^{14}$ eV}},
\newblock \bibinfo{journal}{Astrophys. J. Lett.} \bibinfo{volume}{692}
  (\bibinfo{year}{2009}) \bibinfo{pages}{L130--L133}.
\bibitem[{Abbasi et~al.(2012)}]{IceCube:2012feb}
\bibinfo{author}{R.~Abbasi}, et~al.,
\newblock \bibinfo{title}{{Observation of Anisotropy in the Galactic Cosmic-Ray
  Arrival Directions at 400 TeV with IceCube}},
\newblock \bibinfo{journal}{Astrophys. J.} \bibinfo{volume}{746}
  (\bibinfo{year}{2012}) \bibinfo{pages}{33}.
\bibitem[{Abeysekara et~al.(2014)}]{HAWC:2014dec}
\bibinfo{author}{A.~Abeysekara}, et~al.,
\newblock \bibinfo{title}{{Observation of Small-scale Anisotropy in the Arrival
  Direction Distribution of TeV Cosmic Rays with HAWC}},
\newblock \bibinfo{journal}{Astrophys. J.} \bibinfo{volume}{796}
  (\bibinfo{year}{2014}) \bibinfo{pages}{108}.
\bibitem[{Amenomori et~al.(2007)}]{Tibet:2007aug}
\bibinfo{author}{M.~Amenomori}, et~al.,
\newblock \bibinfo{title}{{Implication of the sidereal anisotropy of \~{}5 TeV
  cosmic ray intensity observed with the Tibet III air shower array}},
\newblock \bibinfo{journal}{American Institute of Physics Conference Series}
  \bibinfo{volume}{932} (\bibinfo{year}{2007}) \bibinfo{pages}{283}.
\bibitem[{Abdo et~al.(2008)}]{Milagro:2008nov}
\bibinfo{author}{A.~Abdo}, et~al.,
\newblock \bibinfo{title}{{Discovery of Localized Regions of Excess 10-TeV
  Cosmic Rays}},
\newblock \bibinfo{journal}{Phys. Rev. Lett.} \bibinfo{volume}{101}
  (\bibinfo{year}{2008}) \bibinfo{pages}{221101}.
\bibitem[{{Tilav} et~al.(2010)}]{Tilav:2010jan}
\bibinfo{author}{S.~{Tilav}}, et~al.,
\newblock \bibinfo{title}{{Atmospheric Variations as observed by IceCube
  (arXiv:1001.0776)}},
\newblock \bibinfo{journal}{Proc. 31st Int. Cosmic Ray Conference, Lodz,
  Poland}  (\bibinfo{year}{2010}).
\bibitem[{{Desiati}(2011)}]{Desiati:2011aug}
\bibinfo{author}{P.~{Desiati}},
\newblock \bibinfo{title}{{Seasonal Variations of High Energy Cosmic Ray Muons
  Observed by the IceCube Observatory as a Probe of Kaon/Pion Ratio}},
\newblock \bibinfo{journal}{Proc. 32nd Int. Cosmic Ray Conference, Beijing,
  China} \bibinfo{volume}{1} (\bibinfo{year}{2011}) \bibinfo{pages}{78
  (arXiv:1111.2735)}.
\bibitem[{Aartsen et~al.(2013)}]{Aartsen:2013llab}
\bibinfo{author}{M.~Aartsen}, et~al.,
\newblock \bibinfo{title}{{Seasonal variation of the muon multiplicity in
  cosmic rays at South Pole (Paper 0763, arXiv:1309.7006v2)}},
\newblock \bibinfo{journal}{Proc. 33rd Int. Cosmic Ray Conference, Rio de
  Janeiro, Brasil}  (\bibinfo{year}{2013}).
\bibitem[{Ahlers and Mertsch(2016)}]{ahlers_review}
\bibinfo{author}{M.~Ahlers}, \bibinfo{author}{P.~Mertsch},
\newblock \bibinfo{title}{{Origin of Small-Scale Anisotropies in Galactic
  Cosmic Rays}},
\newblock \bibinfo{journal}{Progress in Particle and Nuclear Physics}
  (\bibinfo{year}{2016}) \bibinfo{pages}{in preparation}.
\bibitem[{Abreu et~al.(2011)}]{Auger:2011mar}
\bibinfo{author}{P.~Abreu}, et~al.,
\newblock \bibinfo{title}{{Search for first harmonic modulation in the right
  ascension distribution of cosmic rays detected at the Pierre Auger
  Observatory}},
\newblock \bibinfo{journal}{Astropart. Phys.} \bibinfo{volume}{34}
  (\bibinfo{year}{2011}) \bibinfo{pages}{627--639}.
\bibitem[{Chiavassa et~al.(2016)}]{Chiavassa:2015jbg}
\bibinfo{author}{A.~Chiavassa}, et~al.,
\newblock \bibinfo{title}{{A study of the first harmonic of the large scale
  anisotropies with the KASCADE-Grande experiment}},
\newblock \bibinfo{journal}{PoS} \bibinfo{volume}{ICRC2015}
  (\bibinfo{year}{2016}) \bibinfo{pages}{281}.
\bibitem[{Aartsen et~al.(2013)}]{Aartsen:2013llaa}
\bibinfo{author}{M.~Aartsen}, et~al.,
\newblock \bibinfo{title}{{Study of the time-dependence of the cosmic-ray
  anisotropy with IceCube and AMANDA (Paper 0411, arXiv:1309.7006v2)}},
\newblock \bibinfo{journal}{Proc. 33rd Int. Cosmic Ray Conference, Rio de
  Janeiro, Brasil}  (\bibinfo{year}{2013}).
\bibitem[{{Sveshnikova} et~al.(2013){Sveshnikova}, {Strelnikova}, and
  {Ptuskin}}]{Sveshnikova:2013dec}
\bibinfo{author}{L.~G. {Sveshnikova}}, \bibinfo{author}{O.~N. {Strelnikova}},
  \bibinfo{author}{V.~S. {Ptuskin}},
\newblock \bibinfo{title}{{Spectrum and anisotropy of cosmic rays at
  TeV-PeV-energies and contribution of nearby sources}},
\newblock \bibinfo{journal}{Astropart. Phys.} \bibinfo{volume}{50}
  (\bibinfo{year}{2013}) \bibinfo{pages}{33--46}.
\bibitem[{{Blasi} and {Amato}(2012)}]{Blasi:2012jan}
\bibinfo{author}{P.~{Blasi}}, \bibinfo{author}{E.~{Amato}},
\newblock \bibinfo{title}{{Diffusive propagation of cosmic rays from supernova
  remnants in the Galaxy. II: anisotropy}},
\newblock \bibinfo{journal}{JCAP} \bibinfo{volume}{1} (\bibinfo{year}{2012})
  \bibinfo{pages}{11}.
\bibitem[{{Ptuskin}(2012)}]{Ptuskin:2012dec}
\bibinfo{author}{V.~{Ptuskin}},
\newblock \bibinfo{title}{{Propagation of galactic cosmic rays}},
\newblock \bibinfo{journal}{Astropart. Phys.} \bibinfo{volume}{39}
  (\bibinfo{year}{2012}) \bibinfo{pages}{44--51}.
\bibitem[{{Pohl} and {Eichler}(2013)}]{Pohl:2013mar}
\bibinfo{author}{M.~{Pohl}}, \bibinfo{author}{D.~{Eichler}},
\newblock \bibinfo{title}{{Understanding TeV-band Cosmic-Ray Anisotropy}},
\newblock \bibinfo{journal}{Astrophys. J.} \bibinfo{volume}{766}
  (\bibinfo{year}{2013}) \bibinfo{pages}{4}.
\bibitem[{{Erlykin} and {Wolfendale}(2006)}]{Erlykin:2006apr}
\bibinfo{author}{A.~D. {Erlykin}}, \bibinfo{author}{A.~W. {Wolfendale}},
\newblock \bibinfo{title}{{The anisotropy of galactic cosmic rays as a product
  of stochastic supernova explosions}},
\newblock \bibinfo{journal}{Astropart. Phys.} \bibinfo{volume}{25}
  (\bibinfo{year}{2006}) \bibinfo{pages}{183--194}.
\bibitem[{{Kumar} and {Eichler}(2014)}]{Kumar:2014apr}
\bibinfo{author}{R.~{Kumar}}, \bibinfo{author}{D.~{Eichler}},
\newblock \bibinfo{title}{{Large-scale Anisotropy of TeV-band Cosmic Rays}},
\newblock \bibinfo{journal}{Astrophys. J.} \bibinfo{volume}{785}
  (\bibinfo{year}{2014}) \bibinfo{pages}{129}.
\bibitem[{{Mertsch} and {Funk}(2015)}]{Mertsch:2015jan}
\bibinfo{author}{P.~{Mertsch}}, \bibinfo{author}{S.~{Funk}},
\newblock \bibinfo{title}{{Solution to the Cosmic Ray Anisotropy Problem}},
\newblock \bibinfo{journal}{\prl} \bibinfo{volume}{114} (\bibinfo{year}{2015})
  \bibinfo{pages}{021101}.
\bibitem[{{Giacinti} and {Sigl}(2012)}]{Giacinti:2012aug}
\bibinfo{author}{G.~{Giacinti}}, \bibinfo{author}{G.~{Sigl}},
\newblock \bibinfo{title}{{Local Magnetic Turbulence and TeV-PeV Cosmic Ray
  Anisotropies}},
\newblock \bibinfo{journal}{Phys. Rev. Lett.} \bibinfo{volume}{109}
  (\bibinfo{year}{2012}) \bibinfo{pages}{071101}.
\bibitem[{{Biermann} et~al.(2013)}]{Biermann:2013may}
\bibinfo{author}{P.~L. {Biermann}}, et~al.,
\newblock \bibinfo{title}{{Cosmic-Ray Transport and Anisotropies}},
\newblock \bibinfo{journal}{Astrophys. J.} \bibinfo{volume}{768}
  (\bibinfo{year}{2013}) \bibinfo{pages}{124}.
\bibitem[{{Ahlers}(2014)}]{Ahlers:2014jan}
\bibinfo{author}{M.~{Ahlers}},
\newblock \bibinfo{title}{{Anomalous Anisotropies of Cosmic Rays from Turbulent
  Magnetic Fields}},
\newblock \bibinfo{journal}{Phys. Rev. Lett.} \bibinfo{volume}{112}
  (\bibinfo{year}{2014}) \bibinfo{pages}{021101}.
\bibitem[{Ahlers and Mertsch(2015)}]{Ahlers:2015dwa}
\bibinfo{author}{M.~Ahlers}, \bibinfo{author}{P.~Mertsch},
\newblock \bibinfo{title}{{Small-Scale Anisotropies of Cosmic Rays from
  Relative Diffusion}},
\newblock \bibinfo{journal}{{Astrophys. J. Lett.}} \bibinfo{volume}{815}
  (\bibinfo{year}{2015}) \bibinfo{pages}{L2}.
\bibitem[{L\'{o}pez-Barquero et~al.(2016)L\'{o}pez-Barquero, Farber, Xu,
  Desiati, and Lazarian}]{Lopez-Barquero:2015qpa}
\bibinfo{author}{V.~L\'{o}pez-Barquero}, \bibinfo{author}{R.~Farber},
  \bibinfo{author}{S.~Xu}, \bibinfo{author}{P.~Desiati},
  \bibinfo{author}{A.~Lazarian},
\newblock \bibinfo{title}{{Cosmic Ray Small Scale Anisotropies and Local
  Turbulent Magnetic Fields}},
\newblock \bibinfo{journal}{Astrophys. J.} \bibinfo{volume}{830}
  (\bibinfo{year}{2016}) \bibinfo{pages}{19}.
\bibitem[{Harding et~al.(2016)Harding, Fryer, and Mendel}]{Harding:2015pna}
\bibinfo{author}{J.~P. Harding}, \bibinfo{author}{C.~L. Fryer},
  \bibinfo{author}{S.~Mendel},
\newblock \bibinfo{title}{{Explaining TeV Cosmic-Ray Anisotropies with
  Non-Diffusive Cosmic-Ray Propagation}},
\newblock \bibinfo{journal}{Astrophys. J.} \bibinfo{volume}{822}
  (\bibinfo{year}{2016}) \bibinfo{pages}{102}.
\bibitem[{Scherer et~al.(2016)Scherer, Strauss, Ferreira, and
  Fichtner}]{Scherer:2016uyr}
\bibinfo{author}{K.~Scherer}, \bibinfo{author}{R.~D. Strauss},
  \bibinfo{author}{S.~E.~S. Ferreira}, \bibinfo{author}{H.~Fichtner},
\newblock \bibinfo{title}{{Comic ray flux anisotropies caused by
  astrospheres}},
\newblock \bibinfo{journal}{Astropart. Phys.} \bibinfo{volume}{82}
  (\bibinfo{year}{2016}) \bibinfo{pages}{93--98}.
\bibitem[{Ahlers(2016)}]{Ahlers:2016njd}
\bibinfo{author}{M.~Ahlers},
\newblock \bibinfo{title}{{Deciphering the Dipole Anisotropy of Galactic Cosmic
  Rays}},
\newblock \bibinfo{journal}{Phys. Rev. Lett.} \bibinfo{volume}{117}
  (\bibinfo{year}{2016}) \bibinfo{pages}{151103}.
\bibitem[{Ahlers et~al.(2016)}]{Ahlers:2016njl}
\bibinfo{author}{M.~Ahlers}, et~al.,
\newblock \bibinfo{title}{{A New Maximum-Likelihood Technique for
  Reconstructing Cosmic-Ray Anisotropy at All Angular Scales}},
\newblock \bibinfo{journal}{Astrophys. J.} \bibinfo{volume}{823}
  (\bibinfo{year}{2016}) \bibinfo{pages}{10}.
\bibitem[{{Desiati} and {Lazarian}(2013)}]{Desiati:2013jan}
\bibinfo{author}{P.~{Desiati}}, \bibinfo{author}{A.~{Lazarian}},
\newblock \bibinfo{title}{{Anisotropy of TeV Cosmic Rays and Outer Heliospheric
  Boundaries}},
\newblock \bibinfo{journal}{Astrophys. J.} \bibinfo{volume}{762}
  (\bibinfo{year}{2013}) \bibinfo{pages}{44}.
\bibitem[{{Schwadron} et~al.(2014)}]{IBEX:2014feb}
\bibinfo{author}{N.~A. {Schwadron}}, et~al.,
\newblock \bibinfo{title}{{Global Anisotropies in TeV Cosmic Rays Related to
  the Sun's Local Galactic Environment from IBEX}},
\newblock \bibinfo{journal}{Science} \bibinfo{volume}{343}
  (\bibinfo{year}{2014}) \bibinfo{pages}{988--990}.
\bibitem[{{Zhang} et~al.(2014){Zhang}, {Zuo}, and {Pogorelov}}]{Zhang:2014jul}
\bibinfo{author}{M.~{Zhang}}, \bibinfo{author}{P.~{Zuo}},
  \bibinfo{author}{N.~{Pogorelov}},
\newblock \bibinfo{title}{{Heliospheric Influence on the Anisotropy of TeV
  Cosmic Rays}},
\newblock \bibinfo{journal}{Astrophys. J.} \bibinfo{volume}{790}
  (\bibinfo{year}{2014}) \bibinfo{pages}{5}.
\bibitem[{L\'{o}pez-Barquero et~al.(2016)}]{Lopez-Barquero:2016wnt}
\bibinfo{author}{V.~L\'{o}pez-Barquero}, et~al.,
\newblock \bibinfo{title}{{TeV Cosmic-Ray Anisotropy from the Magnetic Field at
  the Heliospheric Boundary}},
\newblock \bibinfo{journal}{arXiv eprints}  (\bibinfo{year}{2016})
  \bibinfo{pages}{astro--ph/1610.03097}.
\bibitem[{{Lazarian} and {Desiati}(2010)}]{Lazarian:2010oct}
\bibinfo{author}{A.~{Lazarian}}, \bibinfo{author}{P.~{Desiati}},
\newblock \bibinfo{title}{{Magnetic Reconnection as the Cause of Cosmic Ray
  Excess from the Heliospheric Tail}},
\newblock \bibinfo{journal}{Astrophys. J.} \bibinfo{volume}{722}
  (\bibinfo{year}{2010}) \bibinfo{pages}{188--196}.
\bibitem[{{Desiati} and {Lazarian}(2012)}]{Desiati:2012jun}
\bibinfo{author}{P.~{Desiati}}, \bibinfo{author}{A.~{Lazarian}},
\newblock \bibinfo{title}{{Cosmic rays and stochastic magnetic reconnection in
  the heliotail}},
\newblock \bibinfo{journal}{Nonlinear Processes in Geophysics}
  \bibinfo{volume}{19} (\bibinfo{year}{2012}) \bibinfo{pages}{351--364}.
\bibitem[{Aartsen et~al.(2014)}]{Aartsen:2014njl}
\bibinfo{author}{M.~G. Aartsen}, et~al.,
\newblock \bibinfo{title}{{IceCube-Gen2: A Vision for the Future of Neutrino
  Astronomy in Antarctica}},
\newblock \bibinfo{journal}{arXiv eprints}  (\bibinfo{year}{2014})
  \bibinfo{pages}{astro--ph/1412.5106}.

\end{thebibliography}







\end{document}